\documentclass[aps,prb,groupedaddress,twocolumn,showpacs]{revtex4}
\usepackage{graphicx,psfrag,amsmath,calc}

\begin{document}

\title{Uniform spin susceptibility tensor 
and quasiparticle density of states \\
in organic quasi-one-dimensional superconductors}

\author{R. D. Duncan}
\author{R. W. Cherng$^\dagger$}
\author{C. A. R. S\'{a} de Melo}
\affiliation{School of Physics, Georgia Institute of Technology,
             Atlanta Georgia 30332}

\date{\today}

\begin{abstract}
We perform calculations relating the order parameter symmetry 
of organic quasi-one-dimensional superconductors 
to the bulk quasiparticle density of 
states and the bulk uniform spin susceptibility tensor at finite 
temperatures.
Current experimental results suggest that some organic 
quasi-one-dimensional superconductors may exhibit
triplet pairing symmetry. The purpose of our analysis
is to attempt to narrow down the number of possibilities 
for the symmetry of the order parameter based on
the current experimental evidence. 
\end{abstract}

\maketitle

%
%
%
%
%
%
%
%
%
%

%
%
\section{Introduction}
\label{introduction}

After the discovery of quasi-one-dimensional organic
superconductors~\cite{jerome-80}, many interesting aspects
were studied~\cite{ishiguro-89,williams-92}. 
The magnetic field versus temperature phase diagram 
for ${ \rm (TMTSF)_2 PF_6}$ under pressure of 6 kbar was recently 
revisited by Lee {\it et.al.}~\cite{lee-97}. 
They found that the upper critical fields along the usual 
${\bf a}$, ${\bf b^{\prime}}$, and ${\bf c}^*$ directions were 
highly anisotropic. The upper critical field along
$ {\bf a}$ and ${\bf b^{\prime}}$ displayed a strong positive curvature 
at lower temperatures.
Furthermore, the slope $[ - dH_{c_2}/dT ]_{T_c}$ for 
${\bf H} \parallel {\bf b^{\prime}}$ was smaller than along 
the ${\bf a}$ direction,
i.e., $H_{c_2}^{({\bf a})} > H_{c_2}^{({\bf b})}$ at high temperature.
While, at low temperatures 
$H_{c_2}^{({\bf b})}$ exceeded $H_{c_2}^{({\bf a})}$, 
after an unusual anisotropy 
inversion at $H^{*} = 1.6 T$. 
These initial results suggested the existence of triplet 
superconductivity in this system.  
An additional boost for the triplet scenario in ${\rm (TMTSF)_2 PF_6}$ 
was given by very recent NMR experiments~\cite{lee-00,lee-02}. 
Lee {\it et. al.}~\cite{lee-00,lee-02} found that
there is {\bf no} $^{77} {\rm Se}$ Knight shift 
in ${ \rm (TMTSF)_2 PF_6 } $ for fields 
${\bf H} \parallel {\bf b}^{\prime}$
($P \approx 6~{\rm kbar}$)~\cite{lee-00}, and 
${\bf H} \parallel {\bf a}$
($P \approx 7~{\rm kbar}$)~\cite{lee-02}.
These results suggest that
$\chi_{b^\prime} \approx \chi_N$, and $\chi_a \approx \chi_N$, 
where $\chi_N$ is the normal 
state susceptibility.
Furthermore, a sharp and narrow (possibly Hebel-Slichter~\cite{hebel-59}) 
peak was observed just below $T_c$ in the 
$^{77} {\rm Se}$ NMR relaxation rate $1/T_1$,
for ${\bf H} \parallel {\bf a}$ $( H = 1.43~{\rm T} )$,
and $P \approx 7~{\rm kbar}$~\cite{lee-02}. 
The combined work of 
Lee {\it et. al.}~\cite{lee-97,lee-00,lee-02} 
suggests the existence of a triplet superconducting phase in 
${ \rm (TMTSF)_2 PF_6 } $. 
However, these recent $^{77} {\rm Se}$ NMR results 
in ${ \rm (TMTSF)_2 PF_6 } $  
should be contrasted with earlier proton NMR results in
${ \rm (TMTSF)_2 ClO_4 }$~\cite{takigawa-87} which indicated the
absence of the coherence peak just below $T_c = 1.03~{\rm K}$ at $H = 0$, 
and a $T^3$ behavior between $T_c/2$ and $T_c$. Based on their experiments,
Takigawa {\it et. al.}~\cite{takigawa-87} argued that 
the superconducting state of ${ \rm (TMTSF)_2 ClO_4 }$ 
has an anisotropic order parameter vanishing along lines on 
the Fermi surface, although they were not able to distinguish
between singlet and triplet states.

The temperature versus magnetic field 
phase diagram of $ {\rm (TMTSF)_2 ClO_4} $ at ambient pressure
was also measured by Lee {\it et. al.}~{\cite{lee-94,lee-95}, for 
${\bf H} \parallel {\bf b}^{\prime}$. 
Their results showed
that the Pauli paramagnetic limit is also exceeded in this compound.
Furthermore, Belin and Behnia~\cite{belin-97} (BB) reported 
measurements of the thermal conductivity in the superconducting state of 
$ {\rm (TMTSF)_2 ClO_4} $, indicating that their data is inconsistent 
with the existence of gap nodes at the Fermi surface as suggested 
by Takigawa {\it et. al.}~\cite{takigawa-87}. BB's argument was that
the $T^3$ behavior of the proton $1/T_1$~\cite{takigawa-87}
was limited only to $T > T_c/2$. Thus, it could not be 
considered convincing
evidence for nodes in the gap because the temperature was not low enough.
Even for conventionally gapped superconductors, the exponential behavior
of $1/T_1$ occurs only at very low temperatures $(T \ll T_c)$.
Therefore, these recent experimental results 
combined~\cite{lee-94,lee-95,belin-97} seem to suggest the existence
of a fully gapped triplet superconducting state 
in ${\rm (TMTSF)_2 ClO_4}$. However, detailed $^{77} {\rm Se}$ NMR 
experiments seem to be lacking for ${\rm (TMTSF)_2 ClO_4}$.

Since both of these materials belong to the same class of compounds and 
have very similar crystal structures, it is feasible that 
they may exhibit similar order parameter symmetries. 
However, additional experiments are necessary in order to make sure
that the order parameter symmetry of the 
$ {\rm (TMTSF)_2 ClO_4} $ is the same
as in $ {\rm (TMTSF)_2 PF_6} $. 
A minimal set of experiments are necessary
to identify the symmetry of the order parameter.  
For instance, data from upper critical field 
${\bf H} \parallel {\bf a}$, $^{77} {\rm Se}$ 
Knight shift and NMR relaxation
experiments are needed for ClO$_4$, 
and information about the thermal conductivity is
needed for PF$_6$.
At zero magnetic field, however, these two materials have very different
phase diagrams.
The compound 
$ {\rm (TMTSF)_2 ClO_4} $ is superconducting at low temperatures and
zero pressure when slowly cooled (R-state), while $ {\rm (TMTSF)_2 PF_6} $ 
is a spin density wave (SDW) 
insulator at low temperatures and low pressures, and 
becomes superconducting at low temperatures only at pressures 
above $5.5~{\rm kbar}$. Although the similar crystal structure
of these systems suggests, from a simple group theoretical point of view, 
that the origin of the pair interaction is the same, the role of the
proximity to an SDW phase in $ {\rm (TMTSF)_2 PF_6} $ 
needs to be investigated both theoretically and experimentally. 
Furthermore, the true nature of the order parameter symmetry can only
be directly evidenced via phase sensitive experiments like those 
performed in the cuprate oxides~\cite{harlingen-93,wellstood-95}.

Abrikosov was the first to suggest the possibility of
triplet superconductivity 
in Bechgaard salts~\cite{abrikosov-83a,abrikosov-83b}
based on the behavior of ${\rm (TMTSF)_2 ClO_4}$ and
${\rm (TMTSF)_2 PF_6}$ under X-ray bombardment. These systems
exhibited a strong suppression of their critical temperature
in the presence of radiation induced non-magnetic 
defects~\cite{choi-82,bouffard-82}.  
Gorkov and Jerome~\cite{gorkov-85} also suggested the possibility
of triplet superconductivity based on a theoretical extrapolation of
the semiclassical upper critical fields calculated near the zero field
critical temperature. In addition, Lebed~\cite{lebed-86} pointed out
that triplet superconductivity would manifest itself in Bechgaard salts
through a remarkable reentrant phase in high magnetic fields. Later, 
Dupuis, Montambaux and S\'a de Melo (DMS)~\cite{dupuis-93} 
studied the field-versus-temperature phase diagram and the vortex lattice
structure of these systems and concluded that these systems could be either
in a inhomogeneous singlet state like the Larkin-Ovchinikov-Fulde-Ferrel
(LOFF) state~\cite{larkin-65,fulde-64} or in a triplet state. 
Further studies of the LOFF state were performed 
by Dupuis~\cite{dupuis-95}, while an equal spin triplet pairing state
(ESTP) was discussed further by S\'a de Melo~\cite{sademelo-96}.

After the upper critical field measurements 
of Lee {\it et. al.}~\cite{lee-97}, Lebed~\cite{lebed-99} was able to show
that the LOFF state was Pauli paramagnetically limited, thus
giving further support for the triplet scenario (at least at high magnetic
fields). The possibility of a magnetic field induced 
singlet (low fields) to triplet (high fields) 
transition was considered by S\'a de Melo~\cite{sademelo-98,sademelo-99} 
and Vaccarella and S\'a de Melo~\cite{vaccarella-00},
but currently there is no experimental evidence of a kink in the 
upper critical field of these systems. Furthermore, reentrant
superconductivity was still expected for the triplet state
at high magnetic fields~\cite{lebed-99,vaccarella-00}. 
All these previous theories neglected the effects of fluctuations 
at high magnetic fields. A first attempt to incorporate fluctuation 
effects was made recently by Vaccarella and 
S\'a de Melo~\cite{vaccarella-01} who showed
that phase fluctuations can suppress the reentrant phase at high
magnetic fields in the ESTP state.
Very recently Lebed, Machida and Ozaki (LMO)~\cite{lebed-00} suggested the 
possibility of a triplet phase with a ${\bf d}$-vector with zero-component
along the ${\bf b}^\prime$ axis and a finite component along the
${\bf a}$ axis. A direct consequence of LMO's proposal is the
existence of an anisotropic  
spin susceptibility at zero temperature and low magnetic fields:
$\chi_{b^\prime} = \chi_N$, and $\chi_{a} \ll \chi_N$,
where $\chi_{a}$ corresponds to ${\bf H} \parallel {\bf a}$, 
$\chi_{b^\prime}$ corresponds to ${\bf H} \parallel {\bf b}^\prime$,  
and $\chi_N$ corresponds to the normal state susceptibility.
However, Lee {\it et. al.}~\cite{lee-00,lee-02} reported new $^{77}{\rm Se}$
NMR (Knight shift) results for ${\rm (TMTSF)_2 PF_6}$ under pressure, 
indicating that $\chi_{b^\prime} \approx \chi_N$, 
and $\chi_{a} \approx \chi_N$. A fully gapped singlet ``d-wave''
order parameter for ${\rm (TMTSF)_2 ClO_4}$ was proposed 
by Shimahara~\cite{shimahara-00}, while gapless 
triplet ``f-wave'' superconductivity for ${\rm (TMTSF)_2 PF_6}$
was proposed by Kuroki, Arita, and Aoki (KAA)~\cite{kuroki-01}.
Duncan, Vaccarella, and S\'a de Melo (DVS)~\cite{duncan-01}
performed a group theoretical analysis and suggested that
a weak spin-orbit fully gapped triplet ``$p_x$-wave'' order
parameter would be a good candidate for superconductivity in 
Bechgaard salts. For this symmetry $\chi_{a} \approx \chi_N$, and
$\chi_{b^\prime} \approx \chi_N$~\cite{sademelo-98,sademelo-99,duncan-01}
in agreement with experimental results~\cite{lee-00,lee-02}.
In addition, this order parameter choice is consistent with the
expectation of weak atomic spin-orbit effects, given that the heaviest
element in these systems is ${\rm Se}$.

This paper is a longer version of the brief report by DVS~\cite{duncan-01}.
Here, we are concerned mostly with the symmetry of the order parameter 
of a triplet quasi-one-dimensional superconductor at zero magnetic field,
and have three main goals.
First, we perform a group theoretical analysis of the 
possible symmetries
of the order parameter for an orthorhombic quasi-one-dimensional 
superconductor. Second, we calculate
the temperature dependence of the bulk quasiparticle density of states 
and of 
the bulk uniform spin susceptibility tensor for various
candidate symmetries of the order parameter consistent with the
group theoretical analysis. Third, we make connections to 
scanning tunneling microscopy (STM)
of quasiparticle density of states and Knight shift measurements of 
the spin susceptibility tensor. 
Thus, the remainder of the paper is organized as follows.
In section \ref{sec:Hamiltonian}, 
we discuss the Hamiltonian used for weak spin-orbit 
(strong spin-orbit) coupling cases for generic singlet (pseudo-singlet)
and triplet (pseudo-triplet) states.
In section \ref{sec:symmetries}, we perform a group theoretical 
analysis of the symmetry of the order parameter, both in the
weak and strong spin-orbit coupling limits.
In section \ref{sec:qdos}, we discuss the quasiparticle
density of states for different symmetries of the order parameter.
In addition, in section \ref{sec:susceptibility},
we analyse the uniform spin susceptibility tensor $\chi_{mn}$
for several order parameter symmetries, and we discuss
finite magnetic field effects and the role of vortices on $\chi_{mn}$.
Finally, we summarize our results in section \ref{sec:summary}. 

\section{Hamiltonian}
\label{sec:Hamiltonian}

In order to analyze the possible different symmetries of the order parameter
we study quasi-one-dimensional 
systems with a single band in an orthorhombic
lattice and allow for singlet or triplet pairing. 
We consider the following 
dispersion relation
\begin{equation}
\label{eqn:dispersion}
\epsilon_{\bf k} = -|t_x| \cos(k_x a) - 
|t_y| \cos(k_y b) - |t_z| \cos(k_z c),
\end{equation}
where $|{t_x}| \gg |{t_y}| \gg |{t_z}|$.
Note that this notation is slightly different from 
the standard notation~\cite{ishiguro-89}. 
The corresponding translation
is $|t_x| \to 2 t_a$; $|t_y| \to 2 t_b$; $|t_z| \to 2 t_c$.
Furthermore, $a$, $b$ and $c$ in our notation correspond to the
unit cell lengths along the crystallographic directions
${\bf a}$, ${\bf b}^{\prime}$, and ${\bf c}^*$ respectively.
In the limit of weak interactions and low
densities these quasi-one-dimensional
systems exhibit a well defined Fermi surface which is open, being formed
of two separate sheets which intersect the Brillouin zone boundaries in
the $k_y$ and $k_z$ directions.
We work with the Hamiltonian
\begin{equation}
\label{eqn:hamiltonian}
H = H_{kin} + H_{int},
\end{equation}
where the kinetic energy part is
\begin{equation}
\label{eqn:hkin}
H_{kin} = \sum_{{\bf k},\alpha} (\epsilon_{\bf k} - \mu) 
\psi_{{\bf k}, \alpha}^{\dagger} \psi_{{\bf k}, \alpha},
\end{equation}
where $\epsilon_{\bf k}$ is the dispersion defined in
Eq.~\ref{eqn:dispersion},
$\mu$ is the chemical potential,
and 
\begin{equation}
\label{eqn:hint}
H_{int} = 
\dfrac{1}{2} \sum_{{\bf k} {\bf k^{\prime}} {\bf q}} 
\sum_{\alpha \beta \gamma \delta}
V_{\alpha \beta \gamma \delta} ({\bf k}, {\bf k^{\prime}})
b_{\alpha \beta}^{\dagger} ({\bf k}, {\bf q})
b_{\gamma \delta} ({\bf k^{\prime}}, {\bf q})
\end{equation}
is the interaction part with
\begin{equation}
b_{\alpha \beta}^{\dagger} ({\bf k}, {\bf q}) = 
\psi_{-{\bf k} + {\bf q}/2, \alpha}^{\dagger}
\psi_{{\bf k} + {\bf q}/2, \beta}^{\dagger},
\end{equation}
where the labels $\alpha$, $\beta$, $\gamma$ and $\delta$ 
are spin indices and 
the labels ${\bf k}$, ${\bf k}^{\prime}$ and ${\bf q}$ represent 
linear momenta. 

In the case of weak spin-orbit coupling
and triplet pairing $(S = 1)$, 
the model interaction tensor can be chosen to be
\begin{equation}
V_{\alpha \beta \gamma \delta} ({\bf k}, {\bf k^{\prime}})
= \Gamma_{\alpha \beta \gamma \delta} 
V_{\Gamma} ({\bf k}, {\bf k^{\prime}})
\phi_{\Gamma} ({\bf k}) \phi^{*}_{\Gamma} ({\bf k^{\prime}}),
\end{equation}
where 
$\Gamma_{\alpha \beta \gamma \delta} = {\bf v}_{\alpha \beta} \cdot
{\bf v}_{\gamma \delta}^{\dagger}/2$ with 
${\rm v}_{\alpha \beta} = (i\sigma \sigma_y)_{\alpha \beta}$. In addition,
the interaction $V_{\Gamma}$ corresponds to the irreducible 
representation $\Gamma$ with basis function
$\phi_{\Gamma} ({\bf k})$ representative of the orthorhombic group. 
In the case of strong spin-orbit coupling and $S = 1$ the interaction 
\begin{equation}
\label{eqn:int-strong}
V_{\alpha \beta \gamma \delta} ({\bf k}, {\bf k^{\prime}})
= 
V_{\Gamma} ({\bf k}, {\bf k^{\prime}})
\left[
\Phi_{\Gamma} ({\bf k}) \cdot {\bf v}_{\alpha \beta}
\right]
\left[
\Phi^{*}_{\Gamma} ({\bf k^{\prime}}) \cdot {\bf v}_{\gamma \delta}^{\dagger}
\right]/2
\end{equation}
where the interaction $V_{\Gamma}$ corresponds to the irreducible 
representation $\Gamma$ with basis function vector
$\Phi_{\Gamma} ({\bf k})$ representative of the orthorhombic group. 
In both weak and strong spin-orbit coupling, 
we use either the equation of motion method~\cite{mineev-99}
or the functional integration method~\cite{sdm-93,sdm-97} 
in the zero center of mass momentum pairing approximation 
(which corresponds to the BCS limit in weak coupling)
to obtain 
the anomalous Green's function
\begin{equation}
\label{eqn:anom-green}
F_{\alpha \beta} ({\bf k}, i\omega_n )  = 
\dfrac{\Delta_{\alpha \beta}({\bf k})} 
{\omega_n^2 + E_{\bf k}^2 },
\end{equation}
and the single particle Green's function
\begin{equation}
\label{eqn:sing-green}
G_{\alpha \beta} ({\bf k}, i\omega_n )  = 
- \dfrac{ i\omega_n + \xi_{\bf k}}{ \omega_n^2 + E_{\bf k}^2 }
\delta_{\alpha \beta},
\end{equation}
where $\xi_{\bf k} = \epsilon_{\bf k} - \mu$
is the dispersion 
$\epsilon_{\bf k}$ relative to the chemical potential $\mu$, 
\begin{equation}
E_{\bf k} = \sqrt{ \xi_{\bf k}^2 + \Delta_{\bf k}^2  }
\end{equation}
is the quasiparticle excitation energy, and 
\begin{equation}
\Delta_{\bf k}^2 \equiv 
{\it Tr} \left[
\tilde\Delta^{\dagger} ({\bf k}) \tilde \Delta ({\bf k})
\right]
/2,
\end{equation}
where the order parameter matrix $\tilde \Delta ({\bf k})$ has elements
$\Delta_{\alpha \beta} ({\bf k})$. The expressions for the 
anomalous (Eq.~\ref{eqn:anom-green}) and for the 
single particle (Eq.~\ref{eqn:sing-green}) Green's functions
are valid only in the unitary case where 
$\tilde\Delta^{\dagger} ({\bf k}) \tilde \Delta ({\bf k})$
is diagonal. 
In this initial work, we do not discuss in detail the non-unitary 
case for triplet superconductivity.
%
%
Using the single particle and anomalous Green's functions defined
above and standard many body methods~\cite{mineev-99,sdm-93,sdm-97},
we derive the order parameter equation in terms of the anomalous 
Green's function as
\begin{equation}
\label{eqn:order-parameter}
\Delta_{\alpha \beta} ({\bf k}) = - T \sum_{ \omega_n, {\bf k}^{\prime} }
V_{\beta \alpha \gamma \delta} ({\bf k}, {\bf k^{\prime}})
F_{\gamma \delta} ({\bf k}^{\prime}),
\end{equation}
while the number equation, that fixes the chemical potential $\mu$, 
can be written as 
\begin{equation}
\label{eqn:number-0}
N = - T \sum_{\omega_n, {\bf k} } 
{\it Tr} 
\left[ 
\tilde G ({\bf k}, i\omega_n)
\dfrac{\partial}{ \partial \mu } \tilde G^{-1} ({\bf k}, i\omega_n)
\right].
\end{equation}
The matrix $\tilde G ({\bf k}, i\omega_n)$ is the $2 \times 2$ block
matrix
\begin{equation}
\tilde G ({\bf k}, i\omega_n) = 
\left(
\begin{array}{cc}
G_{\alpha \beta} ({\bf k}, i\omega_n ) & 
- F_{\alpha \beta} ({\bf k}, i\omega_n ) \\
- F_{\alpha \beta}^{\dagger} ({\bf k}, i\omega_n ) & 
-G_{\beta \alpha} (-{\bf k}, - i\omega_n ) \\ 
\end{array}
\right), 
\end{equation}
with matrix elements which are in turn $2 \times 2$ matrices in spin space.
Performing the Matsubara sums and traces in Eq.~(\ref{eqn:order-parameter})
and Eq.~(\ref{eqn:number-0}) we obtain the more familiar forms
\begin{widetext}
\begin{equation}
\label{eqn:oparam}
\Delta_{\alpha \beta} ({\bf k}) = - \sum_{ {\bf k}^{\prime} }
V_{\beta \alpha \gamma \delta} ({\bf k}, {\bf k^{\prime}})
\dfrac{ \Delta_{\gamma \delta} ({\bf k^{\prime}})}
      { 2 E_{ \bf k^{\prime} } }
\tanh \left( \dfrac{ E_{\bf k^{\prime} } } { 2 T } \right), 
\end{equation}
\begin{equation}
\label{eqn:number}
N \equiv \sum_{ {\bf k} } n_{\bf k} = 
\sum_{ {\bf k} }  
\left[
\left(
1 + \dfrac{ \xi_{\bf k}}{ E_{\bf k} } 
\right) f ( E_{\bf k} )
+
\left(
1 - \dfrac{ \xi_{\bf k}}{ E_{\bf k} } 
\right) 
\left(
1 - f ( E_{\bf k} )
\right)
\right],
\end{equation}
\end{widetext}
for the order parameter and number equations, respectively. 
These two equations must be solved self-consistently away from the
strict BCS limit in order to accommodate particle-hole asymmetries
and strong coupling effects which tend to shift the chemical potential
substantially away from the Fermi energy. Quite generally these
two equations are correct even in the strong coupling (or low density)
regime provided that $T \ll T_c$~\cite{sdm-93,sdm-97}.
%
%

%
\begin{table*}
\caption{Character table for the ${\rm D_{2h}}$ point group.
\label{tab:d2h}
}
\begin{ruledtabular}
\begin{tabular}{cccccccccc}
Representation & $E$ & $C_{2}^{z}$ & $C_{2}^{y}$ & $C_{2}^{x}$ &
$i$ & $i C_{2}^{z}$ & $i C_{2}^{y}$ & $i C_{2}^{x}$ & Basis \\
\hline
$A_{1g}$ & 1 &  1 & 1 & 1 & 1 & 1 & 1 & 1 & 1 \\
$B_{1g}$ & 1 &  1 & -1 & -1 & 1 & 1 & -1 & -1 & XY \\
$B_{2g}$ & 1 & -1 & 1 & -1 & 1 & -1 & 1 & -1 & XZ\\
$B_{3g}$ & 1 & -1 & -1 & 1 & 1 & -1 & -1 & 1 & YZ \\
\hline
\hline
$A_{1u}$ & 1 &  1 & 1 & 1 & -1 & -1 & -1 & -1 & XYZ \\
$B_{1u}$ & 1 &  1 & -1 & -1 & -1 & -1 & 1 & 1 & Z \\
$B_{2u}$ & 1 & -1 & 1 & -1 & -1 & 1 & -1 & 1 & Y \\
$B_{3u}$ & 1 & -1 & -1 & 1 & -1 & 1 & 1 & -1 & X \\
\end{tabular}
\end{ruledtabular}
\end{table*}
\begin{table*}
\caption{Time reversal invariant singlet states in an 
orthorhombic crystal, assuming weak spin-orbit coupling.
\label{tab:singlet}
}
\begin{ruledtabular}
\begin{tabular}{ccccc}
State & Residual Group & Order parameter $\Delta_{si} ({\bf k})$ & 
$E_{\bf k} = 0$ $(\tilde\mu >0)$ & $E_{\bf k} = 0$ $(\tilde\mu <0)$ \\
\hline
$^1A_{1g}$ & $SO_3 \times D_{2h} \times T              $ &  $ 1  $ 
& none  & none \\
$^1B_{1g}$ & $SO_3 \times D_2(C_2^z) \times I \times T $ &  $XY  $ 
& lines  & none \\
$^1B_{2g}$ & $SO_3 \times D_2(C_2^y) \times I \times T $ &  $XZ  $ 
& lines & none \\
$^1B_{3g}$ & $SO_3 \times D_2(C_2^y) \times I \times T $ &  $YZ  $ 
& lines  & none \\
\end{tabular}
\end{ruledtabular}
\end{table*}
\begin{table*}
\caption{Time reversal breaking triplet states in an 
orthorhombic crystal, assuming weak spin-orbit coupling.
\label{tab:triplet1}
}
\begin{ruledtabular}
\begin{tabular}{ccccc}
State & Residual Group & Order parameter ${\bf d} ({\bf k})$ 
& $E_{ {\bf k}\pm} = 0$ $(\tilde\mu > 0)$ & 
$E_{{\bf k}\pm} = 0$ $(\tilde\mu < 0)$ \\
\hline
$^3A_{1u}(b)$ & $D_{\infty} (E)\times D_2 \times I(E)        $ 
& $(1,i,0)XYZ$ & surface (-), lines (+)  & none \\
$^3B_{1u}(b)$ & $D_{\infty} (E)\times D_2(C_2^z) \times I(E) $
& $(1,i,0)Z$   & surface (-), lines (+)   & none \\
$^3B_{2u}(b)$ & $D_{\infty} (E)\times D_2(C_2^y) \times I(E) $
& $(1,i,0)Y$   & surface (-), lines (+)  & none \\
$^3B_{3u}(b)$ & $D_{\infty} (E)\times D_2(C_2^x) \times I(E) $
& $(1,i,0)X$   & surface (-), none (+)  & none \\
\end{tabular}
\end{ruledtabular}
\end{table*}
\begin{table*}
\caption{Time reversal invariant triplet states in an 
orthorhombic crystal, assuming weak spin-orbit coupling.
\label{tab:triplet2}
}
\begin{ruledtabular}
\begin{tabular}{ccccc}
State & Residual Group & Order parameter ${\bf d} ({\bf k})$ 
& $E_{\bf k} = 0$ $(\tilde\mu >0)$ & $E_{\bf k} = 0$ $(\tilde\mu <0)$ \\
\hline
$^3A_{1u}(a)$ & 
$D_{\infty} (C_{\infty})\times D_2 \times I(E) \times T $ 
& $(0,0,1)XYZ$ & lines & none \\
$^3B_{1u}(a)$ & 
$D_{\infty} (C_{\infty})\times D_2(C_2^z)\times I(E)\times T$ 
& $(0,0,1)Z$   & lines & none \\
$^3B_{2u}(a)$ & 
$D_{\infty} (C_{\infty})\times D_2(C_2^y) \times I(E) \times T$ 
& $(0,0,1)Y$   & lines & none \\
$^3B_{3u}(a)$ & 
$D_{\infty} (C_{\infty})\times D_2(C_2^x) \times I(E) \times T$ 
& $(0,0,1)X$& none & none \\
\end{tabular}
\end{ruledtabular}
\end{table*}
\begin{table*}[ht!]
\caption{Time reversal invariant triplet states in an 
orthorhombic crystal, assuming strong spin-orbit coupling.
\label{tab:triplet3}
}
\begin{ruledtabular}
\begin{tabular}{ccccc}
State & Residual Group & Order parameter ${\bf d} ({\bf k})$ & 
$E_{\bf k} = 0$ $(\tilde\mu >0)$ & $E_{\bf k} = 0$ $(\tilde\mu <0)$ \\
\hline
$A_{1u}$ & 
$D_2 \times I(E) \times T        $ & $(AX,BY,CZ)$   
& none, points, lines & none \\
$B_{1u}$ & $D_2(C_2^z) \times I(E) \times T $ & $(AY,BX,CXYZ)$ 
& none, lines & none \\
$B_{2u}$ & $D_2(C_2^y) \times I(E) \times T $ & $(AZ,BXYZ,CX)$
& none, lines & none \\
$B_{3u}$ & $D_2(C_2^x) \times I(E) \times T $ & $(AXYZ,BZ,CY)$ 
& points, lines & none \\
\end{tabular}
\end{ruledtabular}
\end{table*}

\section{Possible symmetries for the order parameter}
\label{sec:symmetries}

We will now discuss the symmetry of the order parameter
$\Delta_{\alpha \beta}$. For this purpose, we consider
an orthorhombic crystal~\cite{triclinic}
with a conventional symmetry normal state, i.e.,
we assume that the normal state does not break 
the full lattice symmetry. By contrast, normal states
that break full lattice symmetry are spatially inhomogeneous. Examples
of such spatially inhomogeneous states are stripe phases that occur
in high-$T_c$ superconductors, magnetic normal phases
(ferro and antiferromagnetic), and normal states in a magnetic field
(where the magnetic translation group needs to be incorporated).
These states, however, will not be considered in the analysis 
performed in this section.

In the case of weak spin-orbit coupling, the most general conventional
symmetry group for the normal state is 
${\rm G_n = SO(3)\times G_c \times U(1) \times T}$, 
where ${\rm SO(3)}$ is the group of rotations in spin space, 
${\rm G_c}$ is the crystal space group, ${\rm U(1)}$ is the gauge group
and ${\rm T}$ corresponds to time reversal symmetry.
Here, rotations of the spin and spatial degrees of freedom are independent. 
Since the superconducting state breaks ${\rm U(1)}$ symmetry, 
the possible order parameters 
correspond to different irreducible representations
of ${\rm G_s = SO(3) \times G_c}$.
However, in the case of strong spin-orbit coupling 
the normal state symmetry group is
${\rm G_n = G_c^{(J)} \times U(1) \times T}$, where now 
${\rm G_c^{(J)}}$ is identical to the the space group 
${\rm G_c}$ except when 
a rotation is involved. In this case, rotations of the spin and 
spatial degrees of 
freedom are not independent, and the possible
order parameters correspond to different irreducible 
representations of ${\rm G_s = G_c^{(J)}}$.
Although spin is not a good quantum number when spin-orbit coupling is 
important, we will still classify the possible states as singlet
or triplet with respect to a pseudo-spin space provided that
the superconducting state does not break inversion symmetry.
The singlet/triplet pseudo-spin classification 
was first introduced by Anderson~\cite{anderson-84} 
in the context of heavy fermions. We will use this classification
here whenever we refer to strong spin-orbit representations.
Considering quasi-one-dimensional superconductors of 
orthorhombic crystal structure~\cite{triclinic}, 
the relevant crystallographic point 
group is ${ \rm D_{2h} }$, which has only one dimensional 
representations~\cite{tinkham-64} (See Table~\ref{tab:d2h}). 

\subsection{General Order Parameter}
\label{sec:general-op}

The general form of the order parameter is
\begin{equation}
\label{eqn:op-matrix}
\tilde \Delta ({\bf k}) = 
i \left( \Delta_{si} ({\bf k}) \tilde {\bf 1}
+ \Delta_{tr} ({\bf k}) {\bf d} ({\bf k}) \cdot {\bf \sigma} \right)
\sigma_y,
\end{equation}
which must transform according to the one dimensional representations
of the orthorhombic point group $D_{2h}$, under the assumption that
the order parameter does not break the crystal 
translational symmetry,
i.e., the order parameter is invariant under all primitive lattice
translations. The matrix $\tilde {\bf 1}$ is the identity 
and the function $\Delta_{si} ({\bf k})$ is symmetric
(even) under the transformation ${\bf k} \to - {\bf k}$.
The three-dimensional vector ${\bf d} ({\bf k})$ 
is antisymmetric (odd) under ${\bf k} \to - {\bf k}$, while 
the function $\Delta_{tr} ({\bf k})$ is symmetric (even) 
under the transformation ${\bf k} \to - {\bf k}$.

We are mostly interested in triplet states 
which do not break time 
reversal symmetry, and our analysis
is confined initially to zero magnetic field. 
However, for completeness, we shall mention briefly
singlet and time-reversal-symmetry-breaking triplet
states. 
In Table~\ref{tab:singlet}, a summary of the
possible order parameter symmetries consistent with
the $D_{2h}$ point group is presented for singlet
systems assuming weak spin-orbit coupling,
while in Table~\ref{tab:triplet1} 
a summary of the possible order parameter
symmetries for time-reversal-symmetry-breaking triplet
states is shown.
Tables~\ref{tab:triplet2} and~\ref{tab:triplet3}
summarize the group theoretical analysis performed 
for the order parameter
matrix $\tilde \Delta ({\bf k})$ for time-reversal-invariant
triplet superconductors in the 
weak and strong spin-orbit coupling cases, respectively.
All the tables mentioned above include the state nomenclature, 
the order parameter for the respective 
singlet or triplet symmetries, and the loci of zeros of the
corresponding quasiparticle excitation spectrum 
$E_{\bf k}$, when $\tilde \mu = \mu - {\rm min}[\epsilon_{\bf k}]$ 
is positive. 
In Table~\ref{tab:triplet2}, the vector (0,0,1) 
is indicated up to an arbitrary 
rotation in spin space. In Table~\ref{tab:triplet3},
the numerical coefficients A, B and C 
define a specific model for the strong spin-orbit interaction 
appearing in Eq.~(\ref{eqn:int-strong}).
It is very important to emphasize here that the basis functions
$X({\bf k}), Y({\bf k})$ and $Z ({\bf k})$ transform like 
$k_x, k_y$ and $k_z$, respectively, under 
the crystallographic point group operations,
but they can not be chosen to be equal to $k_x, k_y$ and $k_z$.
The reason being that the normal state dispersion
$\epsilon_{\bf k}$ intersects the boundary of the Brillouin zone
along the $k_y$ and $k_z$ directions.
Thus, it is necessary to take into account the
periodicity of the order parameter matrix $\tilde \Delta ({\bf k})$
and of the order parameter vector ${\bf d} ({\bf k})$ in
reciprocal (momentum) space. In the case of orthorhombic crystals
with dispersion $\epsilon_{\bf k}$ defined in Eq.~(\ref{eqn:dispersion}),
where $|{t_x}| \gg |{t_y}| \gg |{t_z}|$ a Fermi surface 
exists (for weak attractive interaction or high densities) and does 
intersect the Brillouin zones, 
therefore the minimal basis set must be periodic and may be chosen
to be $X({\bf k}) = \sin (k_x a)$, 
$Y({\bf k}) = \sin (k_y b)$ and $Z ({\bf k}) = \sin (k_z c)$. 

\subsection{Excitation Spectrum Characteristics}
\label{sec:excitation-spectrum}

For the weak spin-orbit coupling (Table~\ref{tab:triplet2}) 
the only candidate for weaker attractive interaction 
($\tilde \mu > 0$) is the state $^3B_{3u} (a)$, where  
the quasiparticle excitation spectrum $E_{\bf k}$  
is fully gapped. 
In the case of strong spin-orbit coupling (Table~\ref{tab:triplet3}) 
there are three candidates for weaker attractive interaction 
($\tilde \mu > 0$), i.e., the states $A_{1u}, B_{1u}$
and $B_{2u}$, where the quasiparticle excitation spectrum 
$E_{\bf k}$ may be fully gapped. 
When $\tilde \mu > 0$, 
the state $A_{1u}$ is fully gapped only
for $A \ne 0$ and for any value of $B$ and $C$; 
the state $B_{1u}$ is fully gapped only 
for $B \ne 0$ and for any value of $A$ and $C$; and
the state $B_{2u}$ is fully gapped only  
for $C \ne 0$ and for any value of $A$ and $B$.

We note in passing that in the case of strong attractive interactions
or small densities, where $\tilde \mu < 0$, 
the excitation spectrum $E_{\bf k}$ is fully
gapped~\cite{duncan-00} for all of the states 
involving Tables~\ref{tab:triplet2} and~\ref{tab:triplet3}, for
example. Therefore, if 
Field Effect Transistor (FET) systems~\cite{batlogg-00a,batlogg-00b}
can be constructed to change the density of carriers $(n)$
in quasi-one-dimensional organic superconductors, it may be possible 
to study systematically the carrier density dependence
of physical properties in these systems. 
The most interesting theoretical case corresponds
to the situation where $E_{\bf k}$ has zeros for $\tilde \mu > 0$
$(n > n_c)$, and $E_{\bf k}$ 
has a full gap for $\tilde \mu < 0$ $(n <  n_c)$.
Here $n_c$ is the critical density where $\tilde \mu = 0$, where
the characteristics of excitation energy $E_{\bf k}$ changes
in a fundamental way. This effect here should correspond to a
Lifshitz transition for superconductors~\cite{duncan-00}
where low temperature properties like
spin susceptibility, specific heat, thermal conductivity,
would change their qualitative behavior depending
on $n$ being greater, equal or less than $n_c$.
The Lifshitz transition should also be accompanied 
by a topological phase transition
in the ground state~\cite{duncan-00}, where
the zero temperature electronic compressibility diverges 
at $n = n_c$. 
However, if the excitation spectrum $E_{\bf k}$
is fully gapped (like in the case of the $^3 B_{3u} (a)$ state)
for $\tilde \mu > 0$  $(n > n_c)$, $E_{\bf k}$ would remain
fully gapped for $\tilde \mu < 0$  $(n < n_c)$, and
only a crossover should occur.

We also note in passing, that the states shown in 
Table~\ref{tab:triplet1} are non-unitary since they break time
reversal symmetry. In this case the excitation spectrum has two
branches and is given by 
$$E_{{\bf k}\pm} = 
\sqrt{ \xi_k^2 + \vert\Delta_{tr}({\bf k})\vert^2
\left[ 
\vert {\bf d} ({\bf k}) \vert^2 \pm \vert {\bf q} ({\bf k}) \vert
\right]},
$$
where ${\bf q} ({\bf k}) = i  {\bf d}^{*} ({\bf k}) \times {\bf d} ({\bf k})$.
For $\tilde \mu > 0$  $(n > n_c)$, 
the regions in momentum space where $E_{{\bf k}-} = 0$ 
correspond to surfaces, while the regions
where $E_{{\bf k}+} = 0$ correspond to lines or the null set, 
as indicated in Table~\ref{tab:triplet1}. 
Thus, for $\tilde \mu > 0$  $(n > n_c)$, the excitation spectrum
is gapless. However, for $\tilde \mu <  0$  $(n < n_c)$, there 
are no regions in momentum space where $E_{{\bf k}\pm} = 0$, 
which means that the superconductor becomes fully gapped. 
This dramatic change in the excitation spectrum (from gapless to 
fully gapped) 
of time reversal breaking (non-unitary) triplet states should also 
correspond to a Lifshitz transition at finite temperatures 
and a topological transition in the ground state~\cite{duncan-00}.
Although, there is currently no experimental evidence that 
quasi-one-dimensional superconductor break time reversal symmetry
(at zero magnetic field), it seems that in the case of Strontium
Ruthenate time reversal symmetry breaking states are potentially
good candidates for the order parameter~\cite{sigrist-01}. 
Strontium Ruthenate is very different from Bechgaard salts
since it is non-organic and has essentially a tetragonal 
quasi-two-dimensional structure. 
If Field Effect Transistor (FET) systems~\cite{batlogg-00a,batlogg-00b}
can be constructed to change the density of carriers $(n)$
in Strontium Ruthenate, it may be also possible:
(a) to study systematically the carrier density dependence
of physical properties; 
(b) to search for a possible Lifshitz transition at finite
temperatures, and
(c) to investigate a possible topological quantum phase transition
in the ground state.

\subsection{Three Dimensional Views of the Order Parameter}
\label{sec:d-vector}

A three dimensional view of the Fermi surface and order
parameter sign for possible weak spin-orbit coupling singlet states
compatible with the orthorhombic $(D_{2h})$ group are shown
shown in Fig.~\ref{fig:singlet} for the case where
$\tilde \mu > 0$.
In Fig.~\ref{fig:singlet}(a), (b), (c) and (d) 
the signs of the orbital component of the order parameter are shown
for the $^1 A_{1g}$ (``$s$-wave''),
$^1 B_{1g}$ (``$d_{xy}$-wave''),
$^1 B_{2g}$ (``$d_{xz}$-wave''),
$^1 B_{1g}$ (``$d_{yz}$-wave''),
respectively. The nodes are indicated by the dark thick lines
separating the order parameter signs.
\begin{figure*}[ht]
\begin{center}
\(
  \begin{array}{c@{\hspace{.5cm}}c}
  \multicolumn{1}{l}{\mbox{\bf (a)}} &
  \multicolumn{1}{l}{\mbox{\bf (b)}} \\ [-0.75cm]
  \includegraphics[width=7cm]{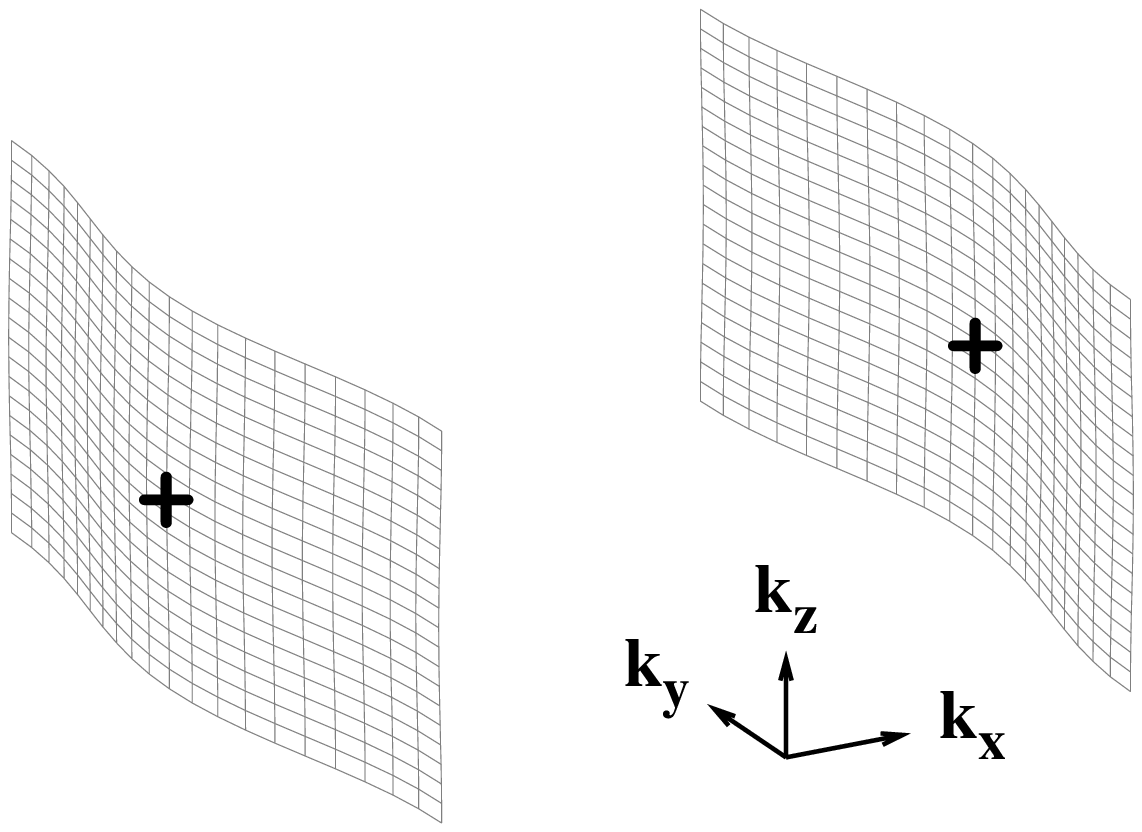} &
  \includegraphics[width=7cm]{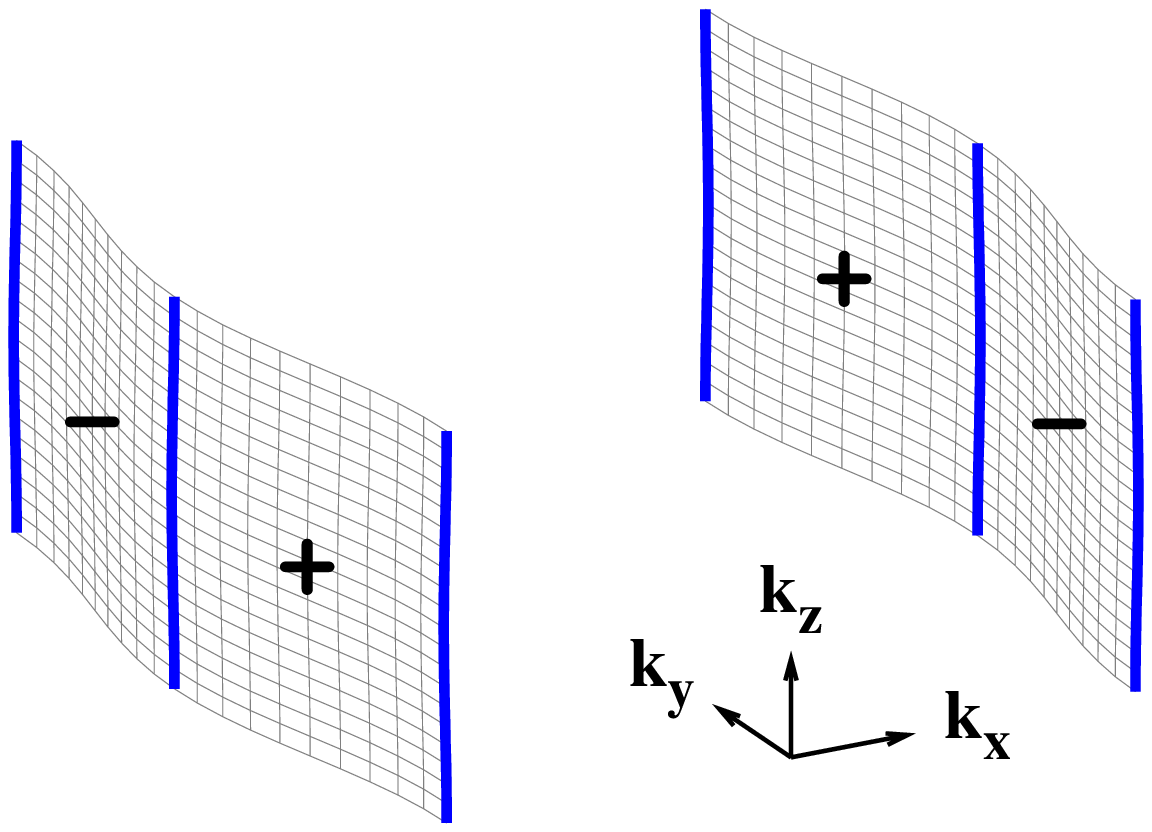} \\
  \multicolumn{1}{l}{\mbox{\bf (c)}} &
  \multicolumn{1}{l}{\mbox{\bf (d)}} \\ [-0.75cm]
  \includegraphics[width=7cm]{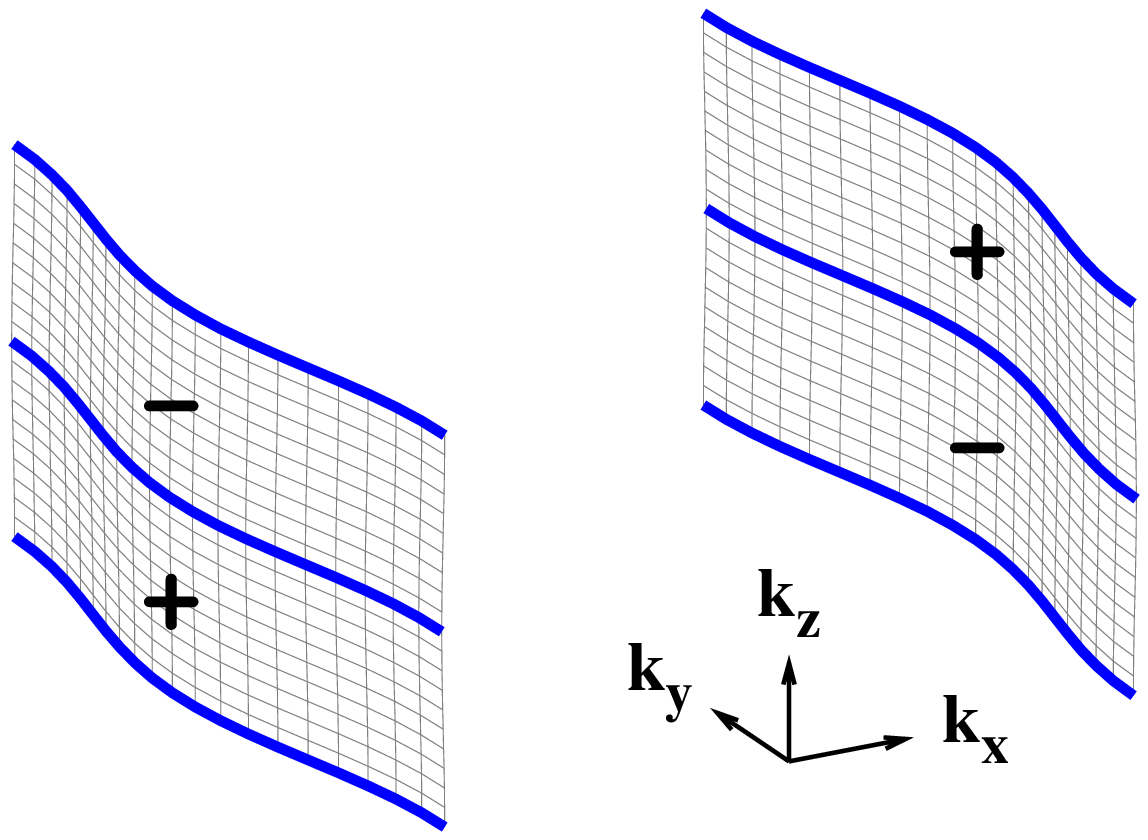} &
  \includegraphics[width=7cm]{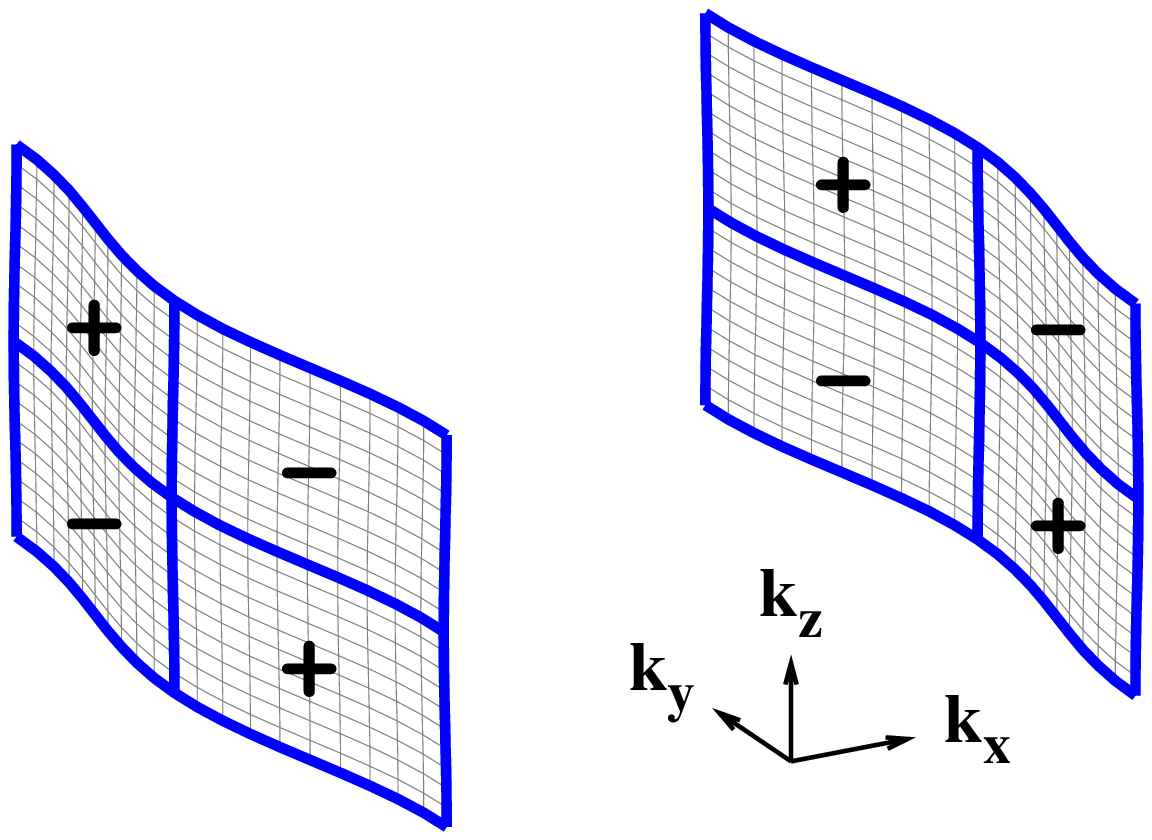} \\
  \end{array}
\)
\end{center}
\caption{Three dimensional views of the sign of the orbital component
of the order parameter.  Various weak spin-orbit coupling 
order parameter symmetries are illustrated (See Table~\ref{tab:singlet}): 
(a) $^1 A_{1g}$ (``$s$-wave'');
(b) $^1 B_{1g}$ (``$d_{xy}$-wave'');
(c) $^1 B_{2g}$ (``$d_{xz}$-wave'');
(d) $^1 B_{3g}$ (``$d_{yz}$-wave'').
The thick lines correspond to the zeros of the
orbital parts of the order parameter as well 
as to the excitation energy node structure $(E_{\bf k} = 0)$ 
on the Fermi surface ($\tilde{\mu} > 0$).
}
\label{fig:singlet}
\end{figure*} 
\begin{figure*}[ht]
\begin{center}
\(
  \begin{array}{c@{\hspace{.5cm}}c}
  \multicolumn{1}{l}{\mbox{\bf (a)}} &
  \multicolumn{1}{l}{\mbox{\bf (b)}} \\ [-0.75cm]
  \includegraphics[width=7cm]{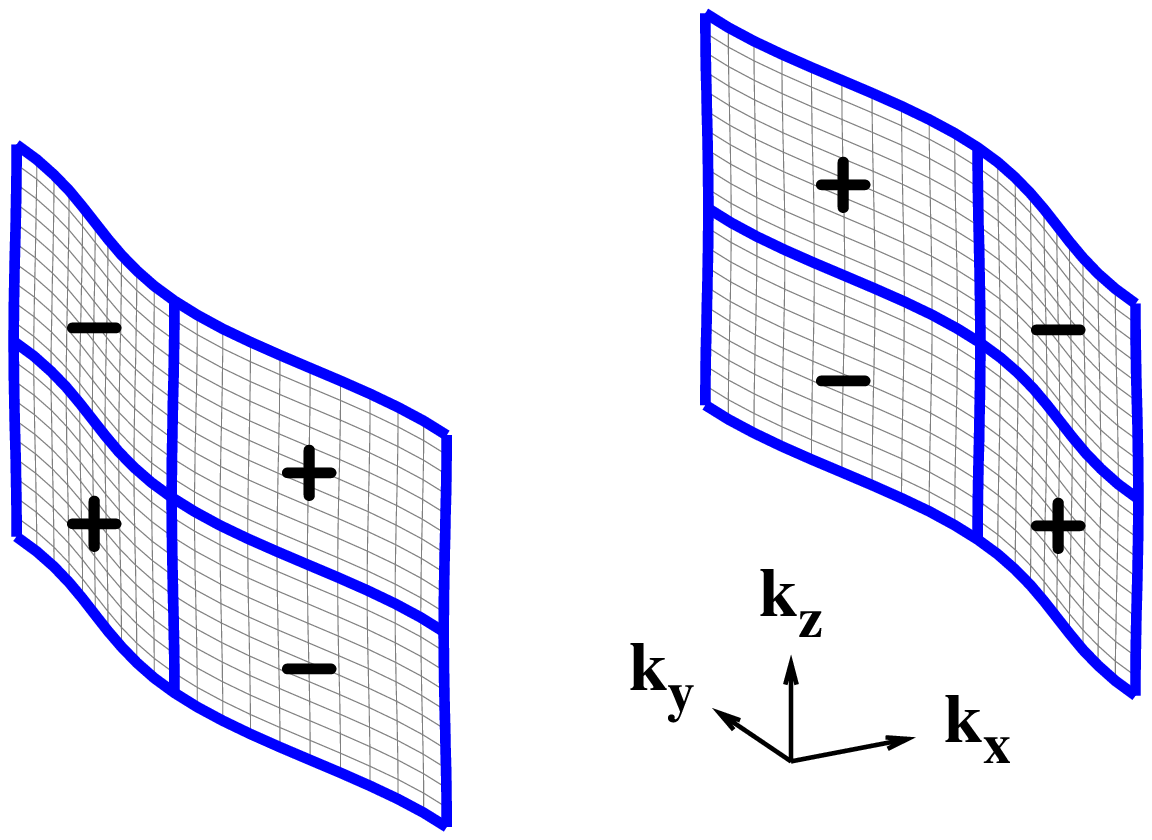} &
  \includegraphics[width=7cm]{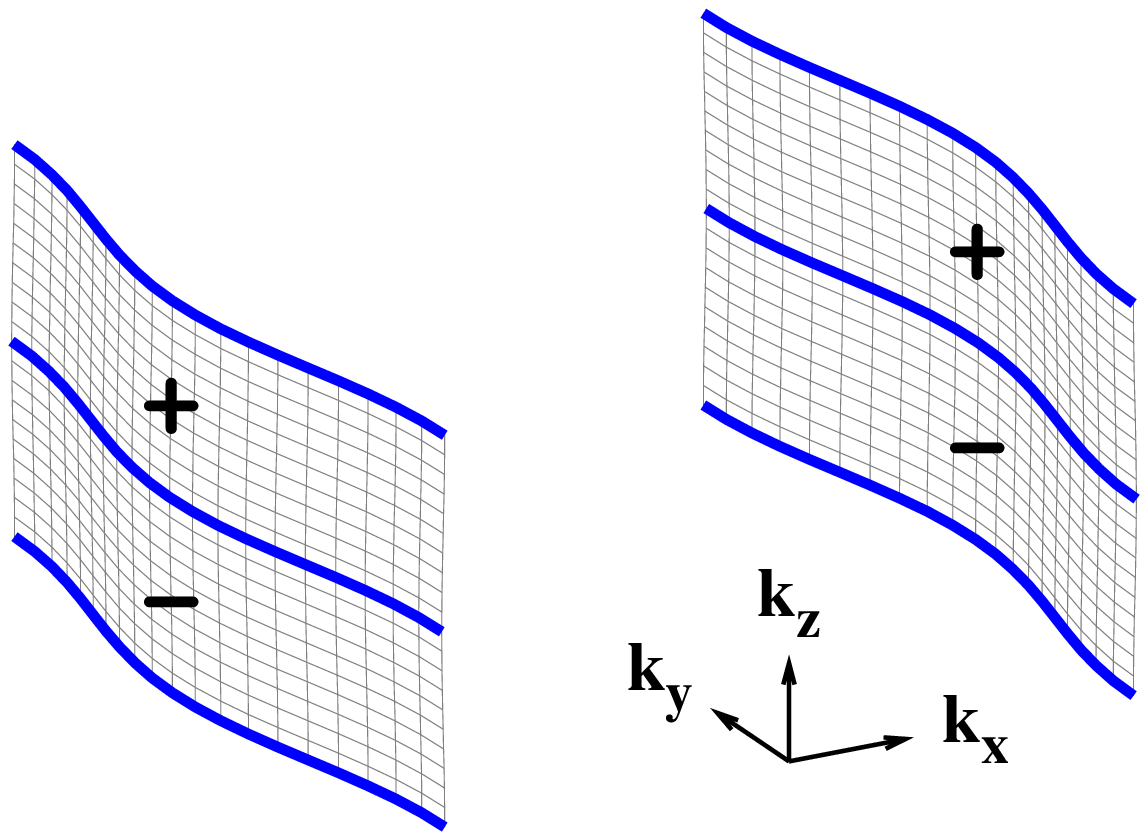}  \\
  \multicolumn{1}{l}{\mbox{\bf (c)}} &
  \multicolumn{1}{l}{\mbox{\bf (d)}} \\ [-0.75cm]
  \includegraphics[width=7cm]{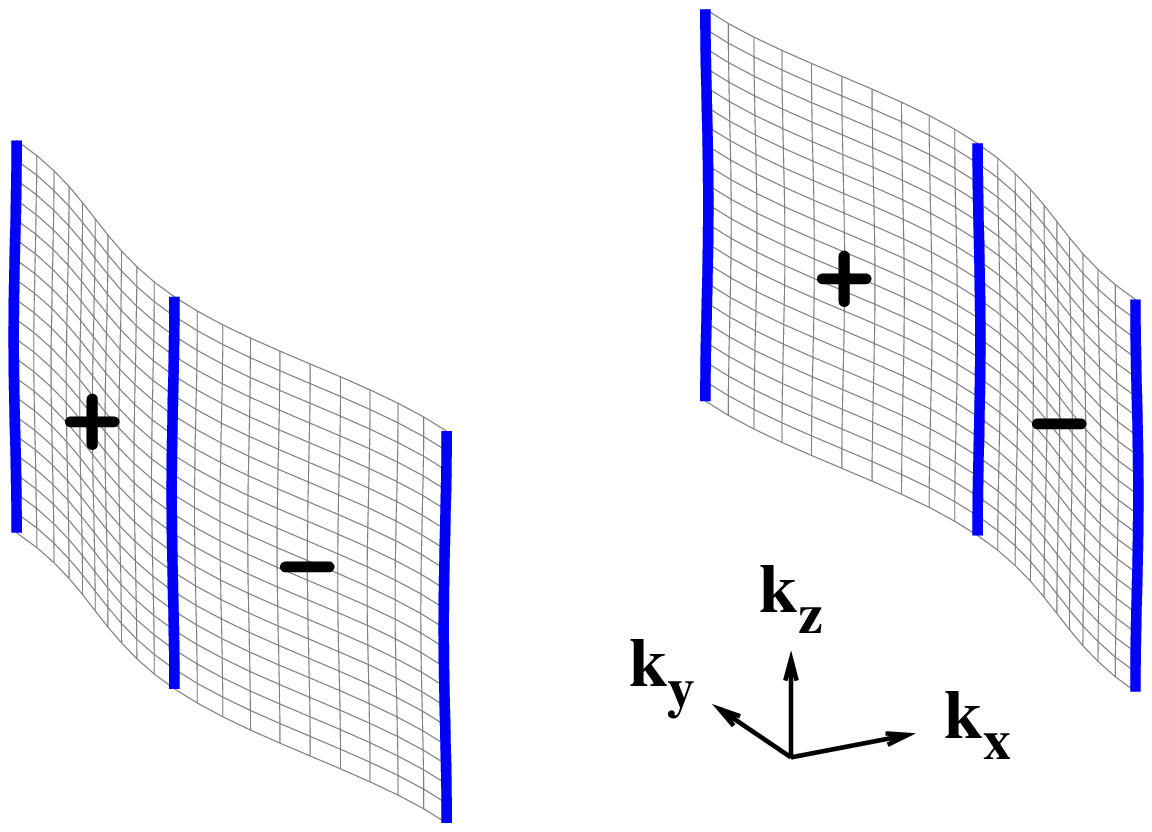} &
  \includegraphics[width=7cm]{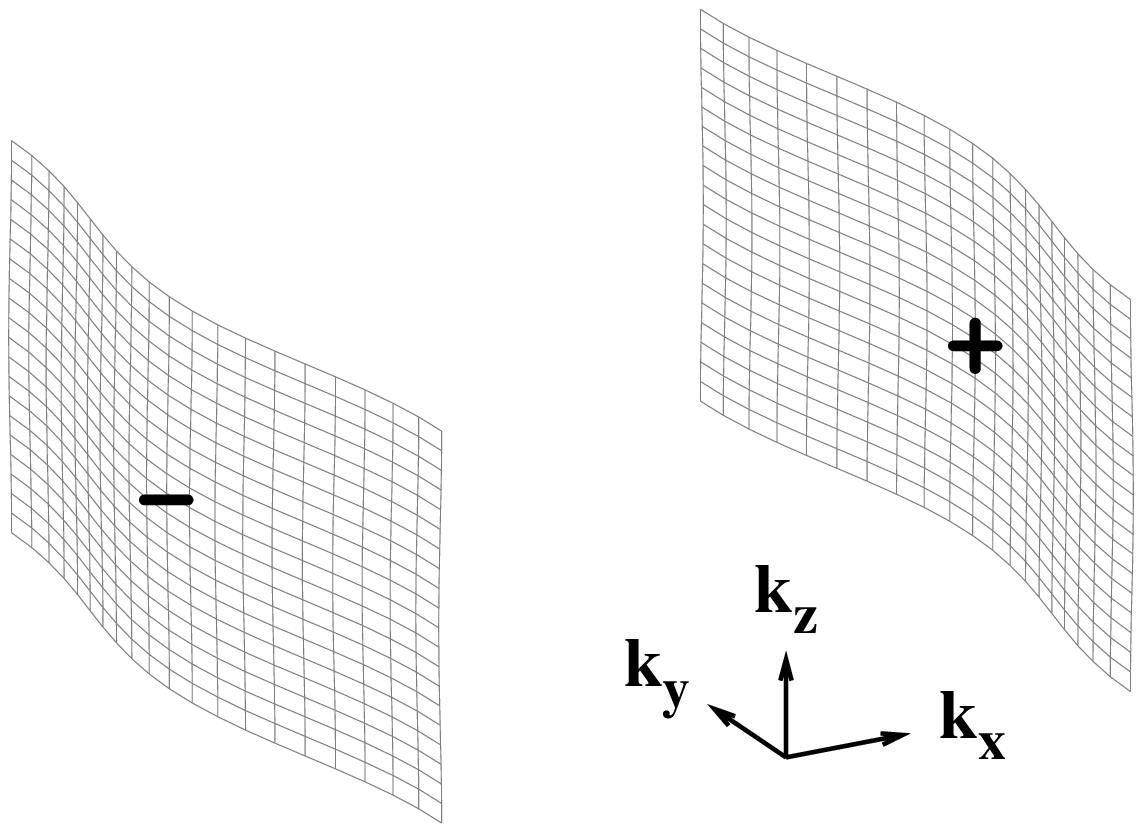}  \\
  \end{array}
\)
\end{center}
\caption{Three dimensional views of the sign of the orbital component
of the ${\bf d}$-vectors.  Various weak spin-orbit coupling 
order parameter symmetries are illustrated (See Table~\ref{tab:triplet2}): 
(a) $^3 A_{1u} (a)$ (``$f_{xyz}$-wave'');
(b) $^3 B_{1u} (a)$ (``$p_{z}$-wave'');
(c) $^3 B_{2u} (a)$ (``$p_{y}$-wave'');
(d) $^3 B_{1u} (a)$ (``$p_{x}$-wave'').
The thick lines correspond to the zeros of the
orbital parts of the ${\bf d}$-vectors as well 
as to the excitation energy node structure $(E_{\bf k} = 0)$ 
on the Fermi surface ($\tilde{\mu} > 0$).
}
\label{fig:triplet2}
\end{figure*} 
In Fig.~\ref{fig:triplet2} a 
three dimensional view of the Fermi surface and order
parameter sign for possible weak-spin orbit coupling
triplet states compatible with the orthorhombic $(D_{2h})$ 
group are shown for the case where
$\tilde \mu > 0$.
In Fig.~\ref{fig:triplet2}(a), (b), (c) and (d) 
the signs of the orbital component of the order parameter are shown
for the $^3 A_{1u} (a)$ (``$f_{xyz}$-wave''),
$^3 B_{1u} (a) $ (``$p_{z}$-wave''),
$^3 B_{2u} (a) $ (``$p_{y}$-wave''),
$^3 B_{1u} (a) $ (``$p_{x}$-wave''),
respectively.


For the weak spin-orbit states is not only useful to know where the
orbital components of the order parameter change sign, but it is also
important to visualize the ${\bf d}$-vector at the Fermi surface. 
In the complete absence of the spin-orbit coupling the ${\bf d}$-vectors 
can point in any direction with equal probability. However
Bechgaard salts are expected to have weak spin-orbit coupling,
leading to the weak pinning of the ${\bf d}$-vector
along a preferred direction. 
Based on recent $^{77} Se$ NMR experiments~\cite{lee-00,lee-02} 
it seems that the ${\bf d}$-vector in ${\rm (TMTSF)_2 PF_6}$ 
is not pointing along ${\bf a}$ or ${\bf b^\prime}$ direction
at least for magnetic fields beyond the 1~T range.
Therefore, in Fig.~\ref{fig:dvec-weak} 
we choose the weak spin-orbit coupling
${\bf d}$-vectors to be pointing along the
${\bf c}$ direction.
We also illustrate in Fig.~\ref{fig:dvec-strong} the ${\bf d}$-vector
structure for the case of strong spin-orbit coupling. 
The ${\bf d}$-vector for the states $A_{1u}$, $B_{1u}$, 
$B_{2u}$, and $B_{3u}$ (see Table~\ref{tab:triplet3} are shown 
in  Figs.~\ref{fig:dvec-strong}a, \ref{fig:dvec-strong}b, 
\ref{fig:dvec-strong}c, and \ref{fig:dvec-strong}d, 
for $A = B = 1$ and $C = 0$, respectively. 
Notice the textured structure of 
the ${\bf d}$-vectors at the Fermi surface.
\begin{figure*}[ht]
\begin{center}
\(
  \begin{array}{c@{\hspace{.5cm}}c}
  \multicolumn{1}{l}{\mbox{\bf (a)}} &
  \multicolumn{1}{l}{\mbox{\bf (b)}} \\ [-0.75cm]
  \includegraphics[width=7cm]{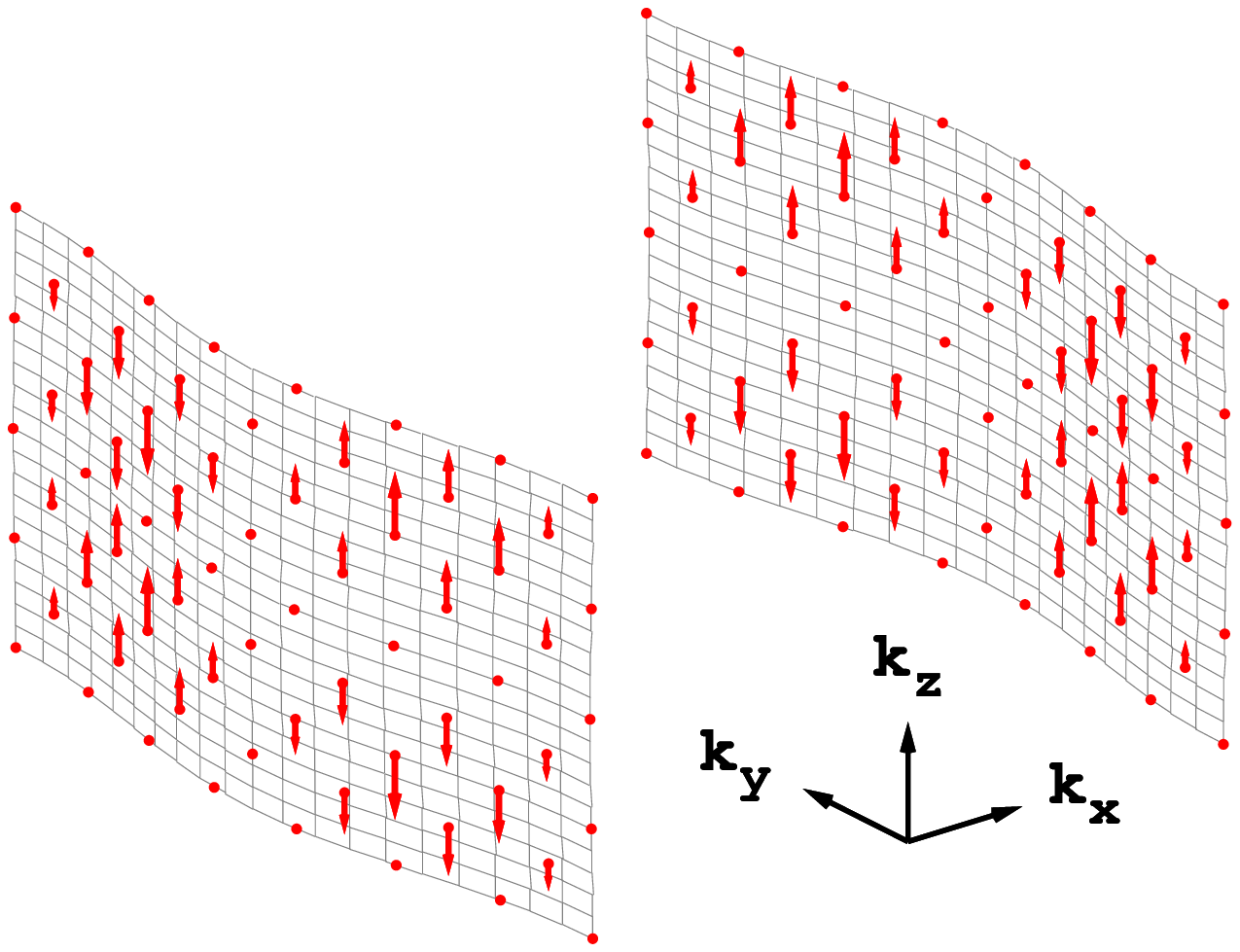} &
  \includegraphics[width=7cm]{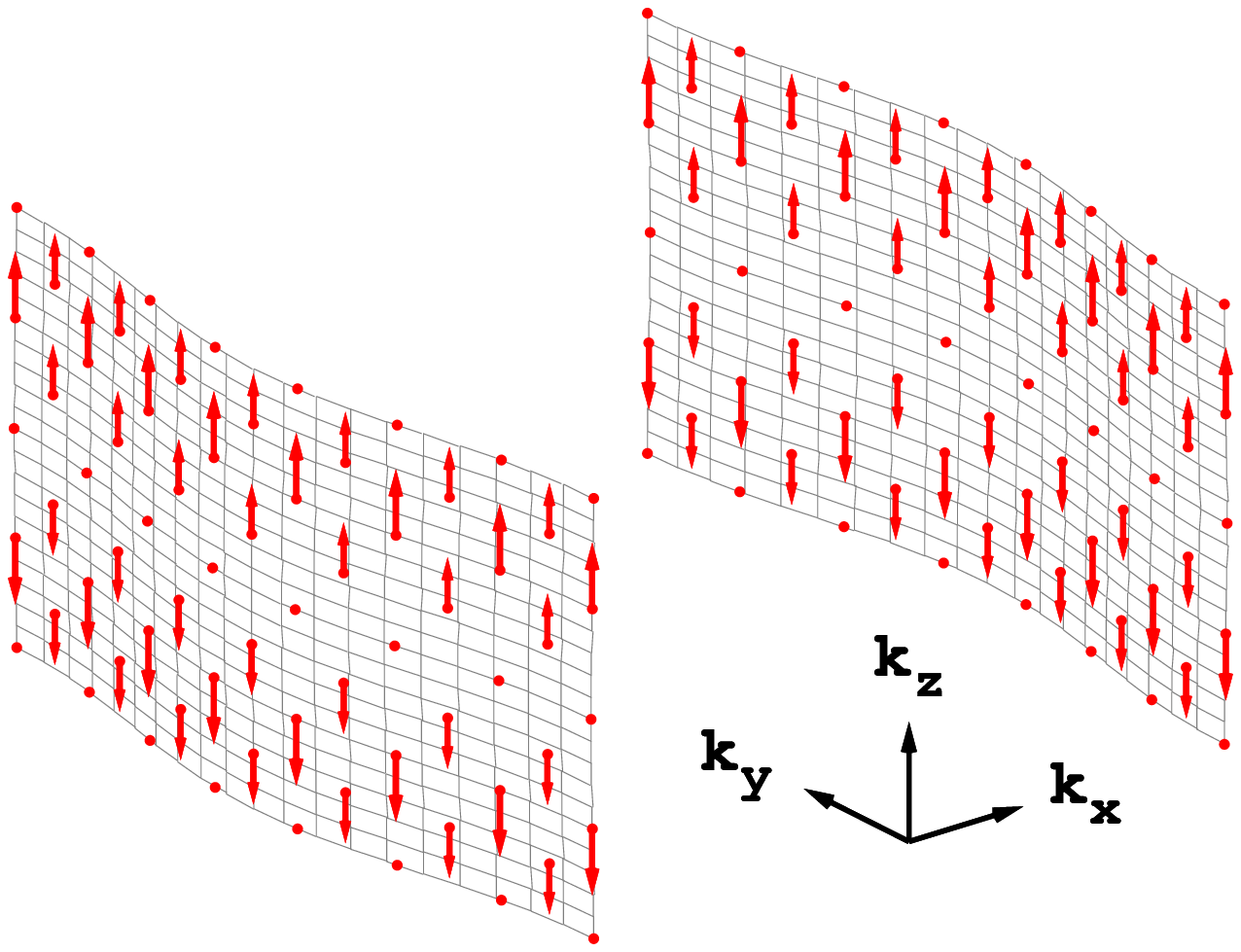} \\ [0.1cm]
  \multicolumn{1}{l}{\mbox{\bf (c)}} &
  \multicolumn{1}{l}{\mbox{\bf (d)}} \\ [-0.75cm]
  \includegraphics[width=7cm]{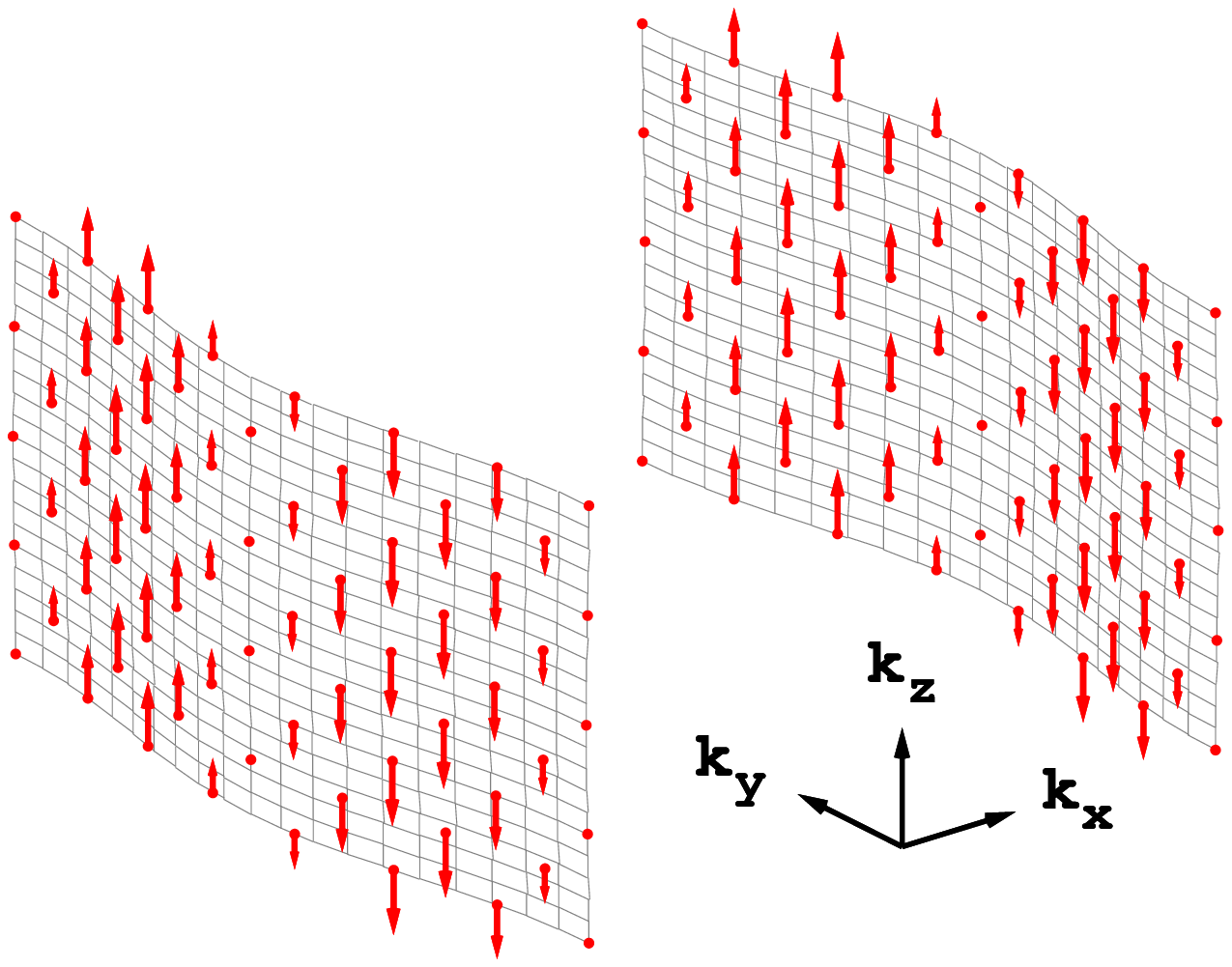} &
  \includegraphics[width=7cm]{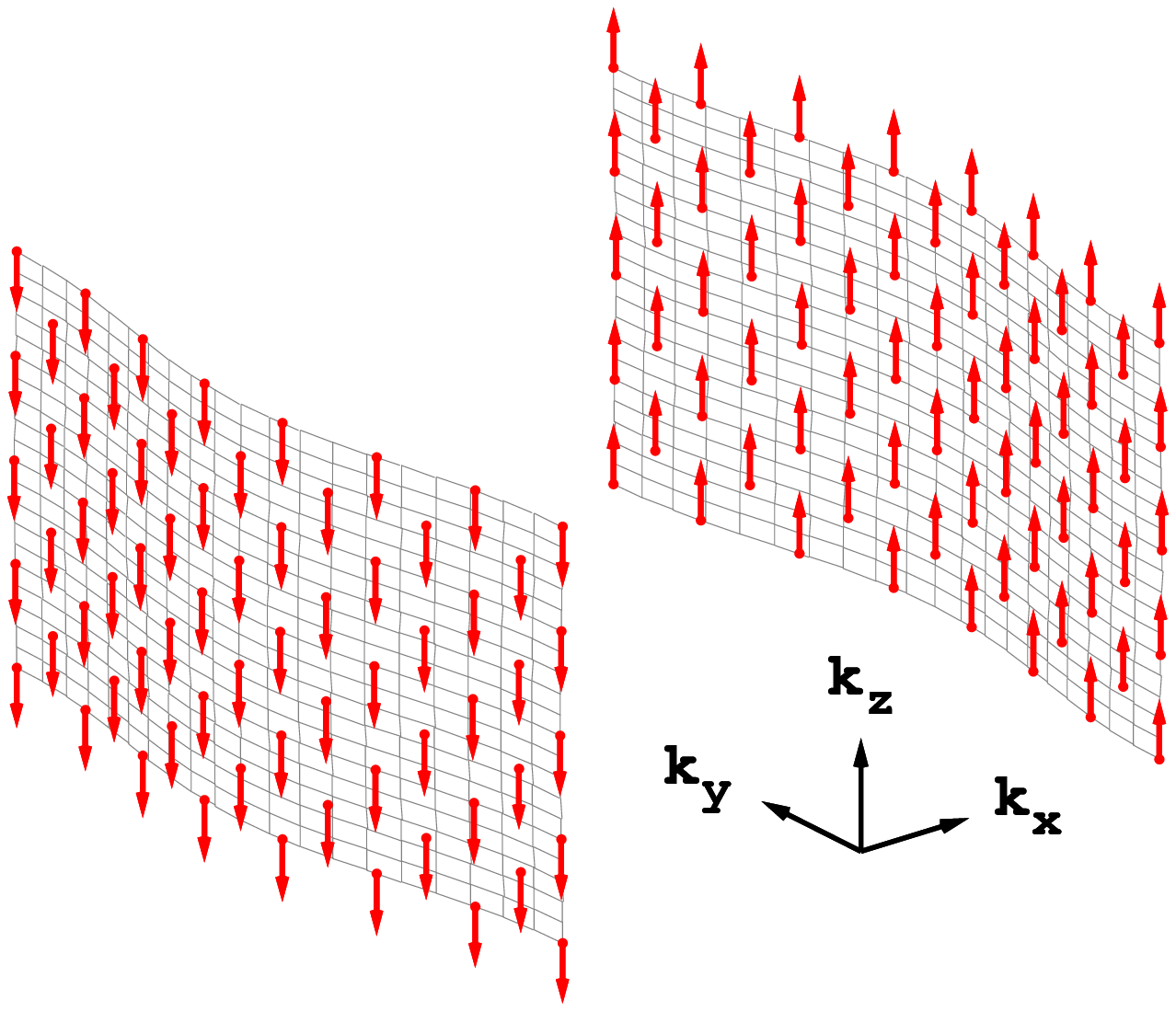} \\
  \end{array}
\)
\end{center}
\caption{Three dimensional
views of ${\bf d}$-vectors at the Fermi surface ($\tilde{\mu}>0$)
for various weak spin-orbit coupling order parameter
symmetries (See Table~\ref{tab:triplet2}): 
(a) $^3 A_{1u} (a)$ (``$f_{xyz}$-wave'');
(b) $^3 B_{1u} (a) $ (``$p_{z}$-wave'');
(c) $^3 B_{2u} (a) $ (``$p_{y}$-wave'');
(d) $^3 B_{1u} (a) $ (``$p_{x}$-wave'').
For definiteness the ${\bf d}$-vectors are chosen to point
along the ${\bf c}$ direction.
}
\label{fig:dvec-weak}
\end{figure*} 
\begin{figure*}[ht]
\begin{center}
\(
  \begin{array}{c@{\hspace{.5cm}}c}
  \multicolumn{1}{l}{\mbox{\bf (a)}} &
  \multicolumn{1}{l}{\mbox{\bf (b)}} \\ [-0.75cm]
  \includegraphics[width=7cm]{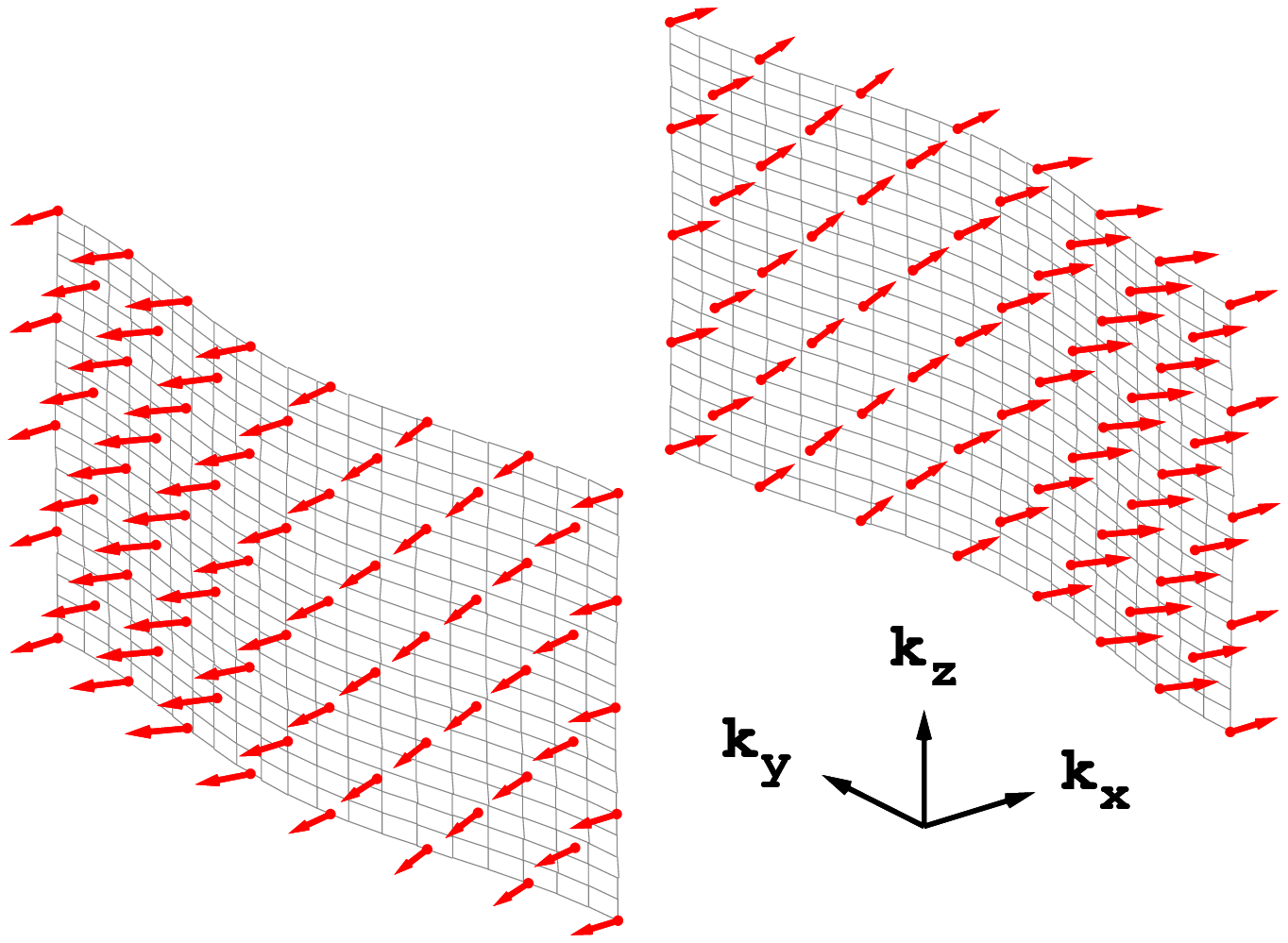} &
  \includegraphics[width=7cm]{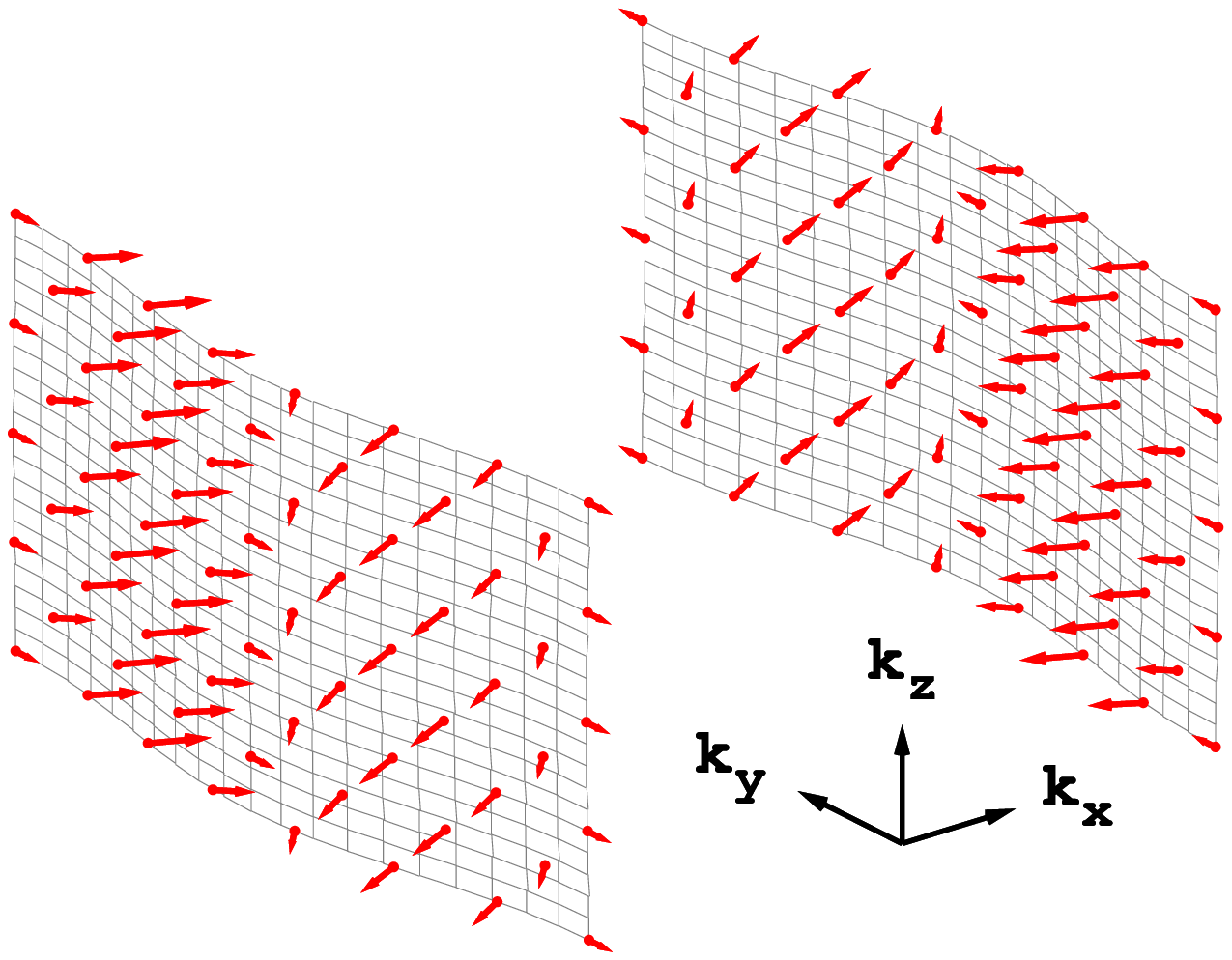} \\ [0.1cm]
  \multicolumn{1}{l}{\mbox{\bf (c)}} &
  \multicolumn{1}{l}{\mbox{\bf (d)}} \\ [-0.75cm]
  \includegraphics[width=7cm]{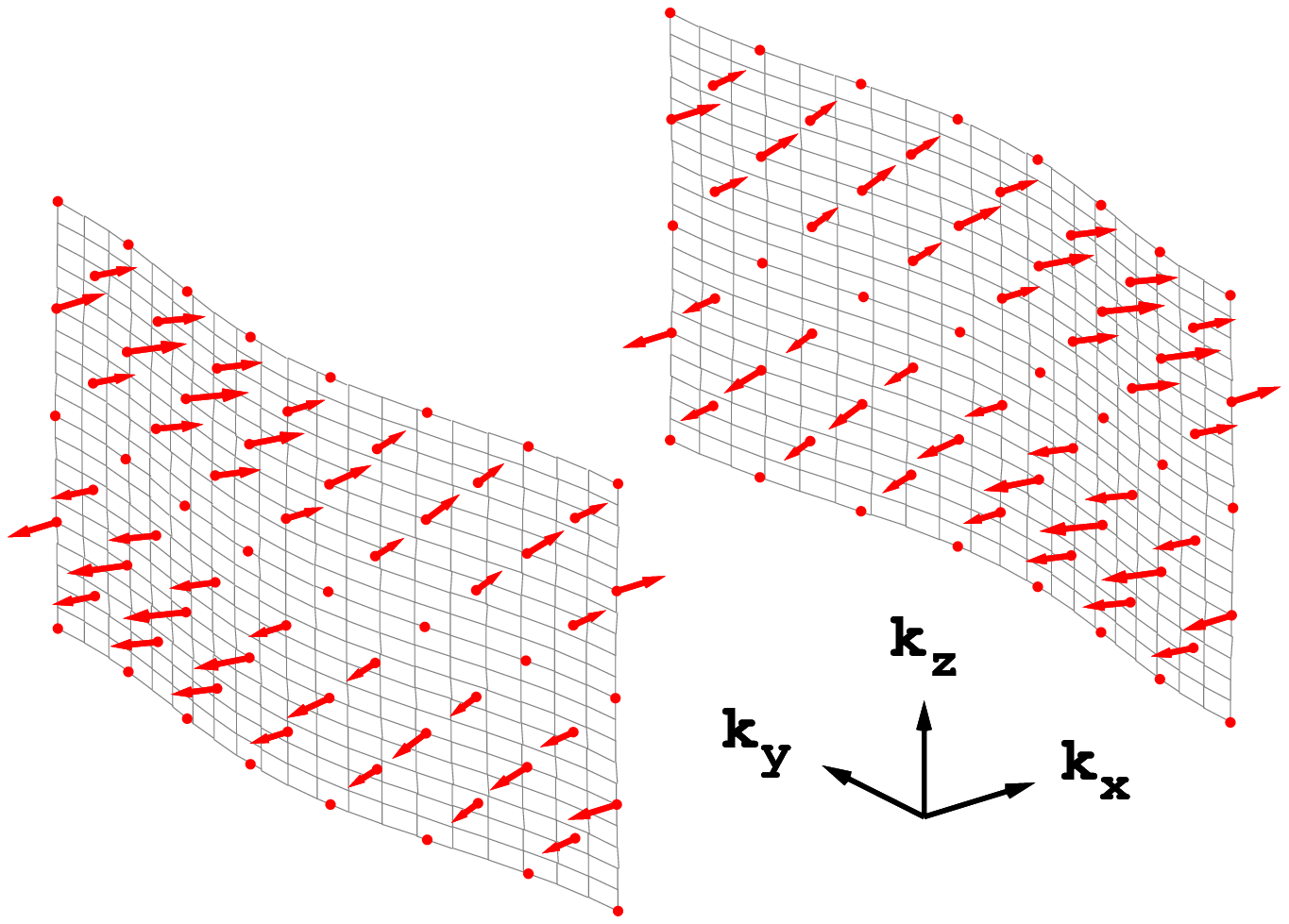} &
  \includegraphics[width=7cm]{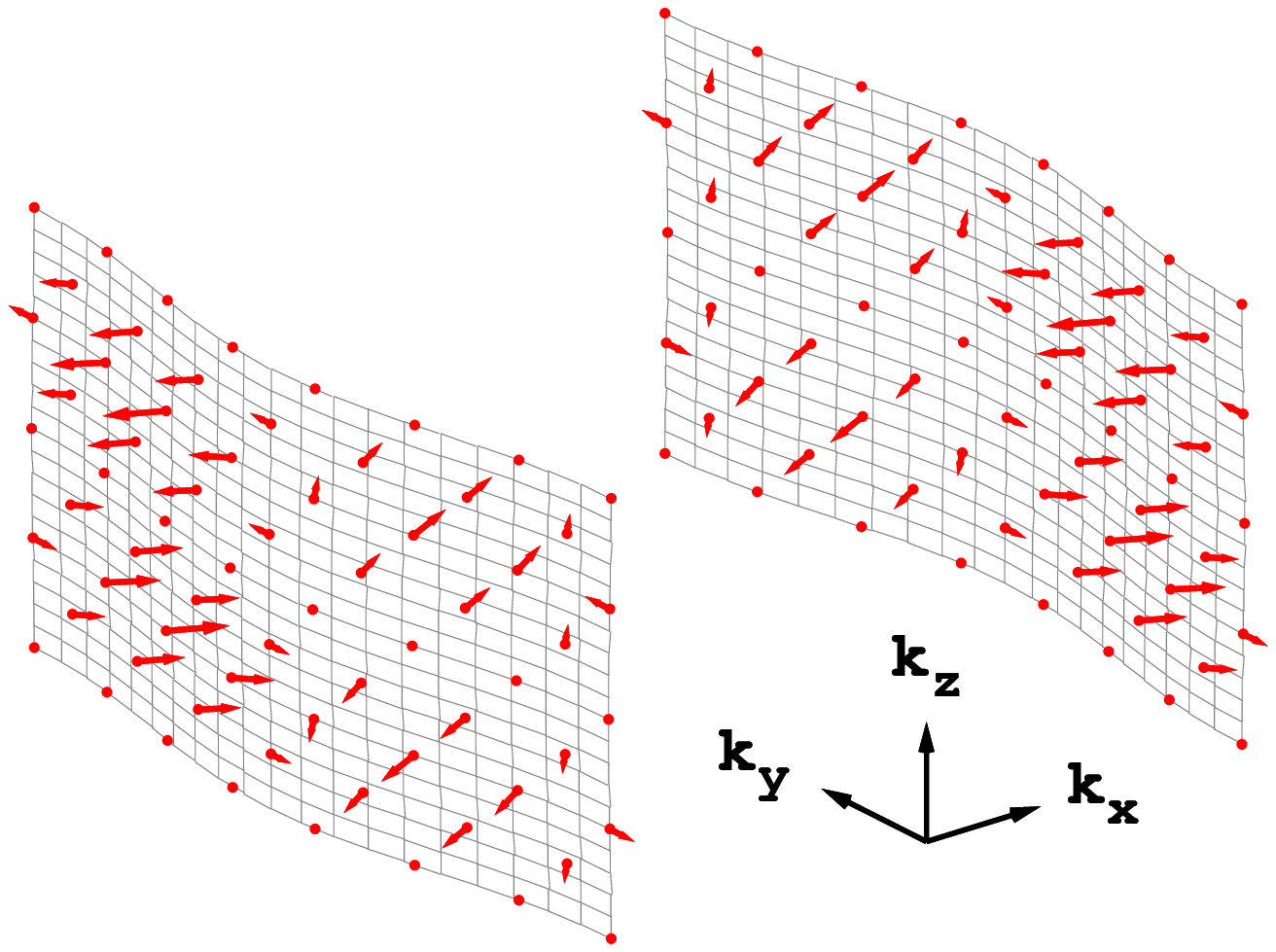} \\
  \end{array}
\)
\end{center}
\caption{Three dimensional
views of ${\bf d}$-vectors at the Fermi surface ($\tilde{\mu}>0$)
for various strong spin-orbit coupling order parameter
symmetries (See Table~\ref{tab:triplet3}): 
(a) $A_{1u}$; 
(b) $B_{1u}$;
(c) $B_{2u}$;
(d) $B_{1u}$, for $A = B = 1$, and $C = 0$.
Notice that the direction of the ${\bf d}$-vectors change
along the Fermi surface, thus producing a textured structure.
}
\label{fig:dvec-strong}
\end{figure*} 

As discussed in the introduction the combined experimental
evidence of the upper critical field measurements of 
Lee {\it et. al.}~\cite{lee-97,lee-94,lee-95}, thermal
conductivity measurements of BB~\cite{belin-97},
and the Knight shift and NMR relaxation experiments
of Lee {\it et. al.}~\cite{lee-00,lee-02}, suggest that 
${\rm (TMTSF)_2 ClO_4}$ and/or 
${\rm (TMTSF)_2 PF_6}$ may be fully gapped triplet 
superconductors. Thus, we consider next 
the temperature dependences of the quasiparticle 
density of states (QDOS) and of the uniform 
spin susceptibility tensor for a few potential 
triplet candidate states, which 
are fully gapped: 
a) the weak spin-orbit coupling
state $^3B_{3u} (a)$ of Table~\ref{tab:triplet2};
b) the strong spin-orbit coupling states $A_{1u}$ of table
~\ref{tab:triplet3} with $A \ne 0$.
Other fully gapped triplet states correspond to the
$B_{1u}$ and $B_{2u}$ of table ~\ref{tab:triplet3} with $B \ne 0$,
and $C \ne 0$, respectively. We will not discuss these states
any further here. However, we will make comparisons of the chosen triplet
candidates with the results of reference 
singlet states $^1 A_{1g}$ (``$s$-wave'')
and $^1 B_{1g}$ (``$d_{xy}$-wave'').

\section{Quasiparticle Density of States}
\label{sec:qdos}
The bulk quasiparticle density of states (QDOS) for these different
symmetries can be obtained from 
the single particle Green's function as
\begin{equation}
{N} (\omega) = - \dfrac{1}{\pi} {\it Tr} \sum_{\bf k} 
{\it Im }
G_{\alpha \beta} ({\bf k}, i\omega_n = \omega + i\delta), 
\end{equation}
where $G_{\alpha \beta} ({\bf k}, i\omega_n)$ is defined in 
Eq.~(\ref{eqn:sing-green}). Features of the QDOS shown 
in Fig.~\ref{fig:dos} could be measured, in principle, during STM or
photoemission experiments, depending on
experimental resolution and material surface cleanness.
(In the case of photoemission, radiation damage due to X-rays,
can also be an issue).  
Although, to our knowledge, these experiments have not yet 
been successfully 
performed in ${\rm (TMTSF)_2 ClO_4}$ and ${\rm (TMTSF)_2 PF_6}$, 
our discussion can serve as qualitative guides for the 
extraction of gaps and 
symmetry dependent features of the experimental results when 
they become available. In particular, STM or photoemission measured
gaps could be compared with gaps measured thermodynamically (e.g.,
thermal conductivity~\cite{belin-97} or specific heat).
However, a note of caution is in place here. Due to the 
expected unconventionality of quasi-one-dimensional superconductors,
surface quasiparticle bound states could be present in these
systems. The presence of these surface quasiparticle bound states
was first noticed by Buchholtz and Zwickangl (BZ)~\cite{buchholtz-81}
in isotropic $p$-wave superconductors and by
Hu~\cite{hu-94} in layered $d$-wave superconductors. Recently,
Sengupta {\it et. al.}~\cite{sengupta-01} followed the ideas
of BZ~\cite{buchholtz-81} and Hu~\cite{hu-94} and discussed the existence
of these surface quasiparticle bound states in quasi-one-dimensional
superconductors. The QDOS that we discuss is
a bulk property, thus the contribution of
these surface states is not reflected in our calculated QDOS.

We show in Fig.~\ref{fig:dos-sd-wave}
the bulk QDOS at $T = 0~{\rm K}$, and $T = 1~{\rm K}$ for the singlet states
$^1 A_{1g}$ (``$s$-wave'') and $^1 B_{1g}$ (``$d_{xy}$-wave'') 
as references.
While in Fig.~\ref{fig:dos} we present 
the QDOS at $T = 0~{\rm K}$ and $T = 1~{\rm K}$, 
with $T_c = 1.5~{\rm K}$,
for the states $^3B_{3u}(a)$ (weak spin-orbit coupling) and 
$A_{1u}$ (strong spin-orbit coupling) for various values
of the constants $A$, $B$, and $C$. 
These constants reflect the strength and the 
anisotropy of the pairing interaction 
$V_{\alpha \beta \gamma \delta} ( {\bf k}, {\bf k}^\prime )$ defined 
in Eq.~\ref{eqn:int-strong} for the strong spin-orbit coupling case.
The symmetry dependent features of the QDOS are manifested through 
the magnitude of the order parameter vector ${\bf d} ({\bf k})$. 
For the $^3B_{3u}(a)$ state the magnitude of ${\bf d} ({\bf k})$ is 
\begin{equation}
\label{eqn:dpx}
|{\bf d} ({\bf k})| \propto |\sin(k_x a)|,
\end{equation}
while for the $A_{1u}$ state it takes the form $|{\bf d} ({\bf k})|$ 
\begin{equation}
\label{eqn:dpxyz}
\propto  
\sqrt{ A^2|\sin(k_x a)|^2 +  B^2|\sin(k_y b)|^2 +  C^2|\sin(k_z c)|^2}.
\end{equation}

In Fig.~\ref{fig:dos}(b), (c) and (d) we show
only the case corresponding to $C = 0$, where $A \gg B$, $A = B$,
and $A \ll B$, respectively. Notice that  
Fig.~\ref{fig:dos}(d) is nearly identical to Fig.~\ref{fig:dos}(a) 
given that $|{\bf d} ({\bf k})|$ is essentially the same in this case. 
Furthermore, notice that while the position of the peaks in 
Figs.~\ref{fig:dos}(a), (b), (c) and (d) are essentially the same
($\omega_{p} \approx \pm 5.3~{\rm K}$), 
the corresponding gap sizes are respectively 
$\omega_{g} = 2.27~{\rm K}; 0.29~{\rm K}; 1.59~{\rm K}; 2.27~{\rm K}$. 
In Fig.~\ref{fig:dos-px-zoom} we show the 
frequency dependence near the gap edge at $T = 0$ for the 
$^3B_{3u}(a)$ (weak spin-orbit coupling) state indicated
in Fig.~\ref{fig:dos}(a).
The characteristic shape exhibited here can also be found in
the gapped states $A_{1u}$ shown in Fig.~\ref{fig:dos}(b), (c), and (d).
In addition, the frequency dependence near the gap edge in these states
obeys a square-root power law.

Although STM measures current (I) versus voltage (V) 
characteristics and not QDOS, the I-V (or $dI/dV$) characteristic
are related to the QDOS. 
In a STM experiment one has to consider also 
the contribution of surface bound states, specially at zero voltage bias 
where a large contribution is expected 
in some cases~\cite{buchholtz-81,hu-94,sengupta-01}.
However, gap sizes, finite energy peaks and 
the general shape of the bulk QDOS
should be, in principle, identifiable in an STM experiment
at finite voltage bias.

It is also desirable to have photoemission experiments
probing QDOS in the superconducting state of ${\rm (TMTSF)_2 ClO_4}$ 
and ${\rm (TMTSF)_2 PF_6}$, however the surfaces and the bulk of these
materials seem to be very sensitive to X-rays. Thus, it is possible
that photoemission may be not able to probe the QDOS without 
introducing disorder at the surface and the bulk of these materials. 
Even if STM and photoemission experiments could
be successfully performed, these experiments alone 
cannot uniquely determine the 
symmetry of the order parameter in 
triplet quasi-one-dimensional superconductors
since the QDOS depends only on the magnitude of ${\bf d} ({\bf k})$.
Therefore, we now turn out attention to the calculation of the spin
susceptibility tensor, which
explicitly depends both on the magnitude
and direction of ${\bf d} ({\bf k})$.

\begin{figure*}[ht!]
\begin{center}
\(
  \psfrag{w}{$\omega$}
  \psfrag{N(w)}{$N(\omega)$}
  \psfrag{T=0}{\hskip -0.2cm {\footnotesize \mbox{$T=0~{\rm K}$}}}
  \psfrag{T=1}{{\footnotesize \mbox{$T=1~{\rm K}$}}}
  \begin{array}{c@{\hspace{.5cm}}c}
  \multicolumn{1}{l}{\mbox{\bf (a)}} & 
  \multicolumn{1}{l}{\mbox{\bf (b)}} \\ [-0.5cm]
  \includegraphics[width=7.0cm]{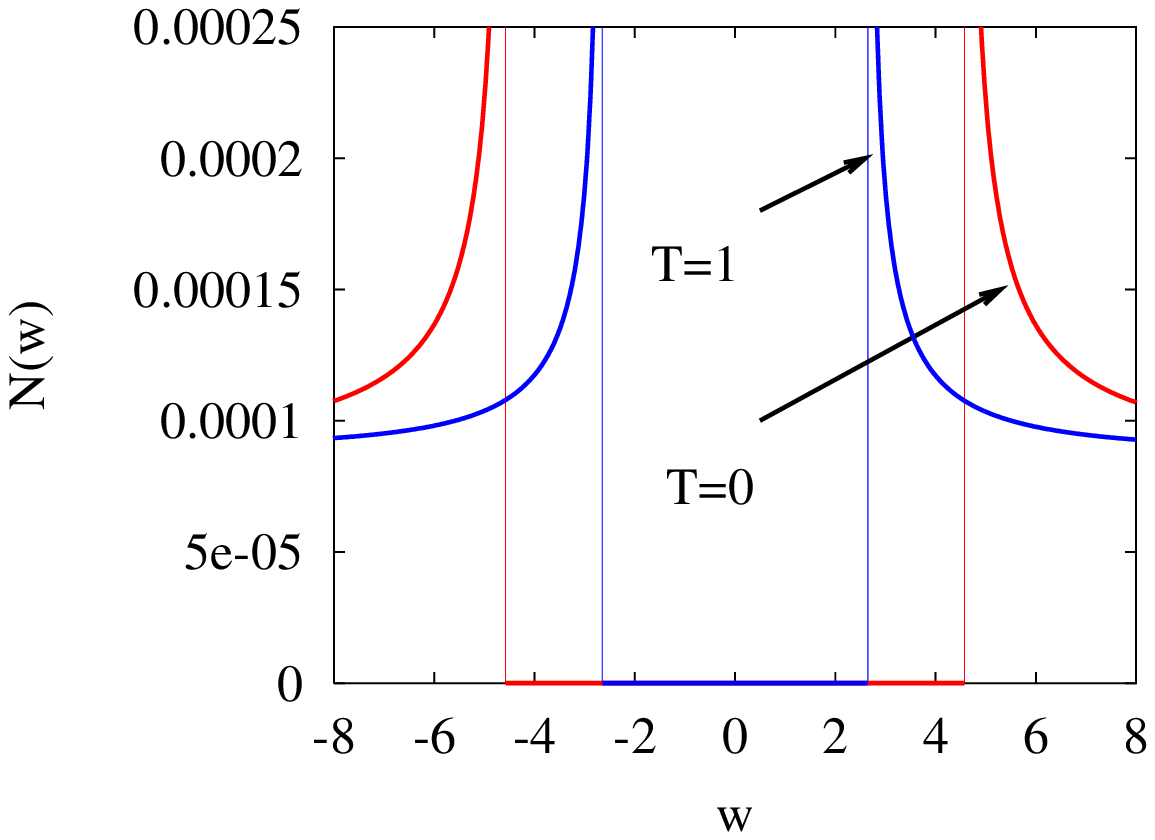} &
  \includegraphics[width=7.0cm]{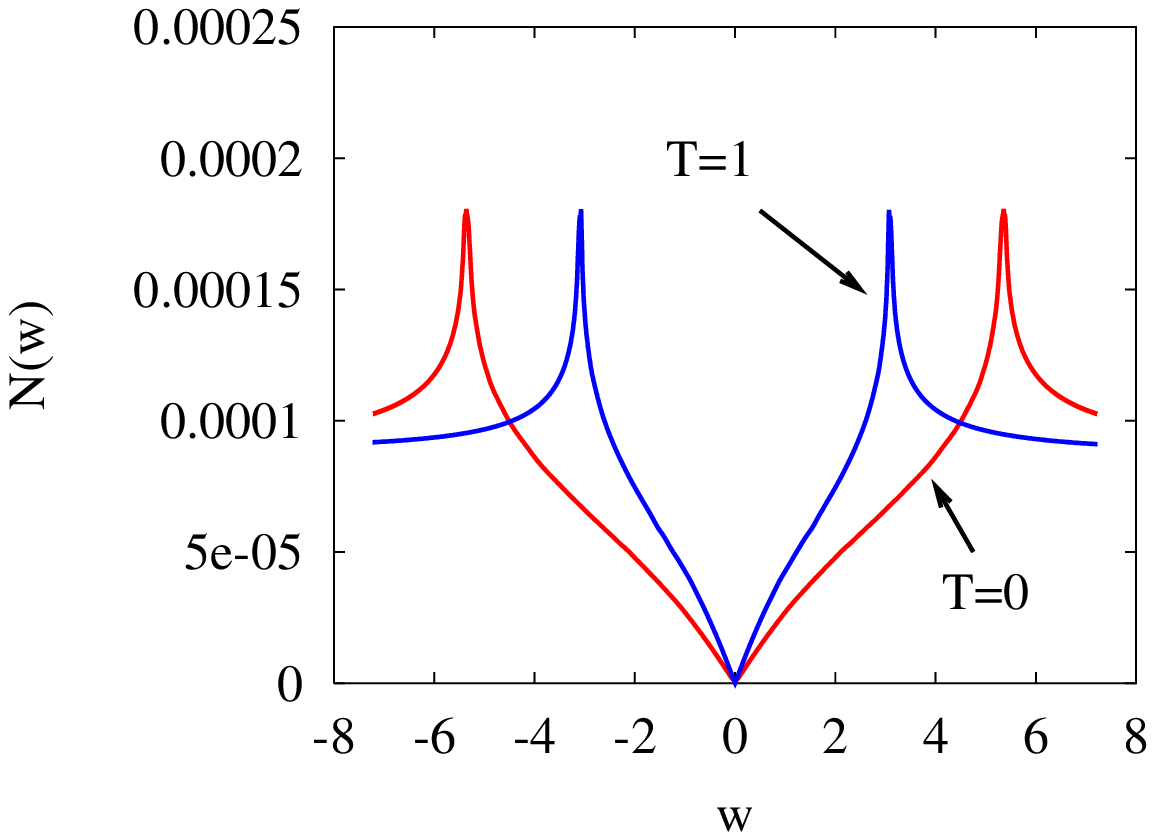}
  \end{array}
\)
\end{center}
\caption{The frequency dependence of the QDOS $N (\omega)$ 
for singlet symmetries at $T=0~{\rm K}$ and $T=1~{\rm K}$. 
States $^1 A_{1g}$ (``$s$-wave'') and $^1 B_{1g}$ (``$d_{xy}$-wave'')
are shown respectively in (a) and (b).
The parameters used are $|t_x| = 5800~{\rm K}$, $|t_y| = 1226~{\rm K}$,
$|t_z| = 48~{\rm K}$, and $\mu = -4101~{\rm K}$, 
with $T_c = 1.5~{\rm K}$.
$N (\omega)$ is in inverse temperature units $({\rm K}^{-1})$
and $\omega$ is in temperature units $({\rm K})$.
}
\label{fig:dos-sd-wave}
\end{figure*}
\begin{figure*}[ht!]
\begin{center}
\(
  \psfrag{w}{$\omega$}
  \psfrag{N(w)}{$N(\omega)$}
  \psfrag{T=0}{\hskip -0.2cm {\footnotesize \mbox{$T=0~{\rm K}$}}}
  \psfrag{T=1}{{\footnotesize \mbox{$T=1~{\rm K}$}}}
  \begin{array}{c@{\hspace{.5cm}}c}
  \multicolumn{1}{l}{\mbox{\bf (a)}} & 
  \multicolumn{1}{l}{\mbox{\bf (b)}} \\ [-0.5cm]
  \includegraphics[width=7.0cm]{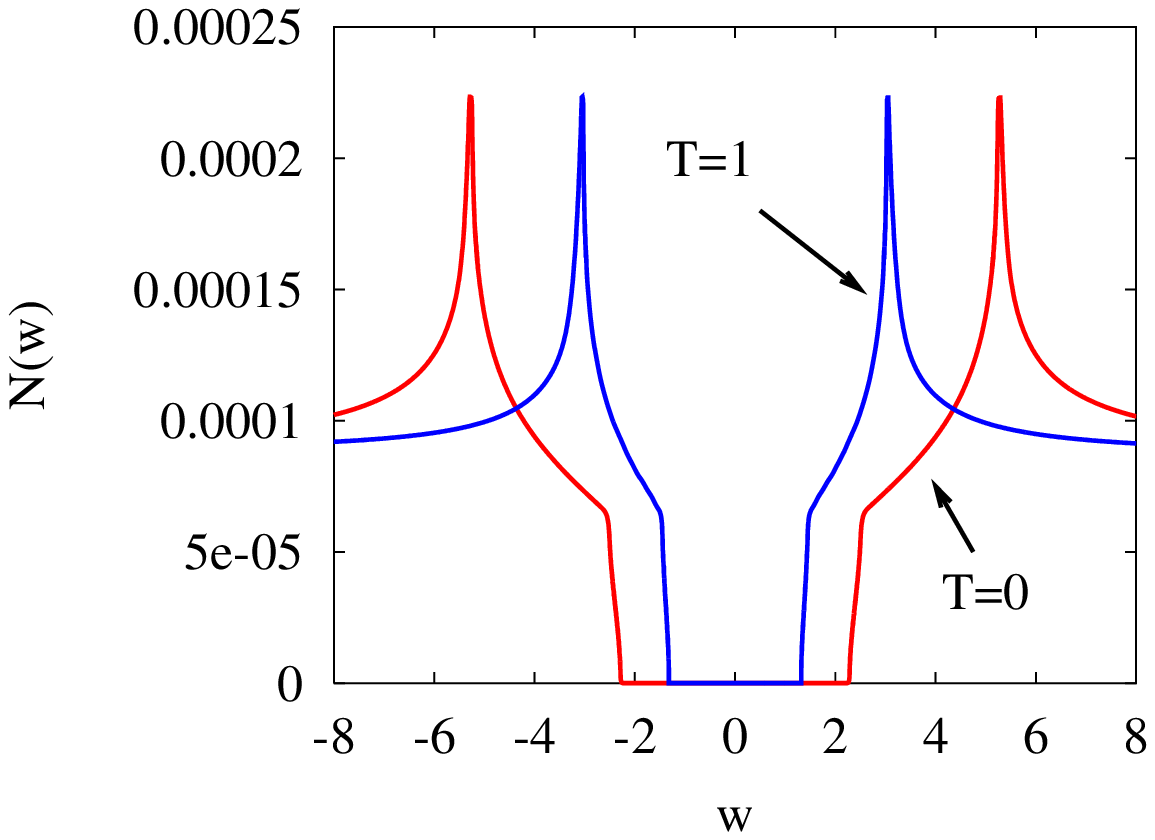} &
  \includegraphics[width=7.0cm]{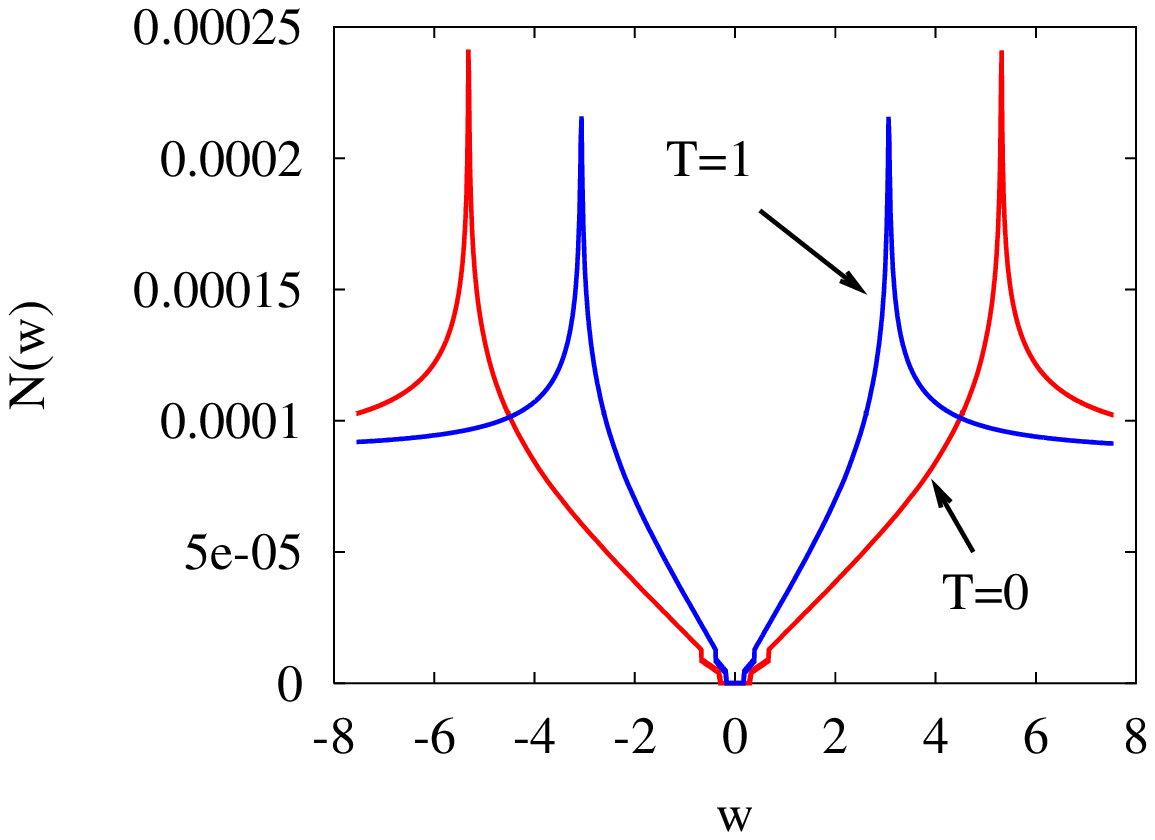} \\
  \multicolumn{1}{l}{\mbox{\bf (c)}} & 
  \multicolumn{1}{l}{\mbox{\bf (d)}} \\ [-0.5cm]
  \includegraphics[width=7.0cm]{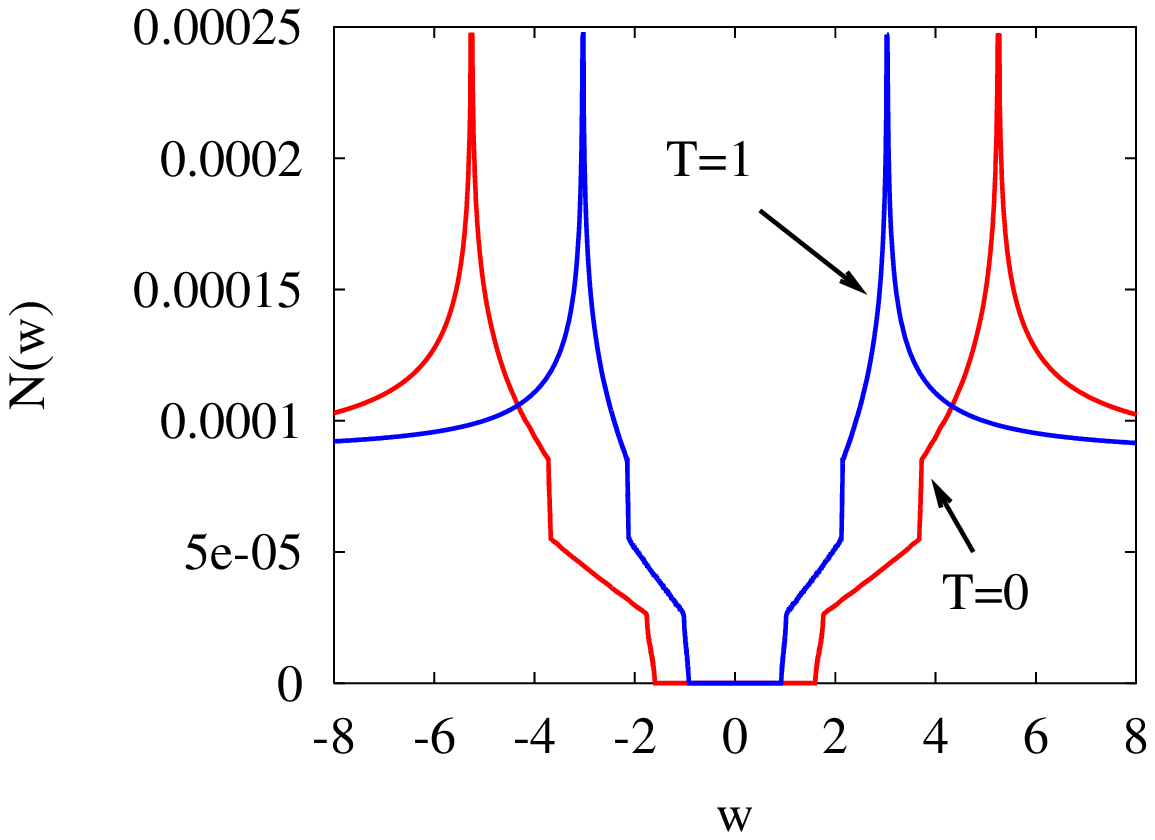} &
  \includegraphics[width=7.0cm]{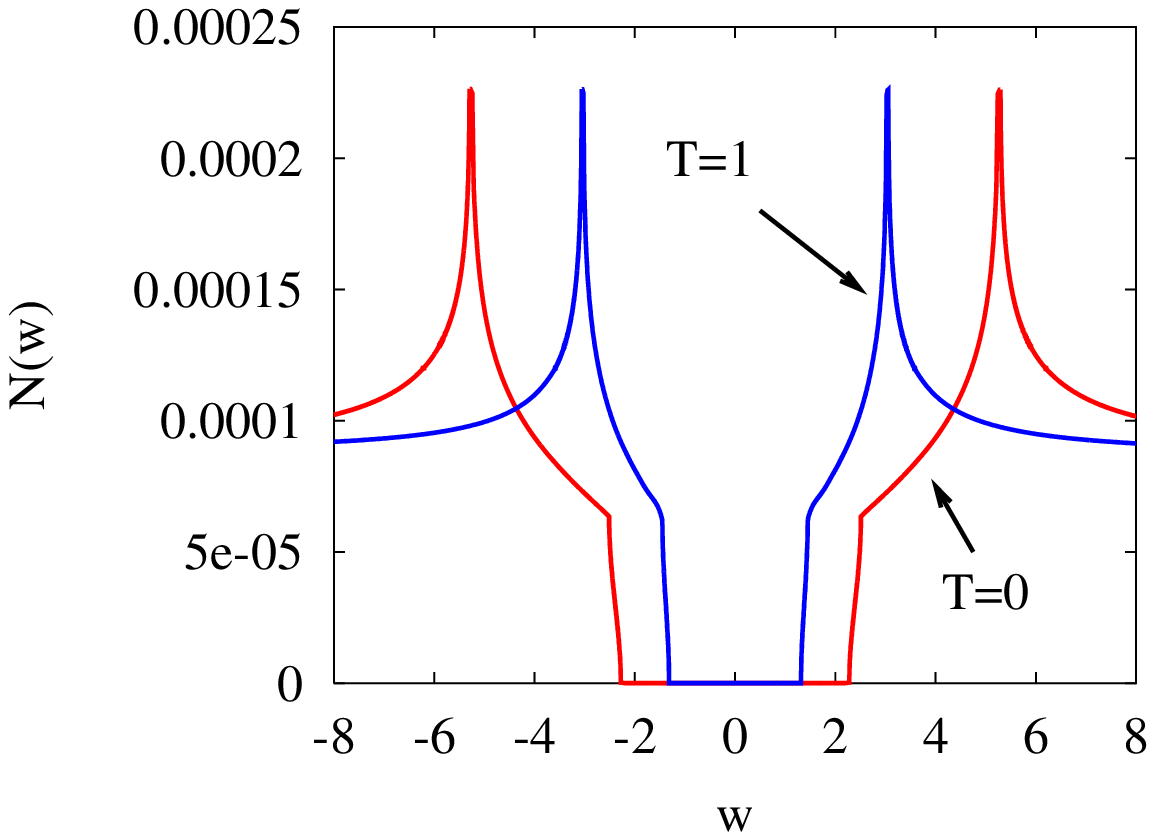} \\
  \end{array}
\)
\end{center}
\caption{Frequency dependence of the QDOS $N(\omega)$ 
for triplet states at $T=0~{\rm K}$ and $T=1~{\rm K}$.
In (a) the weak spin-orbit coupling state $^3B_{3u}$ is shown. 
In (b) strong spin-orbit coupling state $A_{1u}$ is shown
for $A = 0.20$, $B = 1.40$, $C = 0$;
in (c) for $A = 1.00$, $B = 1.00$, $C = 0$; 
in (d) for $A = 1.41$, $B = 0.10$, $C = 0$.
The parameters used are $|t_x| = 5800~{\rm K}$, $|t_y| = 1226~{\rm K}$ and
$|t_z| = 48~{\rm K}$, and $\mu = - 4101~{\rm K}$,
with $T_c = 1.5~{\rm K}$.
$N (\omega)$ is in inverse temperature units $({\rm K}^{-1})$
and $\omega$ is in temperature units $({\rm K})$.
}
\label{fig:dos}
\end{figure*}
\begin{figure}[hb!]
\begin{center}
  \psfrag{w}{$\omega$}
  \psfrag{N(w)}{$N(\omega)$}
  \includegraphics[width=7.0cm]{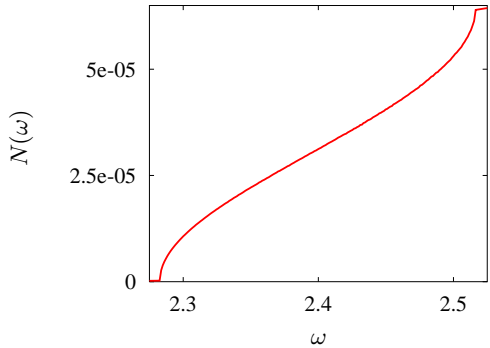} \\
\end{center}
\caption{A closer view of the $T=0$ curve in Fig.~\ref{fig:dos}(a) showing
the frequency dependence of the QDOS $N({\omega})$ near the gap edge 
for the $^3B_{3u}$ state.  
This feature extends over an energy range of about $0.20~{\rm K}$
($17~{\rm \mu eV}$).
$N (\omega)$ is in inverse temperature units $({\rm K}^{-1})$
and $\omega$ is in temperature units $({\rm K})$.
\label{fig:dos-px-zoom}
}
\end{figure}
\section{spin susceptibility}
\label{sec:susceptibility}
We obtain a general form of the
spin susceptibility tensor given by 
\begin{equation}
\label{eqn:chi-qw} 
\chi_{mn} (q_{\mu})
= -\mu_{B}^{2}
(P_{mn})_{\alpha \beta \gamma \delta}
\left[
S_{\alpha \beta \gamma \delta}(q_{\mu})
+
A_{\alpha \beta \gamma \delta}(q_{\mu})
\right],
\end{equation}
where we used Einstein's notation that repeated Greek indices indicate
summation, and the four-vector $q_{\mu} = ({\bf q}, i\nu)$. 
The tensor 
\begin{equation}
(P_{mn})_{\alpha \beta \gamma \delta} = 
{\tilde g}_m {\tilde g}_n 
(\sigma_{m})_{\alpha \beta}(\sigma_{n})_{\gamma \delta},
\end{equation}
contains the Pauli spin matrices and the
scaled gyromagnetic factors $\tilde{g}_{\ell} = g_{\ell}/2$.
The index $\ell$ takes into account the possibility
of anisotropies in the gyromagnetic
factors due to spin-orbit coupling. 
The other two tensors that appear in Eq.~\ref{eqn:chi-qw} are
\begin{widetext}
\begin{equation}
A_{\alpha \beta \gamma \delta}({\bf q}, i\omega) =
{\beta}^{-1} \sum_{{\bf k}, i\omega}
F^{\dagger}_{\alpha \gamma}({\bf k}- {\bf q}, i\nu -i\omega)
F_{\beta \delta}({\bf k},i\omega),
\end{equation}
\begin{equation}
S_{\alpha \beta \gamma \delta}(q_{\mu}) =
{\beta}^{-1} \sum_{{\bf k}, i\omega} 
G_{\delta\alpha}(-{\bf k}+ {\bf q},-i\nu+i\omega)
G_{\beta \gamma}( {\bf k},i\omega)
\delta_{\delta \alpha}\delta_{\beta \gamma},
\end{equation}
\end{widetext}
which contain the Green's functions
described in Eqs.~(\ref{eqn:anom-green}) and~({\ref{eqn:sing-green}),
respectively. 
In the limit of $\omega~\to~0$ and ${\bf q}~\to~{\bf 0}$,
we obtain 
an expression for the spin susceptibility of a triplet superconductor 
(including lattice and particle-hole asymmetry effects),
\begin{equation}
\label{eqn:chi-triplet}
\chi_{mn} ({\bf 0}, 0) = 
\sum_{\bf k} 
\left[\chi_{mn,1} ({\bf k})
+ \chi_{mn,2} ({\bf k})
\right],
\end{equation}
where the ${\bf k}$-dependent tensors have the forms
\begin{equation}
\label{eqn:chi1}
\chi_{mn,1} ({\bf k}) = {\tilde g}_m {\tilde g}_n \chi_{\parallel} ({\bf k}) 
{\it Re}~{\hat d}^{*}_{m} ({\bf k}) {\hat d}_{n} ({\bf k}),
\end{equation}
\begin{equation}
\label{eqn:chi2}
\chi_{mn,2} ({\bf k}) = {\tilde g}_m {\tilde g}_n \chi_{\perp} ({\bf k}) 
\left(
\delta_{mn} - {\it Re}~{\hat d}^{*}_{m} ({\bf k}) {\hat d}_{n} ({\bf k}) 
\right),
\end{equation}
with
${\hat d}_{n} ({\bf k}) = {d}_{n} ({\bf k})/ |d_{n} ({\bf k})| $.
Here, the parallel component is 
\begin{equation}
\label{eqn:chi-parallel}
\chi_{\parallel} ({\bf k}) = 
- 2 \mu_B^2 \dfrac{ \partial f (E_{\bf k})}{ \partial E_{\bf k} }, 
\end{equation}
while the perpendicular component is  
\begin{equation}
\label{eqn:chi-perpendicular}
\chi_{\perp} ({\bf k}) = 
2 \mu_B^2 
\dfrac{ d }{ d\xi_{\bf k} }
\left[ \dfrac{\xi_{\bf k}}{ 2 E_{\bf k}}
\left( 1 - 2 f(E_{\bf k}) \right)
\right]. 
\end{equation}
The expressions
derived in Eqs.~(\ref{eqn:chi-triplet}) to~(\ref{eqn:chi-perpendicular}) 
include some particle-hole 
symmetry effects, and are valid at all temperatures, 
within the BCS limits. They do not include, however, standard 
Fermi liquid corrections~\cite{leggett-75}, since these 
corrections are expected to be small for ${\rm (TMTSF)_2 PF_6}$ 
at lower pressures (near the SDW phase). In this pressure regime 
this compound behaves like a good Fermi liquid. 
However, the same can not
be said at higher pressures, where deviations from 
standard Fermi liquid behavior were reported~\cite{chaikin-95}.
The expression for $\chi_{mn} ({\bf 0}, 0)$ in 
Eq.~(\ref{eqn:chi-triplet}) also allows for anisotropies
in the scaled gyromagnetic factors ${\tilde g}_{\ell}$,
where $\ell = m, n$.
For instance, these anisotropies could
already exist in the normal state of the system. Thus, it
is important to measure experimentally such anisotropies
to determine the strength of the spin-orbit coupling
in the normal state. For the $D_{2h}$ orthorhombic group
$\chi_{mn}$ is diagonal, but in the
presence of spin-orbit coupling the normal state 
$\chi_{mn}^{(N)}$ can have, in principle, all diagonal elements different
from each other i.e., 
$\chi_{11}^{(N)} \ne \chi_{22}^{(N)} \ne \chi_{33}^{(N)} \ne \chi_{11}^{(N)}$.
In the complete absence of
spin-orbit coupling ${\tilde g}_{\ell} = 1$ for all $\ell$, 
and $\chi_{11}^{(N)} = \chi_{22}^{(N)} = \chi_{33}^{(N)} = \chi_N$.

%
\begin{figure*}
\begin{center}
\(
  \psfrag{C}{\hskip -0.8cm $\chi(T)/\chi_N^{}(T_c)$}
  \psfrag{T}{$T/T_c$}
  \psfrag{DT}{$\Delta(T)$}
  \psfrag{chi11}{$\chi_{11}^{}$}
  \psfrag{chi22}{$\chi_{22}^{}$}
  \psfrag{chi33}{$\chi_{33}^{}$}
  \begin{array}{c@{\hspace{.5cm}}c}
  \multicolumn{1}{l}{\hskip -0.3cm \mbox{\bf (a)}} & 
  \multicolumn{1}{l}{\hskip -0.3cm \mbox{\bf (b)}} \\ [-0.5cm]
  \psfrag{chi}{}
  \includegraphics[width=7.0cm]{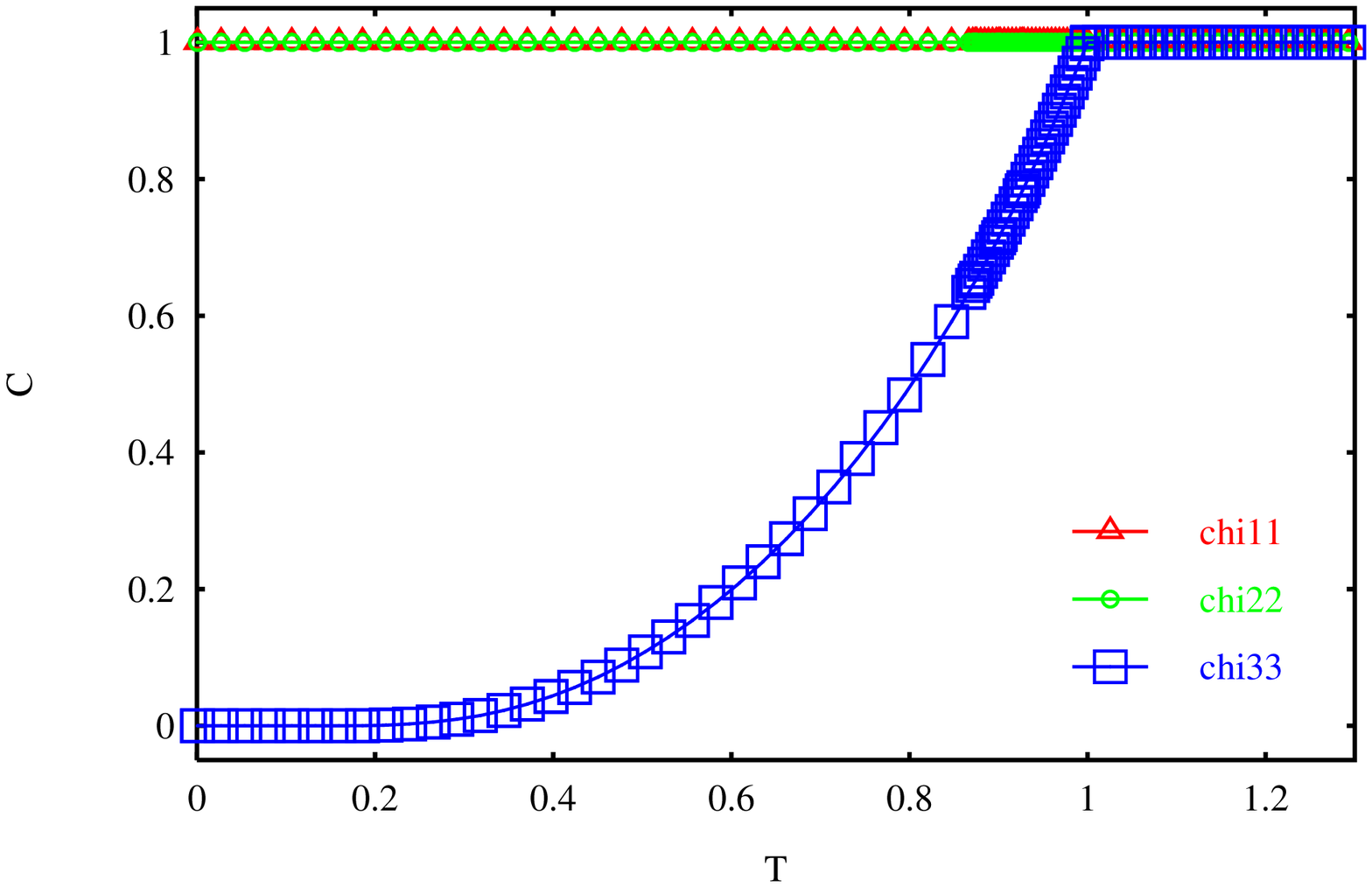} &
  \psfrag{chi}{}
  \includegraphics[width=7.0cm]{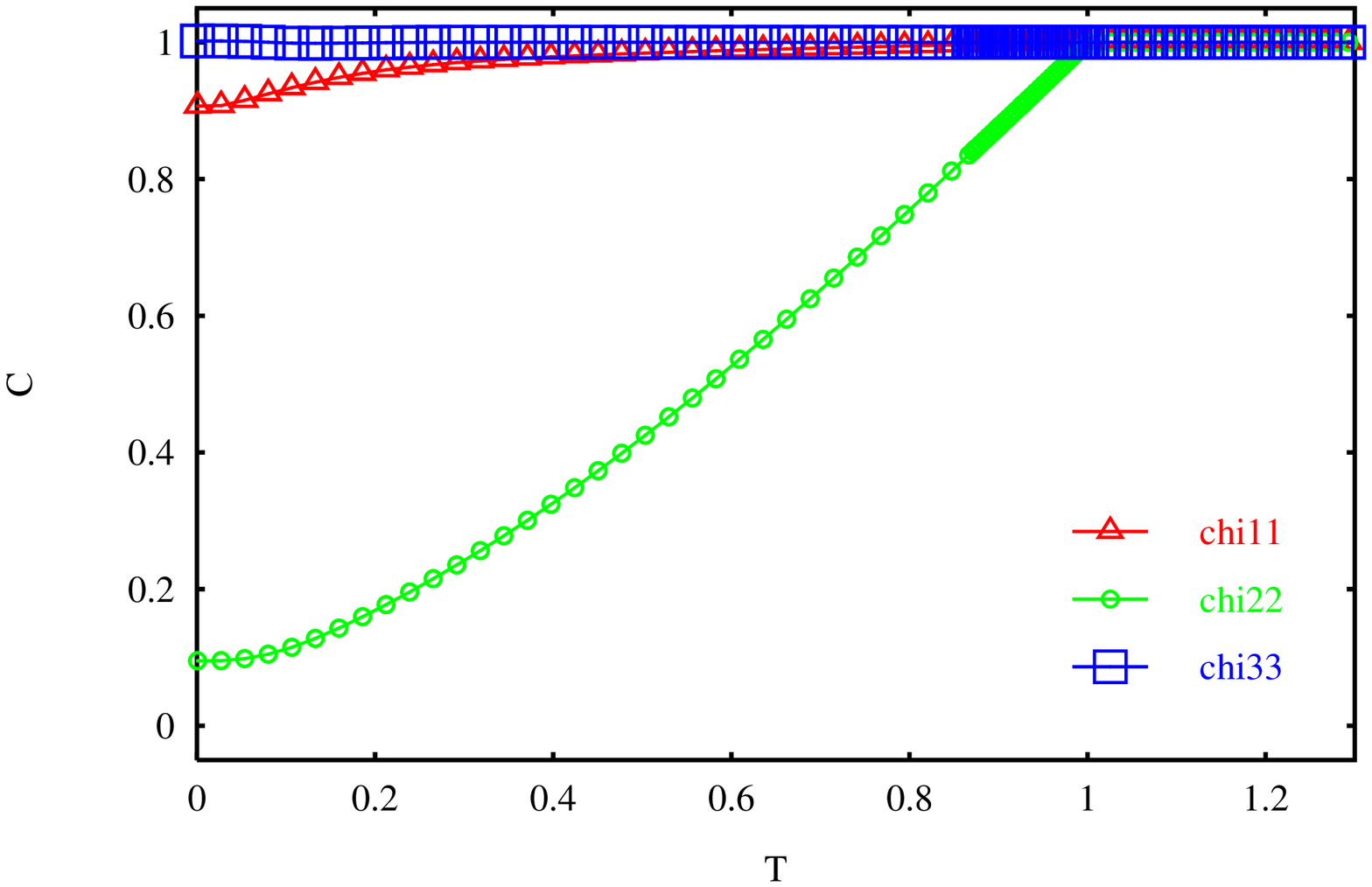} \\
  \multicolumn{1}{l}{\hskip -0.3cm \mbox{\bf (c)}} & 
  \multicolumn{1}{l}{\hskip -0.3cm \mbox{\bf (d)}} \\ [-0.5cm]
  \psfrag{chi}{}
  \includegraphics[width=7.0cm]{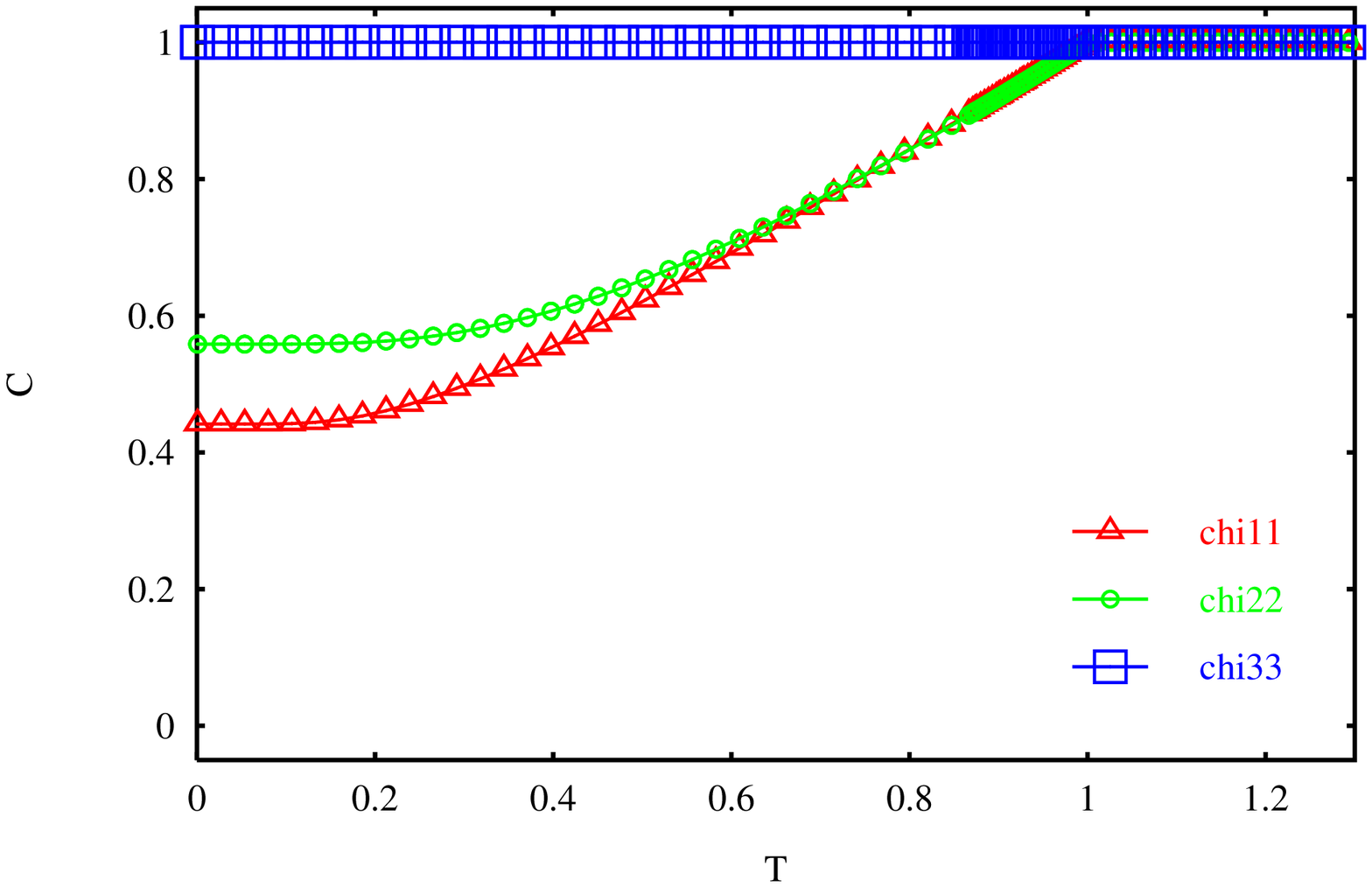} &
  \psfrag{chi}{}
  \includegraphics[width=7.0cm]{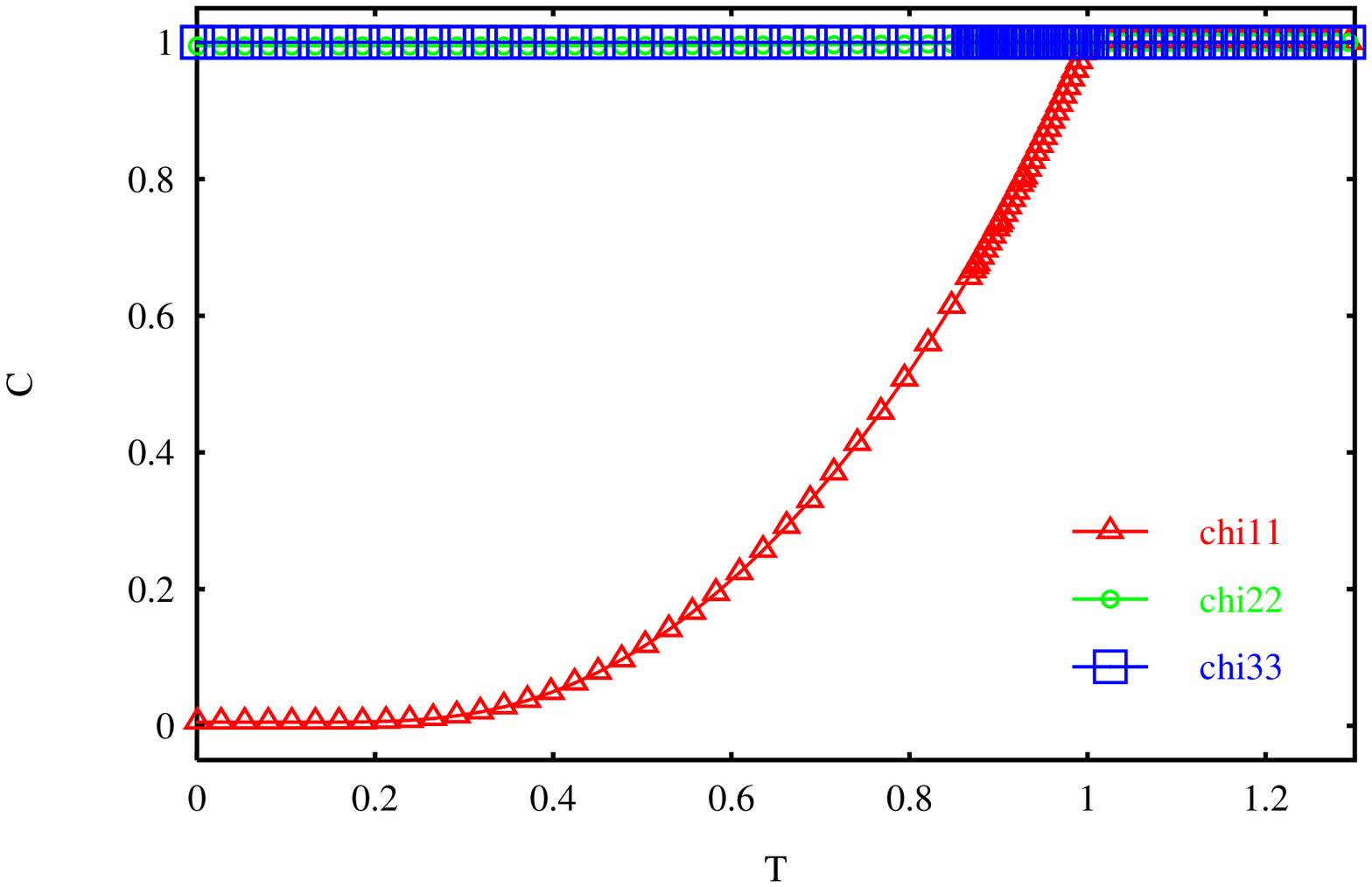}
  \end{array}
\)
\end{center}
\caption{ Plot of the theoretical 
uniform spin susceptibility tensor components
$\chi_{11}^{}$ (triangles); $\chi_{22}^{}$ (circles);
$\chi_{33}^{}$ (squares) at low temperatures. 
The weak spin-orbit coupling state $^3B_{3u}$ is shown in (a);
strong spin-orbit coupling state $A_{1u}$ is shown
in (b) for $A = 0.20$, $B = 1.40$, $C = 0$;
in (c) for $A = 1.00$, $B = 1.00$, $C = 0$; 
in (d) for $A = 1.41$, $B = 0.10$, $C = 0$. 
The parameters used are $|t_x| = 5800~{\rm K}$, $|t_y| = 1226~{\rm K}$, 
$|t_z| = 48~{\rm K}$, and $\mu = - 4101~{\rm K}$,
with $T_c = 1.5~{\rm K}$.
}
\label{fig:chi}
\end{figure*}

In Fig.~\ref{fig:chi},
the theoretical uniform $\chi_{mn}$ is shown  
only for the $^3 B_{3u}$ and $A_{1u}$ states. We do not discuss 
other strong spin-orbit coupling states $B_{1u}$ and $B_{2u}$,
which can also be fully gapped because the atomic spin-orbit coupling
in Bechgaard salts seems to be small (${\rm Se}$ is the heaviest
element).
The triangles correspond to $\chi_{11}$
circles to $\chi_{22}$, and squares to $\chi_{33}$.
The susceptibilities $\chi_{mn}$ are scaled by their
normal state values $\chi_{mn}^{(N)}$.
It is known experimentally (Knight
shift)~\cite{lee-00}
that the spin susceptibility of ${\rm (TMTSF)_2 PF_6}$ 
for ${\bf H} \parallel {\bf b}^{\prime}$ $(H = 2.38~{\rm T})$
is very close to $\chi_{b^{\prime}}^{(N)}$. 
Knight shift experiments with ${\bf H} \parallel {\bf a}$~\cite{lee-02}
$(H = 1.34~{\rm T})$ also indicate that the spin susceptibility is
very close to $\chi_{a}^{(N)}$.
However if experiments indicate that the normal state
susceptibilities obey the relation 
$\chi_{a}^{(N)} \approx \chi_{b^\prime}^{(N)} 
\approx \chi_{c^*}^{(N)} \approx \chi_N$, then
this is strongly suggestive that spin-orbit coupling
effects are small, as expected from the fact that
${\rm (TMTSF)_2 ClO_4}$ and ${\rm (TMTSF)_2 PF_6}$ have reasonably
light atoms.
Furthermore, in recent experiments, Lee {\it et. al.}~\cite{lee-02} observed
a coherence peak in the NMR relaxation rate $1/T_1$  
for ${\bf H} \parallel {\bf a}$, $H = 1.43~{\rm T}$.
In conventional superconductors the enhancement of $1/T_1$ 
near $T_c$ is often associated with a divergence of the
density of states at the edge of the energy gap~\cite{hebel-59}.
Although the weak spin-orbit coupling fully gapped state
$^3B_{3u} (a)$ (``$p_x$-wave'') (see Fig.~\ref{fig:dos})
does not exhibit a divergence in QDOS at the edge of the 
energy gap, the QDOS changes very rapidly at the edge and
may produce a peak or a kink in $1/T_1$ for $^{77} {\rm Se}$.

These results reinforce the evidence for triplet superconductivity
in (TMTSF)$_2$PF$_6$. 
For fields along the ${\bf c}^*$ direction the
superconducting state is easily destroyed, which makes impractical 
the Knight shift measurement for $^{77}$Se. 
This measurement requires magnetic fields of the order of one Tesla,
which is an order of magnitude higher than the upper critical field
along this direction~\cite{lee-97}.
For definiteness, we choose the unit vectors
${\hat 1}$ $(m = 1)$, ${\hat 2}$ $(m = 2)$ and 
$\hat 3$ $(m = 3)$,
to point along the ${\bf a}$, 
${\bf b}^{\prime}$, and ${\bf c}^*$ directions, respectively.

\subsection{Finite Magnetic Field Effects}
\label{sec:magnetic-field-effects}

In this section, we discuss qualitatively some
of the effects of a finite magnetic field upon the
spin susceptibility tensor. We consider first the
effects of vortices on the
spin susceptibility tensor within a simple two fluid model,
and second the possible rotation of the ${\bf d}$-vector in the
presence of a finite magnetic field.

\subsubsection{Role of Vortices}

Here, we would like to comment on the role of vortices
when the spin susceptibility tensor is measured. The Knight
shift measurements of Lee {\it et. al.}~\cite{lee-00,lee-02}
in ${\rm (TMTSF)_2 PF_6}$
had to be performed at reasonably high fields $H = 2.38~{\rm T}$
for ${\bf H} \parallel {\bf b}$ and 
$H = 1.43~{\rm T}$ for ${\bf H} \parallel {\bf a}$ in order to 
obtain a clear NMR signal. In these field regimes the
superconductor ${\rm (TMTSF)_2 PF_6}$ is in an inhomogeneous vortex state.
This vortex state was analysed by DMS~\cite{dupuis-93}, and
later shown by S\'a de Melo~\cite{sademelo-96} to correspond
to a rectangular Abrikosov vortex lattice (in the semiclassical
low field regime) and to a rectangular Josephson vortex lattice 
(in the quantum high field regime) 
for equal spin triplet pairing or singlet pairing.
We will use this result to construct a simple phenomenological
two-fluid model to describe the effect of vortices on the spin
susceptibility tensor, while a more sophisticated calculation is
under way~\cite{note-chi_mn}. 
Let us consider our superconductor at finite $T$ and $H$, 
and assume a semiclassical description of
the vortex state. Assume that the magnetic field is applied 
along the ${\hat n}$ direction $(H_{\hat n})$
and produces a rectangular Abrikosov lattice 
with lattice constants $\ell_{\hat \alpha}$ and $\ell_{\hat\beta}$, where
${\hat \alpha}$ and ${\hat\beta}$ are direction perpendicular to ${\hat n}$.
The density of vortices is 
$n_v = 1/\ell_{\hat \alpha} \ell_{\hat\beta}$. Assuming that the
vortices have  finite core size, characterized by lengths
$\xi_{\hat \alpha}$ and $\xi_{\hat\beta}$, and
that the vortex core is normal~\cite{core-states}, the normal
fraction of the superconductor is 
\begin{equation}
n_f =  
\dfrac{\xi_{\hat{\alpha}}   \xi_{\hat{\beta}}}
      {\ell_{\hat{\alpha}} \ell_{\hat{\beta}}} =
\dfrac{H_{\hat{n}}}
      {H_{c_2 \hat{n}}}
\end{equation}
Vortex core states have been extensively investigated in singlet 
superconductors~\cite{caroli-64,bardeen-69,kramer-74,ullah-90,gygi-91,
sdm-94,ichioka-96,tesanovic-98,sdm-99}; however, these states in
triplet superconductors remain to be investigated~\cite{core-states}.  
We treat core states very crudely as nearly 
{\it particle-in-a-well}-like with characteristic sizes 
controlled by the coherence lengths  
$\xi_{\hat{\alpha}}$, and $\xi_{\hat{\beta}}$ transverse to the
magnetic field direction. This is a crude improvement upon 
treating vortex cores as purely normal, while a more sophisticated
calculation of vortex core states is 
under way~\cite{note-chi_mn}.   
In the present case,
if the two fluids are viewed as nearly independent~\cite{boundaries},
then the total spin susceptibility tensor of the 
superconductor is the sum of the response
of its individual components.
Thus, we write
\begin{equation}
\label{eqn:chi-pheno}
\chi_{mn}^{} (T, H_{\hat n}) = 
  \chi_{mn}^{(C)} (T, H_{\hat n}) n_f + 
  \chi_{mn}^{(S)} (T, H_{\hat n}) ( 1 - n_f),
\end{equation}
where $\chi_{mn}^{(C)} (T, H_{\hat n})$ is the vortex core
contribution, and $\chi_{mn}^{(S)} (T, H_{\hat n})$ is the
superconducting contribution.
However it is best to illustrate the qualitative behavior  
of $\chi_{mn}^{} (T, H_{\hat n})$ via the ratio
\begin{equation}
\label{eqn:chi-magnetic-ratio}
R  (T, H_{\hat n}) = 
R_C^{} (T, H_{\hat n})n_f + R_S^{} (T, H_{\hat n}) ( 1 - n_f )
\end{equation}
Here $R  (T, H_{\hat n}) 
\equiv \chi_{mn}^{} (T, H_{\hat n}) / \chi_{mn}^{(N)} (T, H_{\hat n})$,
and $R_{j} (T, H_{\hat n}) 
\equiv \chi_{mn}^{(j)}  (T, H_{\hat n}) / \chi_{mn}^{(N)} (T, H_{\hat n})$,
with $j = C, S$.
In Fig.~\ref{fig:chi-magnetic-ratio}a the ratio $R  (T, H_{\hat n})$  
for a weak spin-orbit coupling triplet superconductor is illustrated
for fixed temperature $T < T_c$.
It is assumed that the applied magnetic field is perpendicular
to the ${\bf d}$-vector $({\bf H} \perp {\bf d})$,
and that ${\bf d}$-vector direction is not affected by the 
magnetic field. 
In this case, 
$0 \le R_C (T, H_{\hat n}) \le 1$, such that 
$R_C (T, H_{\hat n})$ changes from a smaller value at small $n_f$ 
at fixed temperature, and as $n_f$ tends to $1$, $R_C (T, H_{\hat n})$ 
converges to 1, given that the core spectrum becomes the normal state
continuum at $H_{\hat n} = H_{c_2 \hat n}$.
The superconducting term is assumed to be essentially constant for
$({\bf H} \perp {\bf d})$, and 
$R_S (T, H_{\hat n}) \approx 1$,
implying that the total ratio 
$ R (T, H_{\hat n}) \approx 1.$
These results are largely independent of temperature and field
within the approximation used. However, the discussion 
above serves only as a qualitative guide, while a more
sophisticated analysis is under way~\cite{note-chi_mn}.
In Fig.~\ref{fig:chi-magnetic-ratio}b the qualitatively similar
situations of a weak spin-orbit coupling singlet superconductor 
or weak spin-orbit coupling triplet superconductor
with ${\bf H} \parallel {\bf d}$
are illustrated.
In this case, the core term
$R_C (T, H_{\hat n})$ still satisfies the condition
$0 \le R_C (T, H_{\hat n}) \le 1$, and grows at fixed temperature
from a smaller value at small $n_f$ to $R_C (T, H_{\hat n}) = 1$  
as $n_f$ tends to $1$. In addition, the contribution of the 
superconducting part $R_S (T, H_{\hat n})$ 
depends strongly on magnetic field (at fixed temperature) 
growing from a small value at small $n_f$ to $R_S (T, H_{\hat n}) = 1$
as $n_f \to 1$. The condition
$0 \le R_S  (T, H_{\hat n}) \le 1$ 
is also satisfied. Therefore, $R (T, H_{\hat n})$ would
be very strongly dependent on magnetic field (at fixed temperature) in  
(a) the weak spin-orbit triplet case for ${\bf H} \parallel {\bf d}$ , 
or in (b) the weak spin-orbit singlet case.
\begin{figure}
\begin{center}
\(
  \psfrag{r}{$R$}
  \psfrag{rC}{$R_{C}$}
  \psfrag{rS}{$R_{S}$}
  \psfrag{rT}{$R$}
  \psfrag{H}{$H$}
  \psfrag{Hc2}{$H_{c_2}$}
  \psfrag{0}{$0$}
  \psfrag{1}{$1$}
  \begin{array}{c}
   \multicolumn{1}{l}{\hspace{-0.2cm}\mbox{\bf (a)}} \\
   \includegraphics[width=7.0cm]{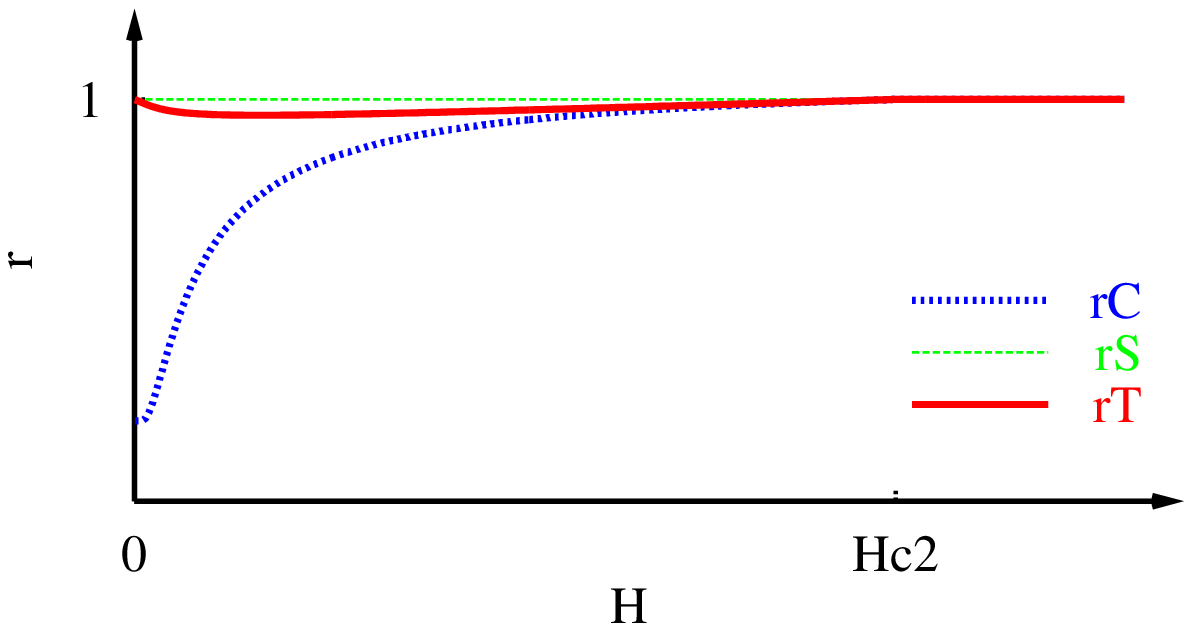} \\
   \multicolumn{1}{l}{\hspace{-0.2cm}\mbox{\bf (b)}} \\
   \includegraphics[width=7.0cm]{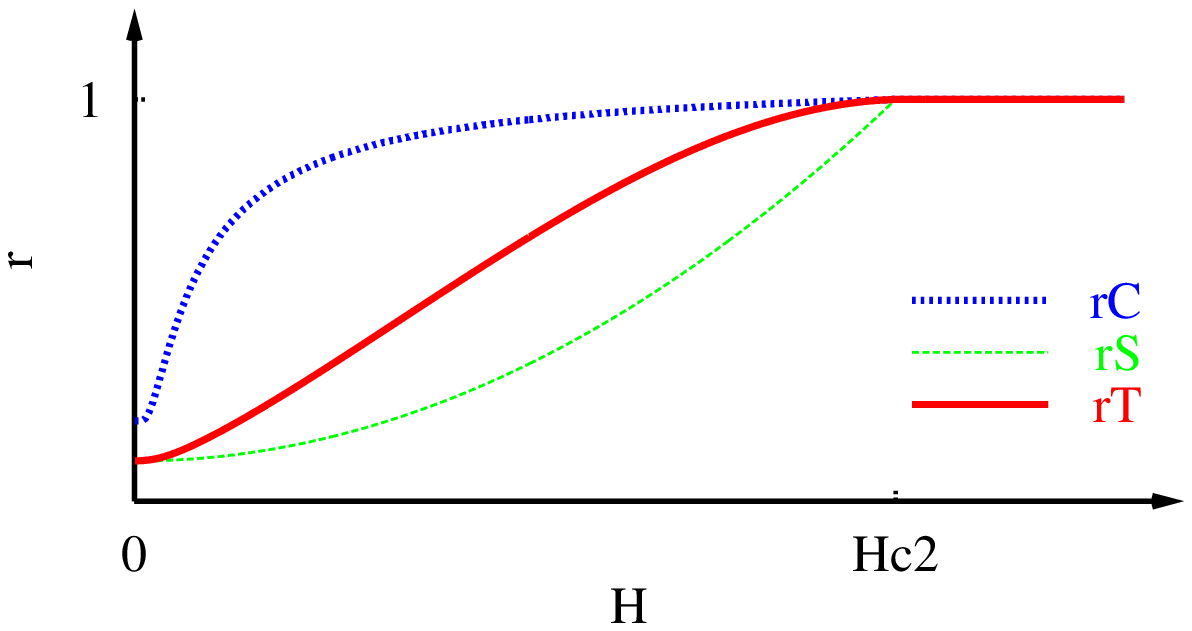}
\end{array}\)
\end{center}
\caption{
Schematic plots of the ratios 
$R$ (solid line);
$R_C$ (dashed line); and 
$R_S$ (dotted line) are shown in 
(a) for the case of a weak spin-orbit coupling triplet superconductor
with ${\bf H} \perp {\bf d}$; and in 
(b) for the case of a weak spin-orbit coupling singlet superconductor or
weak spin-orbit coupling triplet superconductor
with ${\bf H} \parallel {\bf d}$.
The ratio $R$ is connected to $R_C$ and $R_S$ via Eq.~\ref{eqn:chi-magnetic-ratio}.
(See text following Eq.~\ref{eqn:chi-magnetic-ratio} for definition of
ratios $R$, $R_C$ and $R_S$.)
}
\label{fig:chi-magnetic-ratio}
\end{figure}
%
%
Thus, if measurements of temperature and magnetic field
dependence of $\chi_{mn} (T, H_{\hat n})$ in 
${\rm (TMTSF)_2 PF_6}$ are such that
$R_S (T, H_{\hat n}) \approx 1$ (being largely 
independent of magnetic field at fixed temperature), 
then a weak spin-orbit triplet state would be consistent with 
such finding for ${\bf H} \perp {\bf d}$.
It is possible, however, that a small magnetic field
can rotate the ${\bf d}$-vector in the weak spin-orbit
coupling limit, and this effect is discussed next.

\subsubsection{ Rotation of {\bf d}-vector }

In what follows, we will discuss the possible rotation
of ${\bf d}({\bf k})$, not including 
vortex core contributions, i.e., we will consider effectively
only $\chi_{mn}^{(S)} (T, H_{\hat n})$. However, similar
effects are expected for the core contribution. 

For the orthorhombic symmetry 
$\chi_{mn}$ is diagonal and
is calculated under the assumption that
${\bf d}({\bf k})$ is constant, i.e., the direction of ${\bf d}({\bf k})$
is assumed not to change upon the application of a small magnetic field. 
In the case of state $A_{1u}$ (strong spin-orbit coupling) 
a small magnetic field cannot rotate ${\bf d}({\bf k})$
which is pinned to a particular lattice direction. 
In the strong spin-orbit coupling case $\chi_{mn}$ is still
diagonal, however the diagonal components are not equal in general.
Thus, the experimentally measured $\chi_{mn}^{ex}$ and the
theoretically calculated $\chi_{mn}^{th}$ (at constant ${\bf d}({\bf k})$)
should agree for small enough magnetic fields 
(see Fig.~\ref{fig:chi}b~\ref{fig:chi}c, and ~\ref{fig:chi}d).
However, in the case of state $^3B_{3u}(a)$ (weak spin-orbit coupling) 
a small magnetic field can easily rotate ${\bf d}({\bf k})$
to be perpendicular to ${\bf H}$ (barring
any additional orbital effects due to the coupling
of ${\bf H}$ with charge degrees of freedom), 
thus minimizing the magnetic free-energy
$F_{mag} = - H_m \chi_{mn}H_n/2$. In the weak spin-orbit coupling case, 
$\chi_{mn}^{ex} \approx \chi_{N} \delta_{mn}$, where
$\chi_N$ is the normal state value, for any direction of ${\bf H}$.
Thus, what experimentalists measure, $\chi_{mn}^{ex}$, and 
what we calculate, $\chi_{mn}^{th}$, 
(shown in Fig.~\ref{fig:chi}a)
would be different. (This point was considered 
by Leggett~\cite{leggett-75} in his discussion 
of the phases of liquid $^3$He).
Another important distinguishing feature between 
strong spin-orbit and weak spin-orbit coupling scenarios 
is the magnitude of the flop field $H_{f}$ required to rotate 
the ${\bf d}$-vector from a preferred lattice direction.
In the case of strong spin-orbit coupling the flop fields
are in the Tesla range, while for weak spin-orbit coupling
the flop fields are in the Gauss range.
For instance, based on the discussion above, 
the state $^3B_{3u}(a)$ (weak spin-orbit coupling ``$p_x$-wave'')
would present no Knight shift for any direction 
of the externally applied magnetic field provided
that the magnitude of ${\bf H}$ is larger than
the small spin-orbit pinning field $H_f$ for any
given direction. 
Under these conditions, no Knight shift should
be observed for fields along the ${\bf a}$, 
${\bf b}^{\prime}$ or ${\bf c}^*$-axis or 
for any set of angles with respect to these axis.
Furthermore, in the weak spin-orbit coupling state
$^3B_{3u}(a)$ (``$p_x$-wave''), the normal state
susceptibilities should be nearly identical, i.e. 
$\chi_{11}^{(N)} \approx \chi_{22}^{(N)} \approx \chi_{33}^{(N)} = \chi_N$,
since the scaled gyromagnetic factors 
${\tilde g}_{\ell} \approx 1$ for all $\ell$. 

The spin susceptibility tensor for a triplet
superconductor is very sensitive to magnetic
fields both in the weak or strong spin-orbit coupling scenarios 
since the applied field can depin the ${\bf d}$-vector from 
a preferred lattice direction 
and cause a flop transition with a corresponding change in the
spin susceptibility.  
This means that with increasing magnetic field at fixed
temperature $T$, $H > H_{c_1}$, and with 
${\bf H} \parallel {\bf d}$ initially, the ${\bf d}$-vector flops
at $ H = H_{f} < H_{c_2}$. Therefore, beyond $H_f$ the
${\bf d}$-vector becomes perpendicular to the applied 
magnetic field $({\bf d} \perp {\bf H})$.
This flop transition produces a discontinuous 
change in the spin susceptibility from smaller 
to larger (normal) values, as the ${\bf d}$-vector
rearranges itself from ${\bf d} \parallel {\bf H}$ to 
${\bf d} \perp {\bf H}$.

%
\begin{figure}
\begin{center}
\(
  \begin{array}{c@{\hspace{.01cm}}c}
   \multicolumn{1}{c}{\mbox{\bf (a)}} & 
   \multicolumn{1}{c}{\mbox{\bf (d)}} \\ [-0.25cm]
   \psfrag{K}{}
   \psfrag{1}{$1$}
   \psfrag{Hexp}{$H_{min}$}
   \psfrag{Hc2}{$H_{c_2}$}
   \psfrag{Hf}{$H_{f}$}
   \includegraphics[width=4.4cm]{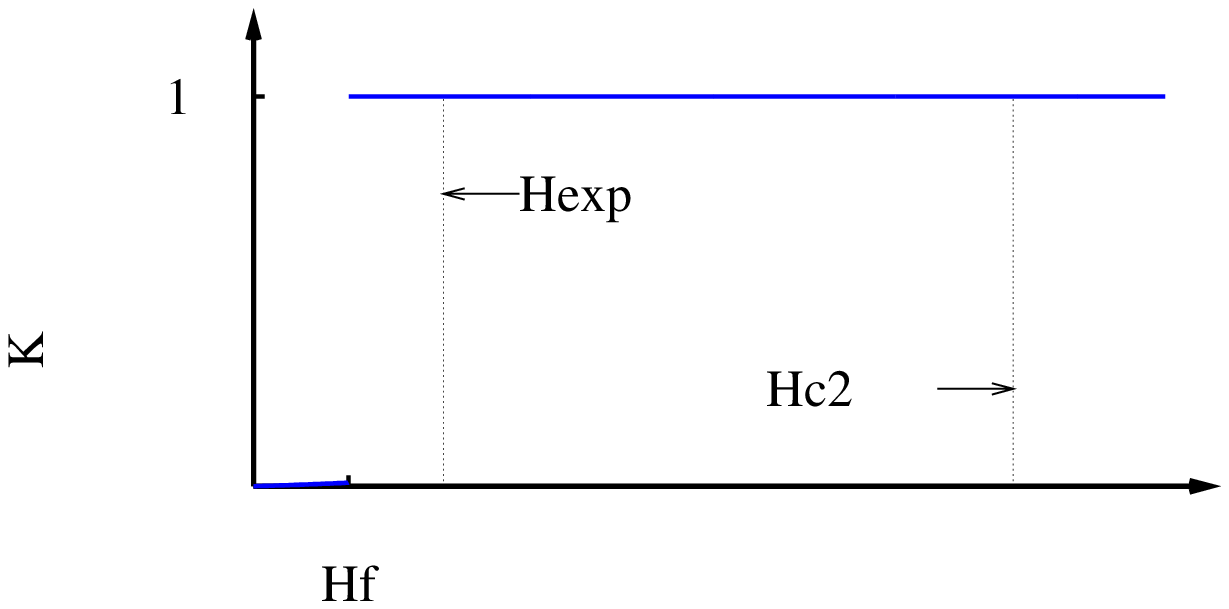} &
   \psfrag{K}{}
   \psfrag{1}{$1$}
   \psfrag{Hexp}{$H_{min}$}
   \psfrag{Hc2}{$H_{c_2}$}
   \psfrag{Hf}{$H_{f}$}
   \includegraphics[width=4.4cm]{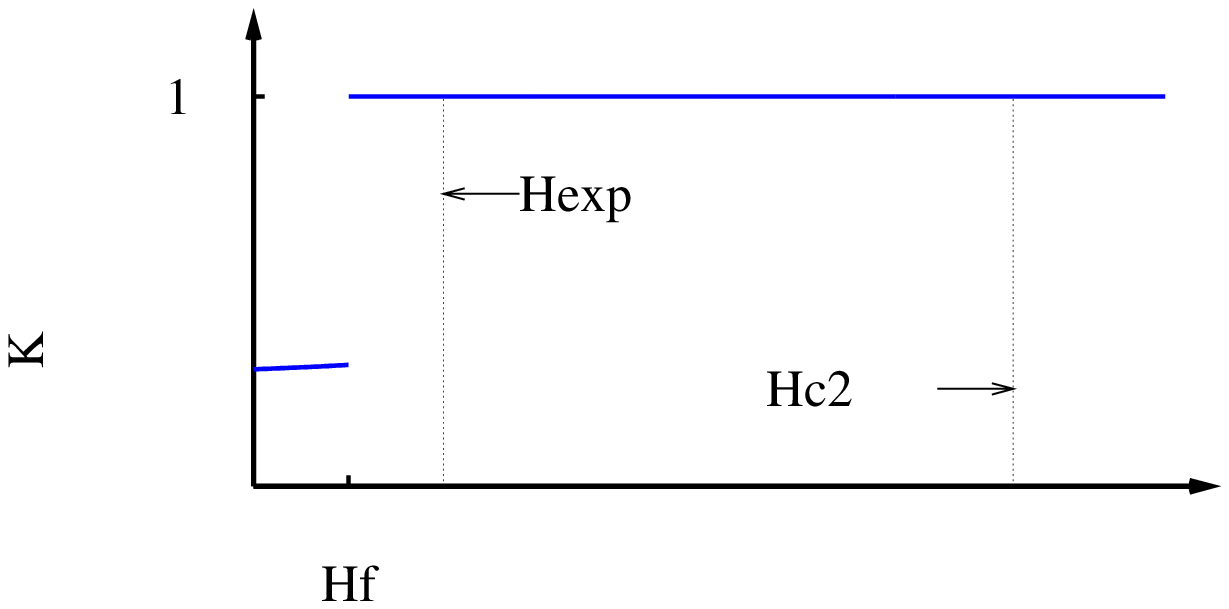} \\ [0.1cm]
   \multicolumn{1}{c}{\mbox{\bf (b)}} & 
   \multicolumn{1}{c}{\mbox{\bf (e)}} \\ [-0.25cm]
   \psfrag{K}{$\chi_{nn}/\chi_{nn}^{(N)}$}
   \psfrag{1}{$1$}
   \psfrag{Hf}{$H_{f}$}
   \includegraphics[width=4.4cm]{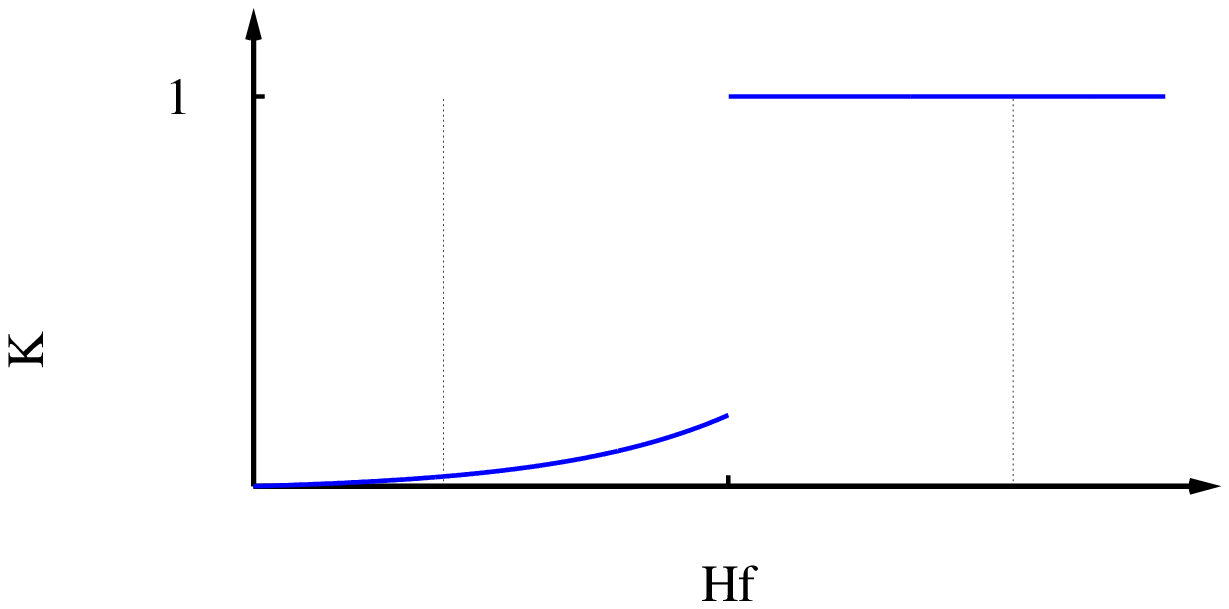} &
   \psfrag{K}{}
   \psfrag{1}{$1$}
   \psfrag{Hf}{$H_{f}$}
   \includegraphics[width=4.4cm]{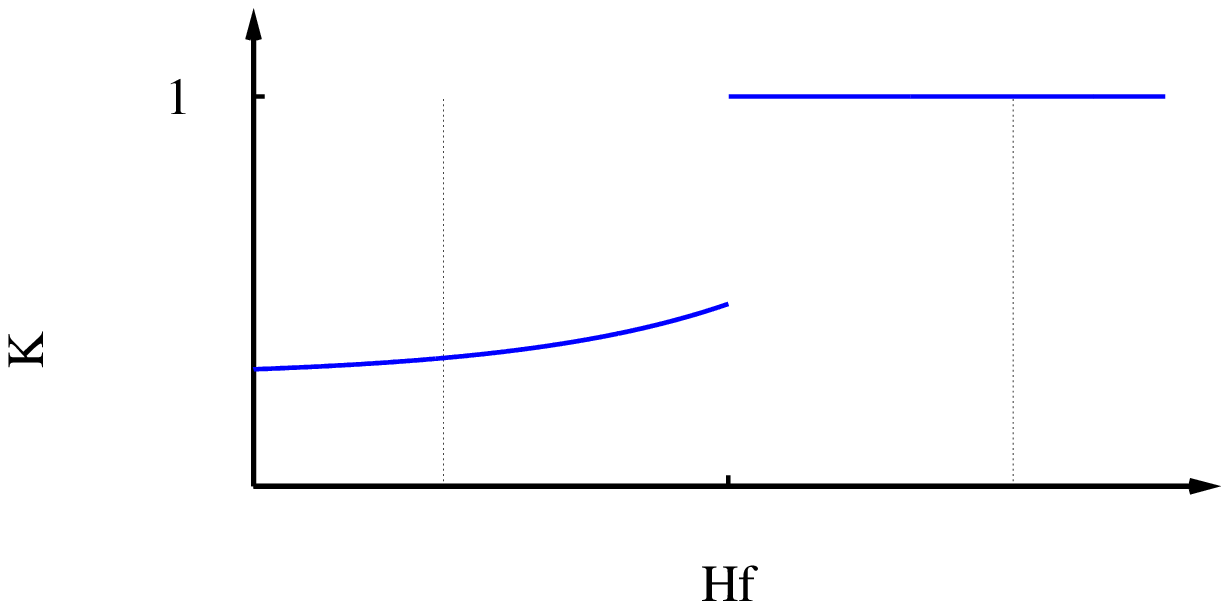} \\ [0.1cm]
   \multicolumn{1}{c}{\mbox{\bf (c)}} & 
   \multicolumn{1}{c}{\mbox{\bf (f)}} \\ [-0.25cm]
   \psfrag{K}{}
   \psfrag{1}{$1$}
   \psfrag{Hf}{$H_{f}$}
   \includegraphics[width=4.4cm]{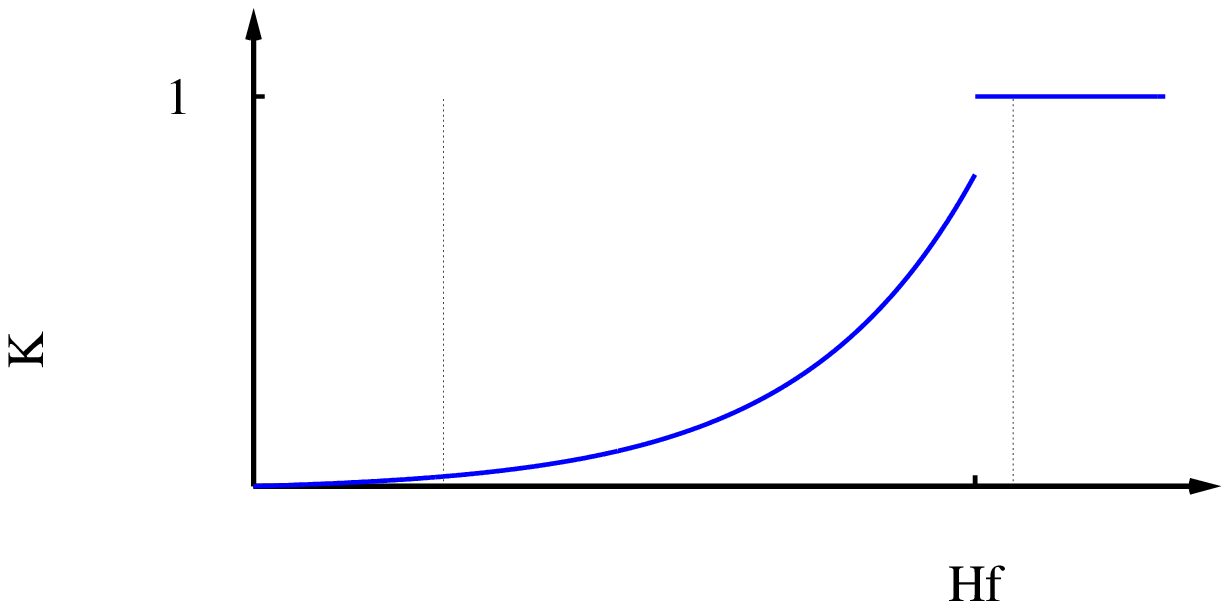} &
   \psfrag{K}{}
   \psfrag{1}{$1$}
   \psfrag{Hf}{$H_{f}$}
   \includegraphics[width=4.4cm]{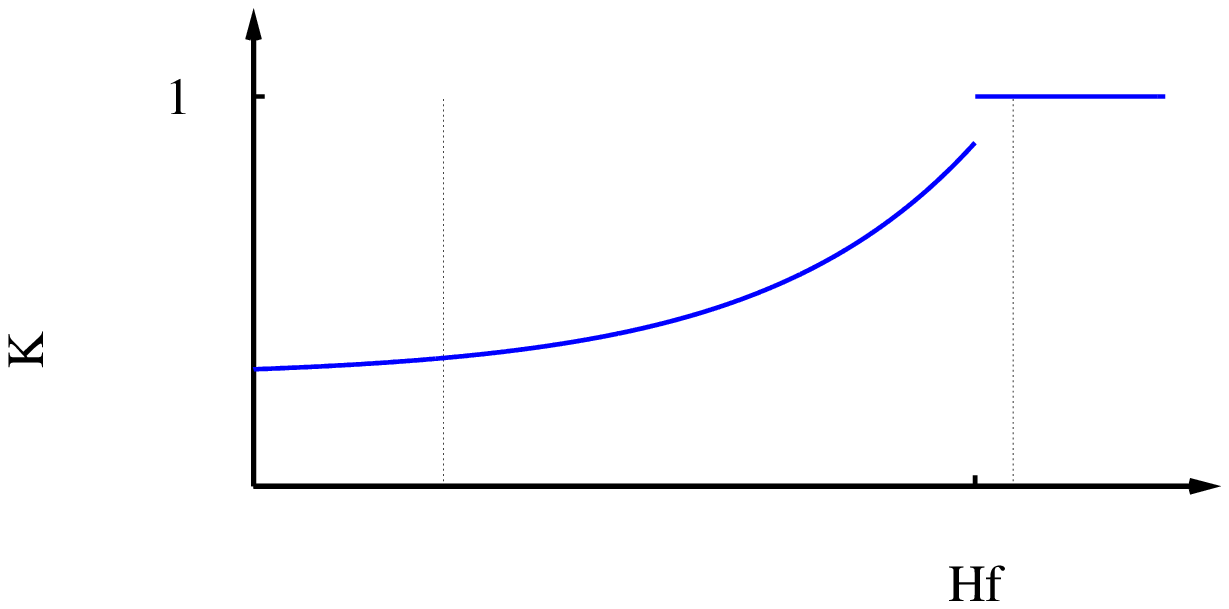}
   \end{array}
\)
\end{center}
\caption[a,b]{
Scenarios for the behavior of the experimental 
$\chi_{nn}/\chi_{nn}^{(N)}$ considering
that the ${\bf d}$-vector is pinned along the ${\hat n}$ direction.
The ratio $\chi_{nn}/\chi_{nn}^{(N)}$ is schematically plotted
as a function of applied magnetic field 
at a constant $T=0$ (left set) and $T > 0$ (right set).
The lower upper critical field 
$H_{c_1}^{(\hat n)}$
is assumed to be very small and it is not shown.
In each of these graphs, the left and right vertical dotted lines
represent the minimum field $(H_{min}^{(\hat n)})$ 
required to obtain Knight shift experimental
data and the upper critical field $(H_{c_2}^{(\hat n)})$ 
corresponding a fixed temperature T.
Three possible flop transition scenarios
at $T = 0$: 
(a) $H_f^{(\hat n)} < H_{min}^{(\hat n)} < H_{c_2}^{(\hat n)}$ 
(flop transition
beyond present experimental reach);
(b) $H_{min}^{(\hat n)} < H_{f}^{(\hat n)} < H_{c_2}^{(\hat n)}$ 
(flop transition
can be observed if jump in susceptibility is large enough);
(c) $H_{min}^{(\hat n)} < H_{f}^{(\hat n)} \lesssim H_{c_2}^{(\hat n)}$ 
(flop transition
may be difficult to be observed since jump in susceptibility
may be too small). Similar scenarios are illustrated 
in (d), (e) and (f) for $T > 0$.
}
\label{fig:flop}
\end{figure}
%

%
%

Consider for definiteness that the ${\bf d}$-vector is
pinned along a given lattice direction ${\hat n}$, and that
${\bf H} \parallel {\bf d}$. 
The observability of this flop transition depends on
how the magnitude $H_{f}^{(\hat n)}$ compares to 
the magnitudes of the minimum field required to perform a 
Knight shift experiment, $H_{min}^{(\hat n)}$,
the lower $H_{c_1}^{(\hat n)}$ and 
the upper $H_{c_2}^{(\hat n)}$ 
critical fields along the given direction $\hat n$.
Three possible scenarios are illustrated in Fig.~\ref{fig:flop},
where $H_{c_1}^{(\hat n)}$ is considered to be very small and it is not
indicated in the figure. 
If $H_{f}^{(\hat n)} \ll H_{min}^{(\hat n)}$ 
then the transition cannot be observed in
a Knight shift experiment. 
If $H_{f}^{(\hat n)} \lesssim H_{c_2}^{(\hat n)}$ then, depending on 
experimental resolution, the transition may not
be resolvable and the data would seem characteristic of a singlet
response.
However, the flop
transition could be easily measured 
in the intermediate regime 
$H_{min}^{(\hat n)} < H_f^{(\hat n)} < H_{c_2}^{(\hat n)}$.  
If ${\bf d}$ was pinned along ${\hat n} = {\bf a}$, 
then for ${\bf H} \parallel {\bf a}$ such that 
$H > H_f^{({\bf a})}$ a flop transition could take place.
Similarly situations would occur if the ${\bf d}$-vectors 
were pinned along ${\hat n} = {\bf b^{\prime}}$ $({\bf c^*})$,
then for ${\bf H} \parallel {\bf b^\prime}$ $({\bf c^*})$ such that 
$H > H_f^{({\bf b^\prime})}$ $( H > H_f^{({\bf c})})$ 
a flop transition could take place.
So far there is no experimental evidence of this
possible ${\bf d}$-vector flop transition.

\subsubsection{Connection to Anisotropy Inversion in $H_{c_2}$}
It is also important to make a connection between 
spin susceptibility measurements~\cite{lee-00,lee-02} and
upper critical field measurements~\cite{lee-97} 
in ${\rm (TMTSF)_2 PF_6}$. 
Lee {\it et. al.}~\cite{lee-97} observed an anisotropy inversion
between $H_{c_2}^{({\bf a})}$ and $H_{c_2}^{({\bf b^\prime})}$ 
at $H^* \approx 1.6~T$.
It was suggested theoretically~\cite{lebed-00} that this
anisotropy inversion was related to the presence of
a component of the ${\bf d}$-vector along the ${\bf a}$ direction, 
such that $H_{c_2}^{({\bf a})}$  $({\bf H} \parallel {\bf a})$ would be 
paramagnetically limited. In addition, it was suggested that~\cite{lebed-00}
the ${\bf d}$-vector would have zero component 
along the ${\bf b^\prime}$ direction, 
such that $H_{c_2}^{({\bf b^\prime})}$  $({\bf H} \parallel {\bf b^\prime})$ 
would not be paramagnetically limited. While this suggestion is
somewhat appealing, it seems to indicate that the ${\bf d}$-vector
has a strongly pinned component along the ${\bf a}$ axis.

A comparison between 
the upper critical field experiments of Lee {\it et. al.}~\cite{lee-97}
for ${\rm (TMTSF)_2 PF_6}$, with the upper critical
field experiments of Shivaram {\it et. al.}~\cite{shivaram-86}
for ${\rm UPt_3}$ can be illuminating. In ${\rm UPt_3}$ an
anisotropy inversion occurs between the upper critical
fields between $H_{c_2}^{({\bf a})}$ and $H_{c_2}^{({\bf c})}$ 
at $H^* \approx 1.9~T$. This anisotropy inversion in ${\rm UPt_3}$
was successfully explained by Choi and Sauls~\cite{sauls-91} 
under the assumption that $H_{c_2}^{({\bf c})}$ is paramagnetically
limited, while $H_{c_2}^{({\bf a})}$ is not. The nearly perfect fitting
of the experimental curves of ${\rm UPt_3}$ required the assumption that the
${\bf d}$-vector was locked to the ${\bf c}$ axis by a strong
spin-orbit coupling. Considering that the values of the anisotropy 
inversion field $H^*$ for ${\rm (TMTSF)_2 PF_6}$
and ${\rm UPt_3}$ are very similar in magnitude,
the theoretical interpretation~\cite{lebed-00} of the
anisotropy inversion in ${\rm (TMTSF)_2 PF_6}$
seems to assume implicitly a strong spin-orbit coupling, 
which is very unlikely for Bechgaard salts given that 
the heaviest element is ${\rm Se}$.
The ratio $r_{so}$ between (atomic) spin-orbit couplings in 
${\rm (TMTSF)_2 PF_6}$ (for ${\rm Se}$) and 
${\rm UPt_3}$ (for ${\rm U}$) can be 
estimated to be small: $r_{so} \lesssim 0.15$.
This implies that the physical origin of the anisotropy inversion
in ${\rm (TMTSF)_2 PF_6}$ should not be implicitly or 
explicitly the same as in ${\rm UPt_3}$.
Furthermore, the theoretical suggestion connecting the anisotropy
inversion with spin susceptibility measurements~\cite{lebed-00}
would lead to $\chi_{a} \approx 0.2 \chi_N$ at $T = 0$. 
However, experiments indicate that $\chi_{a} \approx \chi_N$~\cite{lee-02}
in ${\rm (TMTSF)_2 PF_6}$ $(T_c = 1.2~K)$ down to $T = 0.32~K$.  

It is important to note that the 
related compound ${\rm (TMTSF)_2 AsF_6}$ has an 
SDW phase below $T_c = 12~K$ with easy, intermediate and
hard axis along the ${\bf b^\prime}$, ${\bf a}$, and ${\bf c^*}$
directions, respectively~\cite{mortensen-82}.
This system also presents a spin-flop transition where the
spin orientation flops from the ${\bf b^\prime}$ axis to an 
orientation predominantly parallel to ${\bf a}$~\cite{mortensen-82}. 
Given that the SDW phase in ${\rm (TMTSF)_2 PF_6}$ 
is very similar to the SDW phase 
in ${\rm (TMTSF)_2 AsF_6}$~\cite{mortensen-82}, 
we infer that the connection between the experimental results 
$\chi_{a} \approx \chi_N$ and 
$\chi_{b^\prime} \approx \chi_N$~\cite{lee-00,lee-02} with
the anisotropy inversion of the upper critical fields
$H_{c_2}^{({\bf a})}$ and $H_{c_2}^{({\bf b^\prime})}$ in 
${\rm (TMTSF)_2 PF_6}$ may require a deeper understanding
of the interplay between the spin-density-wave
and the superconducting phases. Thus, 
it is important to search for 
an $H_{c_2}$ anisotropy inversion in the sister compound ${\rm (TMTSF)_2 ClO_4}$
in order to have a more complete picture of the superconducting state 
in the Bechgaard salt family ${\rm (TMTSF)_2 X}$.
To our knowledge detailed experimental studies of $H_{c_2}^{({\bf a})}$ (high fields and
low temperatures) have not yet been performed for ${\rm (TMTSF)_2 ClO_4}$.
%
%
Having made this last experimental connection, 
we are ready to summarize our results.

\section{Summary}
\label{sec:summary}

We performed a group theoretical analysis of the possible
symmetries compatible with the ${\rm D_{2h}}$ (orthorhombic) point group, and
focused on the weak and strong spin-orbit coupling triplet cases
at zero magnetic field. 
We also discussed order parameter symmetry features 
and temperature dependence of the 
quasiparticle density of states and spin susceptibility tensor 
of an orthorhombic quasi-one-dimensional superconductor~\cite{triclinic}.
Based on current experimental evidence and the 
assumption that the origin of superconductivity 
in ${\rm (TMTSF)_2 ClO_4}$ and ${\rm (TMTSF)_2 PF_6}$
is essentially the same,
we suggested that the weak spin-orbit coupling state
$^3B_{3u} (a)$ (``$p_x$-wave'') is a very good 
candidate for the order parameter
symmetry for these systems since this state is:
(1) fully gapped and consistent with thermal 
conductivity measurements~\cite{belin-97};
(2) characterized by weak spin-orbit coupling
and consistent with weak spin-orbit scattering fits of 
$T_c (H)$ for ${\rm (TMTSF)_2 PF_6}$~\cite{lee-98} at low magnetic fields;
(3) consistent with no observable 
Knight shift when ${\bf H} 
\parallel {\bf a}$~\cite{lee-02} 
or $\parallel {\bf b}^{\prime}$~\cite{lee-97},
and predicted to have no observable Knight shift for any
direction ${\hat n}$ of ${\bf H}$ provided that the magnitude of
the applied field is larger than the small spin-orbit pining
field $H_f^{(\hat n)}$. 
We would like to thank NSF (Grant No. 
DMR-9803111) and its REU program for support. 
One of us (C. A. R. S{\'a} de Melo) would like to thank 
the Aspen Center for Physics for its hospitality}.

$^\dagger$ Present address: Department of Physics, 
Massachusetts Institute of Technology,
Cambridge, MA 02139-4307.

\bibliography{symmetry-long.12}

\begin{thebibliography}{64}
\expandafter\ifx\csname natexlab\endcsname\relax\def\natexlab#1{#1}\fi
\expandafter\ifx\csname bibnamefont\endcsname\relax
  \def\bibnamefont#1{#1}\fi
\expandafter\ifx\csname bibfnamefont\endcsname\relax
  \def\bibfnamefont#1{#1}\fi
\expandafter\ifx\csname citenamefont\endcsname\relax
  \def\citenamefont#1{#1}\fi
\expandafter\ifx\csname url\endcsname\relax
  \def\url#1{\texttt{#1}}\fi
\expandafter\ifx\csname urlprefix\endcsname\relax\def\urlprefix{URL }\fi
\providecommand{\bibinfo}[2]{#2}
\providecommand{\eprint}[2][]{\url{#2}}

\bibitem[{\citenamefont{Jerome et~al.}(1980)\citenamefont{Jerome, Mazaud,
  Ribault, and Bechgaard}}]{jerome-80}
\bibinfo{author}{\bibfnamefont{D.}~\bibnamefont{Jerome}},
  \bibinfo{author}{\bibfnamefont{A.}~\bibnamefont{Mazaud}},
  \bibinfo{author}{\bibfnamefont{M.}~\bibnamefont{Ribault}}, \bibnamefont{and}
  \bibinfo{author}{\bibfnamefont{K.}~\bibnamefont{Bechgaard}},
  \bibinfo{journal}{J. Phys. Lett.} \textbf{\bibinfo{volume}{41}},
  \bibinfo{pages}{L95} (\bibinfo{year}{1980}).

\bibitem[{\citenamefont{Ishiguro and Yamaji}(1989)}]{ishiguro-89}
\bibinfo{author}{\bibfnamefont{T.}~\bibnamefont{Ishiguro}} \bibnamefont{and}
  \bibinfo{author}{\bibfnamefont{K.}~\bibnamefont{Yamaji}},
  \emph{\bibinfo{title}{Organic Superconductors}}
  (\bibinfo{publisher}{Springer-Verlag}, \bibinfo{year}{1989}).

\bibitem[{\citenamefont{Williams et~al.}(1992)\citenamefont{Williams, Ferraro,
  Thorn, Carlson, Geiser, Wang, Kini, and Whangboo}}]{williams-92}
\bibinfo{author}{\bibfnamefont{J.~M.} \bibnamefont{Williams}},
  \bibinfo{author}{\bibfnamefont{J.~R.} \bibnamefont{Ferraro}},
  \bibinfo{author}{\bibfnamefont{R.~J.} \bibnamefont{Thorn}},
  \bibinfo{author}{\bibfnamefont{K.~G.} \bibnamefont{Carlson}},
  \bibinfo{author}{\bibfnamefont{U.}~\bibnamefont{Geiser}},
  \bibinfo{author}{\bibfnamefont{H.~H.} \bibnamefont{Wang}},
  \bibinfo{author}{\bibfnamefont{A.~M.} \bibnamefont{Kini}}, \bibnamefont{and}
  \bibinfo{author}{\bibfnamefont{M.~H.} \bibnamefont{Whangboo}},
  \emph{\bibinfo{title}{Organic Superconductors (Including Fullerenes):
  Synthesis, Structure, Properties and Theory}} (\bibinfo{publisher}{Prentice
  Hall}, \bibinfo{year}{1992}).

\bibitem[{\citenamefont{Lee et~al.}(1997)\citenamefont{Lee, Naughton, Danner,
  and Chaikin}}]{lee-97}
\bibinfo{author}{\bibfnamefont{I.~J.} \bibnamefont{Lee}},
  \bibinfo{author}{\bibfnamefont{M.~J.} \bibnamefont{Naughton}},
  \bibinfo{author}{\bibfnamefont{G.~M.} \bibnamefont{Danner}},
  \bibnamefont{and} \bibinfo{author}{\bibfnamefont{P.~M.}
  \bibnamefont{Chaikin}}, \bibinfo{journal}{Phys. Rev. Lett.}
  \textbf{\bibinfo{volume}{78}}, \bibinfo{pages}{3555} (\bibinfo{year}{1997}).

\bibitem[{\citenamefont{Lee et~al.}(2000)\citenamefont{Lee, Chow, Clark,
  Strouse, Naughton, Chaikin, and Brown}}]{lee-00}
\bibinfo{author}{\bibfnamefont{I.~J.} \bibnamefont{Lee}},
  \bibinfo{author}{\bibfnamefont{D.~S.} \bibnamefont{Chow}},
  \bibinfo{author}{\bibfnamefont{W.~G.} \bibnamefont{Clark}},
  \bibinfo{author}{\bibfnamefont{M.~J.} \bibnamefont{Strouse}},
  \bibinfo{author}{\bibfnamefont{M.~J.} \bibnamefont{Naughton}},
  \bibinfo{author}{\bibfnamefont{P.~M.} \bibnamefont{Chaikin}},
  \bibnamefont{and} \bibinfo{author}{\bibfnamefont{S.~E.} \bibnamefont{Brown}}
  (\bibinfo{year}{2000}), \eprint{cond-mat/0001332}.

\bibitem[{\citenamefont{Lee et~al.}(2002)\citenamefont{Lee, Brown, Clark,
  Strouse, Naughton, Kang, and Chaikin}}]{lee-02}
\bibinfo{author}{\bibfnamefont{I.~J.} \bibnamefont{Lee}},
  \bibinfo{author}{\bibfnamefont{S.~E.} \bibnamefont{Brown}},
  \bibinfo{author}{\bibfnamefont{W.~G.} \bibnamefont{Clark}},
  \bibinfo{author}{\bibfnamefont{M.~J.} \bibnamefont{Strouse}},
  \bibinfo{author}{\bibfnamefont{M.~J.} \bibnamefont{Naughton}},
  \bibinfo{author}{\bibfnamefont{W.}~\bibnamefont{Kang}}, \bibnamefont{and}
  \bibinfo{author}{\bibfnamefont{P.~M.} \bibnamefont{Chaikin}},
  \bibinfo{journal}{Phys. Rev. Lett.} \textbf{\bibinfo{volume}{88}},
  \bibinfo{pages}{017004} (\bibinfo{year}{2002}).

\bibitem[{\citenamefont{Hebel and Slichter}(1959)}]{hebel-59}
\bibinfo{author}{\bibfnamefont{L.~C.} \bibnamefont{Hebel}} \bibnamefont{and}
  \bibinfo{author}{\bibfnamefont{C.~P.} \bibnamefont{Slichter}},
  \bibinfo{journal}{Phys. Rev.} \textbf{\bibinfo{volume}{113}},
  \bibinfo{pages}{1504} (\bibinfo{year}{1959}).

\bibitem[{\citenamefont{Takigawa et~al.}(1987)\citenamefont{Takigawa, Yasuoka,
  and Saito}}]{takigawa-87}
\bibinfo{author}{\bibfnamefont{M.}~\bibnamefont{Takigawa}},
  \bibinfo{author}{\bibfnamefont{H.}~\bibnamefont{Yasuoka}}, \bibnamefont{and}
  \bibinfo{author}{\bibfnamefont{G.}~\bibnamefont{Saito}},
  \bibinfo{journal}{Journ. Phys. Soc. Japan} \textbf{\bibinfo{volume}{56}},
  \bibinfo{pages}{873} (\bibinfo{year}{1987}).

\bibitem[{\citenamefont{Lee et~al.}(1994)\citenamefont{Lee, Hope, Leone, and
  Naughton}}]{lee-94}
\bibinfo{author}{\bibfnamefont{I.~J.} \bibnamefont{Lee}},
  \bibinfo{author}{\bibfnamefont{A.~P.} \bibnamefont{Hope}},
  \bibinfo{author}{\bibfnamefont{M.~J.} \bibnamefont{Leone}}, \bibnamefont{and}
  \bibinfo{author}{\bibfnamefont{M.~J.} \bibnamefont{Naughton}},
  \bibinfo{journal}{Applied Superconductivity} \textbf{\bibinfo{volume}{2}},
  \bibinfo{pages}{753} (\bibinfo{year}{1994}).

\bibitem[{\citenamefont{Lee et~al.}(1995)\citenamefont{Lee, Hope, Leone, and
  Naughton}}]{lee-95}
\bibinfo{author}{\bibfnamefont{I.~J.} \bibnamefont{Lee}},
  \bibinfo{author}{\bibfnamefont{A.~P.} \bibnamefont{Hope}},
  \bibinfo{author}{\bibfnamefont{M.~J.} \bibnamefont{Leone}}, \bibnamefont{and}
  \bibinfo{author}{\bibfnamefont{M.~J.} \bibnamefont{Naughton}},
  \bibinfo{journal}{Synth. Metals} \textbf{\bibinfo{volume}{70}},
  \bibinfo{pages}{747} (\bibinfo{year}{1995}).

\bibitem[{\citenamefont{Belin and Behnia}(1997)}]{belin-97}
\bibinfo{author}{\bibfnamefont{S.}~\bibnamefont{Belin}} \bibnamefont{and}
  \bibinfo{author}{\bibfnamefont{K.}~\bibnamefont{Behnia}},
  \bibinfo{journal}{Phys. Rev. Lett.} \textbf{\bibinfo{volume}{79}},
  \bibinfo{pages}{2125} (\bibinfo{year}{1997}).

\bibitem[{\citenamefont{Wollman et~al.}(1993)\citenamefont{Wollman, Harlingen,
  Lee, Ginsberg, and Leggett}}]{harlingen-93}
\bibinfo{author}{\bibfnamefont{D.~A.} \bibnamefont{Wollman}},
  \bibinfo{author}{\bibfnamefont{D.~J.~V.} \bibnamefont{Harlingen}},
  \bibinfo{author}{\bibfnamefont{W.~C.} \bibnamefont{Lee}},
  \bibinfo{author}{\bibfnamefont{D.~M.} \bibnamefont{Ginsberg}},
  \bibnamefont{and} \bibinfo{author}{\bibfnamefont{A.~J.}
  \bibnamefont{Leggett}}, \bibinfo{journal}{Phys. Rev. Lett.}
  \textbf{\bibinfo{volume}{71}}, \bibinfo{pages}{2134} (\bibinfo{year}{1993}).

\bibitem[{\citenamefont{Mathai et~al.}(1995)\citenamefont{Mathai, Gim, Black,
  Amar, and Wellstood}}]{wellstood-95}
\bibinfo{author}{\bibfnamefont{A.}~\bibnamefont{Mathai}},
  \bibinfo{author}{\bibfnamefont{Y.}~\bibnamefont{Gim}},
  \bibinfo{author}{\bibfnamefont{R.~C.} \bibnamefont{Black}},
  \bibinfo{author}{\bibfnamefont{A.}~\bibnamefont{Amar}}, \bibnamefont{and}
  \bibinfo{author}{\bibfnamefont{F.~C.} \bibnamefont{Wellstood}},
  \bibinfo{journal}{Phys. Rev. Lett.} \textbf{\bibinfo{volume}{74}},
  \bibinfo{pages}{4523} (\bibinfo{year}{1995}).

\bibitem[{\citenamefont{Abrikosov}(1983{\natexlab{a}})}]{abrikosov-83a}
\bibinfo{author}{\bibfnamefont{A.~A.} \bibnamefont{Abrikosov}},
  \bibinfo{journal}{JETP Lett.} \textbf{\bibinfo{volume}{37}},
  \bibinfo{pages}{503} (\bibinfo{year}{1983}{\natexlab{a}}).

\bibitem[{\citenamefont{Abrikosov}(1983{\natexlab{b}})}]{abrikosov-83b}
\bibinfo{author}{\bibfnamefont{A.~A.} \bibnamefont{Abrikosov}},
  \bibinfo{journal}{J. Low Temp. Phys.} \textbf{\bibinfo{volume}{53}},
  \bibinfo{pages}{359} (\bibinfo{year}{1983}{\natexlab{b}}).

\bibitem[{\citenamefont{Choi et~al.}(1982)\citenamefont{Choi, Chaikin, Huang,
  Haen, Engler, and Greene}}]{choi-82}
\bibinfo{author}{\bibfnamefont{M.~Y.} \bibnamefont{Choi}},
  \bibinfo{author}{\bibfnamefont{P.~M.} \bibnamefont{Chaikin}},
  \bibinfo{author}{\bibfnamefont{S.~Z.} \bibnamefont{Huang}},
  \bibinfo{author}{\bibfnamefont{P.}~\bibnamefont{Haen}},
  \bibinfo{author}{\bibfnamefont{E.~M.} \bibnamefont{Engler}},
  \bibnamefont{and} \bibinfo{author}{\bibfnamefont{R.~L.}
  \bibnamefont{Greene}}, \bibinfo{journal}{Phys. Rev. B}
  \textbf{\bibinfo{volume}{25}}, \bibinfo{pages}{6208} (\bibinfo{year}{1982}).

\bibitem[{\citenamefont{Bouffard et~al.}(1982)\citenamefont{Bouffard, Ribault,
  Brussetti, Jerome, and Bechgaard}}]{bouffard-82}
\bibinfo{author}{\bibfnamefont{S.}~\bibnamefont{Bouffard}},
  \bibinfo{author}{\bibfnamefont{M.}~\bibnamefont{Ribault}},
  \bibinfo{author}{\bibfnamefont{R.}~\bibnamefont{Brussetti}},
  \bibinfo{author}{\bibfnamefont{D.}~\bibnamefont{Jerome}}, \bibnamefont{and}
  \bibinfo{author}{\bibfnamefont{K.}~\bibnamefont{Bechgaard}},
  \bibinfo{journal}{J. Phys. C: Solid State Phys.}
  \textbf{\bibinfo{volume}{15}}, \bibinfo{pages}{2951} (\bibinfo{year}{1982}).

\bibitem[{\citenamefont{Gorkov and Jerome}(1985)}]{gorkov-85}
\bibinfo{author}{\bibfnamefont{L.~P.} \bibnamefont{Gorkov}} \bibnamefont{and}
  \bibinfo{author}{\bibfnamefont{D.}~\bibnamefont{Jerome}},
  \bibinfo{journal}{J. Physique Lett.} \textbf{\bibinfo{volume}{46}},
  \bibinfo{pages}{L643} (\bibinfo{year}{1985}).

\bibitem[{\citenamefont{Lebed}(1986)}]{lebed-86}
\bibinfo{author}{\bibfnamefont{A.~G.} \bibnamefont{Lebed}},
  \bibinfo{journal}{Sov. Phys. JETP Lett.} \textbf{\bibinfo{volume}{44}},
  \bibinfo{pages}{89} (\bibinfo{year}{1986}).

\bibitem[{\citenamefont{Dupuis et~al.}(1993)\citenamefont{Dupuis, Montambaux,
  and S{\'a}~de Melo}}]{dupuis-93}
\bibinfo{author}{\bibfnamefont{N.}~\bibnamefont{Dupuis}},
  \bibinfo{author}{\bibfnamefont{G.}~\bibnamefont{Montambaux}},
  \bibnamefont{and} \bibinfo{author}{\bibfnamefont{C.~A.~R.}
  \bibnamefont{S{\'a}~de Melo}}, \bibinfo{journal}{Phys. Rev. Lett.}
  \textbf{\bibinfo{volume}{70}}, \bibinfo{pages}{2613} (\bibinfo{year}{1993}).

\bibitem[{\citenamefont{Larkin and Ovchinikov}(1965)}]{larkin-65}
\bibinfo{author}{\bibfnamefont{A.~I.} \bibnamefont{Larkin}} \bibnamefont{and}
  \bibinfo{author}{\bibfnamefont{Y.~N.} \bibnamefont{Ovchinikov}},
  \bibinfo{journal}{Sov. Phys. JETP} \textbf{\bibinfo{volume}{20}},
  \bibinfo{pages}{762} (\bibinfo{year}{1965}).

\bibitem[{\citenamefont{Fulde and Ferrel}(1964)}]{fulde-64}
\bibinfo{author}{\bibfnamefont{P.}~\bibnamefont{Fulde}} \bibnamefont{and}
  \bibinfo{author}{\bibfnamefont{R.~A.} \bibnamefont{Ferrel}},
  \bibinfo{journal}{Phys. Rev.} \textbf{\bibinfo{volume}{135}},
  \bibinfo{pages}{A550} (\bibinfo{year}{1964}).

\bibitem[{\citenamefont{Dupuis}(1995)}]{dupuis-95}
\bibinfo{author}{\bibfnamefont{N.}~\bibnamefont{Dupuis}},
  \bibinfo{journal}{Phys. Rev. B} \textbf{\bibinfo{volume}{51}},
  \bibinfo{pages}{9074} (\bibinfo{year}{1995}).

\bibitem[{\citenamefont{S{\'a}~de Melo}(1996)}]{sademelo-96}
\bibinfo{author}{\bibfnamefont{C.~A.~R.} \bibnamefont{S{\'a}~de Melo}},
  \bibinfo{journal}{Physica C} \textbf{\bibinfo{volume}{260}},
  \bibinfo{pages}{224} (\bibinfo{year}{1996}).

\bibitem[{\citenamefont{Lebed}(1999)}]{lebed-99}
\bibinfo{author}{\bibfnamefont{A.~G.} \bibnamefont{Lebed}},
  \bibinfo{journal}{Phys. Rev. B} \textbf{\bibinfo{volume}{59}},
  \bibinfo{pages}{R721} (\bibinfo{year}{1999}).

\bibitem[{\citenamefont{S{\'a}~de Melo}(1998)}]{sademelo-98}
\bibinfo{author}{\bibfnamefont{C.~A.~R.} \bibnamefont{S{\'a}~de Melo}},
  \emph{\bibinfo{title}{The Superconducting State in Magnetic Fields: Special
  Topics and New Trends}} (\bibinfo{publisher}{World Scientific, Singapore},
  \bibinfo{year}{1998}), chap.~\bibinfo{chapter}{15}, pp.
  \bibinfo{pages}{296--324}, \bibinfo{note}{(Edited by C. A. R. S{\'a} de
  Melo)}.

\bibitem[{\citenamefont{S{\'a}~de Melo}(1999{\natexlab{a}})}]{sademelo-99}
\bibinfo{author}{\bibfnamefont{C.~A.~R.} \bibnamefont{S{\'a}~de Melo}},
  \bibinfo{journal}{J. Supercond.} \textbf{\bibinfo{volume}{12}},
  \bibinfo{pages}{459} (\bibinfo{year}{1999}{\natexlab{a}}).

\bibitem[{\citenamefont{Vaccarella and S{\'a}~de Melo}(2000)}]{vaccarella-00}
\bibinfo{author}{\bibfnamefont{C.~D.} \bibnamefont{Vaccarella}}
  \bibnamefont{and} \bibinfo{author}{\bibfnamefont{C.~A.~R.}
  \bibnamefont{S{\'a}~de Melo}}, \bibinfo{journal}{Physica C}
  \textbf{\bibinfo{volume}{341-348}}, \bibinfo{pages}{293}
  (\bibinfo{year}{2000}).

\bibitem[{\citenamefont{Vaccarella and S{\'a}~de Melo}(2001)}]{vaccarella-01}
\bibinfo{author}{\bibfnamefont{C.~D.} \bibnamefont{Vaccarella}}
  \bibnamefont{and} \bibinfo{author}{\bibfnamefont{C.~A.~R.}
  \bibnamefont{S{\'a}~de Melo}}, \bibinfo{journal}{Phys. Rev. B}
  \textbf{\bibinfo{volume}{63}}, \bibinfo{pages}{R180505}
  (\bibinfo{year}{2001}).

\bibitem[{\citenamefont{Lebed et~al.}(2000)\citenamefont{Lebed, Machida, and
  Ozaki}}]{lebed-00}
\bibinfo{author}{\bibfnamefont{A.~G.} \bibnamefont{Lebed}},
  \bibinfo{author}{\bibfnamefont{K.}~\bibnamefont{Machida}}, \bibnamefont{and}
  \bibinfo{author}{\bibfnamefont{M.}~\bibnamefont{Ozaki}},
  \bibinfo{journal}{Phys. Rev. B} \textbf{\bibinfo{volume}{62}},
  \bibinfo{pages}{R795} (\bibinfo{year}{2000}).

\bibitem[{\citenamefont{Shimahara}(2000)}]{shimahara-00}
\bibinfo{author}{\bibfnamefont{H.}~\bibnamefont{Shimahara}},
  \bibinfo{journal}{Phys. Rev. B} \textbf{\bibinfo{volume}{61}},
  \bibinfo{pages}{R14936} (\bibinfo{year}{2000}).

\bibitem[{\citenamefont{Kuroki et~al.}(2001)\citenamefont{Kuroki, Arita, and
  Aoki}}]{kuroki-01}
\bibinfo{author}{\bibfnamefont{K.}~\bibnamefont{Kuroki}},
  \bibinfo{author}{\bibfnamefont{R.}~\bibnamefont{Arita}}, \bibnamefont{and}
  \bibinfo{author}{\bibfnamefont{H.}~\bibnamefont{Aoki}},
  \bibinfo{journal}{Phys. Rev. B} \textbf{\bibinfo{volume}{63}},
  \bibinfo{pages}{094509} (\bibinfo{year}{2001}).

\bibitem[{\citenamefont{Duncan et~al.}(2001)\citenamefont{Duncan, Vaccarella,
  and S{\'a}~de Melo}}]{duncan-01}
\bibinfo{author}{\bibfnamefont{R.~D.} \bibnamefont{Duncan}},
  \bibinfo{author}{\bibfnamefont{C.~D.} \bibnamefont{Vaccarella}},
  \bibnamefont{and} \bibinfo{author}{\bibfnamefont{C.~A.~R.}
  \bibnamefont{S{\'a}~de Melo}}, \bibinfo{journal}{Phys. Rev. B}
  \textbf{\bibinfo{volume}{64}}, \bibinfo{pages}{172503}
  (\bibinfo{year}{2001}).

\bibitem[{\citenamefont{Mineev and Samokhin}(1999)}]{mineev-99}
\bibinfo{author}{\bibfnamefont{V.~P.} \bibnamefont{Mineev}} \bibnamefont{and}
  \bibinfo{author}{\bibfnamefont{K.~V.} \bibnamefont{Samokhin}},
  \emph{\bibinfo{title}{Introduction to Unconventional Superconductivity}}
  (\bibinfo{publisher}{Gordon and Breach, Amsterdam}, \bibinfo{year}{1999}).

\bibitem[{\citenamefont{S{\'a}~de Melo et~al.}(1993)\citenamefont{S{\'a}~de
  Melo, Randeria, and Engelbrecht}}]{sdm-93}
\bibinfo{author}{\bibfnamefont{C.~A.~R.} \bibnamefont{S{\'a}~de Melo}},
  \bibinfo{author}{\bibfnamefont{M.}~\bibnamefont{Randeria}}, \bibnamefont{and}
  \bibinfo{author}{\bibfnamefont{J.~R.} \bibnamefont{Engelbrecht}},
  \bibinfo{journal}{Phys. Rev. Lett.} \textbf{\bibinfo{volume}{71}},
  \bibinfo{pages}{3202} (\bibinfo{year}{1993}).

\bibitem[{\citenamefont{Engelbrecht et~al.}(1997)\citenamefont{Engelbrecht,
  Randeria, and S{\'a}~de Melo}}]{sdm-97}
\bibinfo{author}{\bibfnamefont{J.~R.} \bibnamefont{Engelbrecht}},
  \bibinfo{author}{\bibfnamefont{M.}~\bibnamefont{Randeria}}, \bibnamefont{and}
  \bibinfo{author}{\bibfnamefont{C.~A.~R.} \bibnamefont{S{\'a}~de Melo}},
  \bibinfo{journal}{Phys. Rev. B} \textbf{\bibinfo{volume}{55}},
  \bibinfo{pages}{15153} (\bibinfo{year}{1997}).

\bibitem[{tri()}]{triclinic}
\bibinfo{note}{The Bechgaard salts $ {\rm (TMTSF)_2 ClO_4} $ and ${ \rm
  (TMTSF)_2 PF_6 } $ have truly a triclinic lattice structure.}

\bibitem[{\citenamefont{Anderson}(1984)}]{anderson-84}
\bibinfo{author}{\bibfnamefont{P.~W.} \bibnamefont{Anderson}},
  \bibinfo{journal}{Phys. Rev. B} \textbf{\bibinfo{volume}{30}},
  \bibinfo{pages}{4000} (\bibinfo{year}{1984}).

\bibitem[{\citenamefont{Tinkham}(1964)}]{tinkham-64}
\bibinfo{author}{\bibfnamefont{M.}~\bibnamefont{Tinkham}},
  \emph{\bibinfo{title}{Group Theory and Quantum Mechanics}}
  (\bibinfo{publisher}{McGraw-Hill, New York}, \bibinfo{year}{1964}).

\bibitem[{\citenamefont{Duncan and S{\'a}~de Melo}(2000)}]{duncan-00}
\bibinfo{author}{\bibfnamefont{R.~D.} \bibnamefont{Duncan}} \bibnamefont{and}
  \bibinfo{author}{\bibfnamefont{C.~A.~R.} \bibnamefont{S{\'a}~de Melo}},
  \bibinfo{journal}{Phys. Rev. B.} \textbf{\bibinfo{volume}{62}},
  \bibinfo{pages}{9675} (\bibinfo{year}{2000}).

\bibitem[{\citenamefont{Schon et~al.}(2000{\natexlab{a}})\citenamefont{Schon,
  Kloc, and Batlogg}}]{batlogg-00a}
\bibinfo{author}{\bibfnamefont{J.~H.} \bibnamefont{Schon}},
  \bibinfo{author}{\bibfnamefont{C.}~\bibnamefont{Kloc}}, \bibnamefont{and}
  \bibinfo{author}{\bibfnamefont{B.}~\bibnamefont{Batlogg}},
  \bibinfo{journal}{Nature} \textbf{\bibinfo{volume}{406}},
  \bibinfo{pages}{702} (\bibinfo{year}{2000}{\natexlab{a}}).

\bibitem[{\citenamefont{Schon et~al.}(2000{\natexlab{b}})\citenamefont{Schon,
  Kloc, and Batlogg}}]{batlogg-00b}
\bibinfo{author}{\bibfnamefont{J.~H.} \bibnamefont{Schon}},
  \bibinfo{author}{\bibfnamefont{C.}~\bibnamefont{Kloc}}, \bibnamefont{and}
  \bibinfo{author}{\bibfnamefont{B.}~\bibnamefont{Batlogg}},
  \bibinfo{journal}{Nature} \textbf{\bibinfo{volume}{408}},
  \bibinfo{pages}{549} (\bibinfo{year}{2000}{\natexlab{b}}).

\bibitem[{\citenamefont{Maeno et~al.}(2001)\citenamefont{Maeno, Rice, and
  Sigrist}}]{sigrist-01}
\bibinfo{author}{\bibfnamefont{Y.}~\bibnamefont{Maeno}},
  \bibinfo{author}{\bibfnamefont{T.~M.} \bibnamefont{Rice}}, \bibnamefont{and}
  \bibinfo{author}{\bibfnamefont{M.}~\bibnamefont{Sigrist}},
  \bibinfo{journal}{Physics Today} \textbf{\bibinfo{volume}{54}},
  \bibinfo{pages}{42} (\bibinfo{year}{2001}).

\bibitem[{\citenamefont{Buchholtz and Zwicknagl}(1981)}]{buchholtz-81}
\bibinfo{author}{\bibfnamefont{L.~J.} \bibnamefont{Buchholtz}}
  \bibnamefont{and}
  \bibinfo{author}{\bibfnamefont{G.}~\bibnamefont{Zwicknagl}},
  \bibinfo{journal}{Phys. Rev. B} \textbf{\bibinfo{volume}{23}},
  \bibinfo{pages}{5788} (\bibinfo{year}{1981}).

\bibitem[{\citenamefont{Hu}(1994)}]{hu-94}
\bibinfo{author}{\bibfnamefont{C.~R.} \bibnamefont{Hu}},
  \bibinfo{journal}{Phys. Rev. Lett.} \textbf{\bibinfo{volume}{72}},
  \bibinfo{pages}{1526} (\bibinfo{year}{1994}).

\bibitem[{\citenamefont{Sengupta et~al.}(2000)\citenamefont{Sengupta, Zutic,
  Kwon, Yakovenko, and Sarma}}]{sengupta-01}
\bibinfo{author}{\bibfnamefont{K.}~\bibnamefont{Sengupta}},
  \bibinfo{author}{\bibfnamefont{I.}~\bibnamefont{Zutic}},
  \bibinfo{author}{\bibfnamefont{H.~J.} \bibnamefont{Kwon}},
  \bibinfo{author}{\bibfnamefont{V.~M.} \bibnamefont{Yakovenko}},
  \bibnamefont{and} \bibinfo{author}{\bibfnamefont{S.~D.} \bibnamefont{Sarma}},
  \bibinfo{journal}{Phys. Rev. B} \textbf{\bibinfo{volume}{63}},
  \bibinfo{pages}{144531} (\bibinfo{year}{2000}).

\bibitem[{\citenamefont{Leggett}(1975)}]{leggett-75}
\bibinfo{author}{\bibfnamefont{A.~J.} \bibnamefont{Leggett}},
  \bibinfo{journal}{Rev. Mod. Phys.} \textbf{\bibinfo{volume}{47}},
  \bibinfo{pages}{331} (\bibinfo{year}{1975}).

\bibitem[{\citenamefont{Danner and Chaikin}(1996)}]{chaikin-95}
\bibinfo{author}{\bibfnamefont{G.~M.} \bibnamefont{Danner}} \bibnamefont{and}
  \bibinfo{author}{\bibfnamefont{P.~M.} \bibnamefont{Chaikin}},
  \bibinfo{journal}{Phys. Rev. Lett.} \textbf{\bibinfo{volume}{75}},
  \bibinfo{pages}{4690} (\bibinfo{year}{1996}).

\bibitem[{not()}]{note-chi_mn}
\bibinfo{note}{In the calculation of $\chi_{m n} (T,H)$ the full effect of the
  presence of vortices need to be incorporated. A microscopic calculation of
  $\chi_{m n} (T,H)$ taking into account the spatial inhomogeneities introduced
  by the magnetic field and vortex core states is underway.}

\bibitem[{cor()}]{core-states}
\bibinfo{note}{Vortex core states are not as normal as the normal state of the
  superconductor, they have a lot of structure even in conventional singlet
  ``$s$-wave''superconductors. Vortex core states in singlet ``$d$-wave''
  superconductors have also a rich structure. So it is expected that vortex
  core states in triplet ``$p$-wave'' superconductors are even richer in
  structure due to the additional spin degrees of freedom present. The vortex
  core structure will be discussed in a later publication.}

\bibitem[{\citenamefont{Caroli et~al.}(1964)\citenamefont{Caroli, de~Gennes,
  and Matricon}}]{caroli-64}
\bibinfo{author}{\bibfnamefont{C.}~\bibnamefont{Caroli}},
  \bibinfo{author}{\bibfnamefont{P.~G.} \bibnamefont{de~Gennes}},
  \bibnamefont{and} \bibinfo{author}{\bibfnamefont{J.}~\bibnamefont{Matricon}},
  \bibinfo{journal}{Phys. Lett.} \textbf{\bibinfo{volume}{9}},
  \bibinfo{pages}{307} (\bibinfo{year}{1964}).

\bibitem[{\citenamefont{Bardeen et~al.}(1969)\citenamefont{Bardeen, Kummel,
  Jacobs, and Tewordt}}]{bardeen-69}
\bibinfo{author}{\bibfnamefont{J.}~\bibnamefont{Bardeen}},
  \bibinfo{author}{\bibfnamefont{R.}~\bibnamefont{Kummel}},
  \bibinfo{author}{\bibfnamefont{A.~E.} \bibnamefont{Jacobs}},
  \bibnamefont{and} \bibinfo{author}{\bibfnamefont{L.}~\bibnamefont{Tewordt}},
  \bibinfo{journal}{Phys. Rev.} \textbf{\bibinfo{volume}{187}},
  \bibinfo{pages}{556} (\bibinfo{year}{1969}).

\bibitem[{\citenamefont{Kramer and Pesch}(1974)}]{kramer-74}
\bibinfo{author}{\bibfnamefont{L.}~\bibnamefont{Kramer}} \bibnamefont{and}
  \bibinfo{author}{\bibfnamefont{W.}~\bibnamefont{Pesch}}, \bibinfo{journal}{Z.
  Phys.} \textbf{\bibinfo{volume}{269}}, \bibinfo{pages}{59}
  (\bibinfo{year}{1974}).

\bibitem[{\citenamefont{Ullah et~al.}(1990)\citenamefont{Ullah, Dorsey, and
  Buchholtz}}]{ullah-90}
\bibinfo{author}{\bibfnamefont{S.}~\bibnamefont{Ullah}},
  \bibinfo{author}{\bibfnamefont{A.~T.} \bibnamefont{Dorsey}},
  \bibnamefont{and} \bibinfo{author}{\bibfnamefont{L.~J.}
  \bibnamefont{Buchholtz}}, \bibinfo{journal}{Phys. Rev. B}
  \textbf{\bibinfo{volume}{42}}, \bibinfo{pages}{9950} (\bibinfo{year}{1990}).

\bibitem[{\citenamefont{Gygi and Schl{\"u}ter}(1991)}]{gygi-91}
\bibinfo{author}{\bibfnamefont{F.}~\bibnamefont{Gygi}} \bibnamefont{and}
  \bibinfo{author}{\bibfnamefont{M.}~\bibnamefont{Schl{\"u}ter}},
  \bibinfo{journal}{Phys. Rev. B} \textbf{\bibinfo{volume}{43}},
  \bibinfo{pages}{7609} (\bibinfo{year}{1991}).

\bibitem[{\citenamefont{S{\'a}~de Melo}(1994)}]{sdm-94}
\bibinfo{author}{\bibfnamefont{C.~A.~R.} \bibnamefont{S{\'a}~de Melo}},
  \bibinfo{journal}{Phys. Rev. Lett.} \textbf{\bibinfo{volume}{73}},
  \bibinfo{pages}{1978} (\bibinfo{year}{1994}).

\bibitem[{\citenamefont{Ichioka et~al.}(1996)\citenamefont{Ichioka, Hayashi,
  Enomoto, and Machida}}]{ichioka-96}
\bibinfo{author}{\bibfnamefont{M.}~\bibnamefont{Ichioka}},
  \bibinfo{author}{\bibfnamefont{N.}~\bibnamefont{Hayashi}},
  \bibinfo{author}{\bibfnamefont{N.}~\bibnamefont{Enomoto}}, \bibnamefont{and}
  \bibinfo{author}{\bibfnamefont{K.}~\bibnamefont{Machida}},
  \bibinfo{journal}{Phys. Rev. B} \textbf{\bibinfo{volume}{53}},
  \bibinfo{pages}{15316} (\bibinfo{year}{1996}).

\bibitem[{\citenamefont{Franz and Tesanovic}(1998)}]{tesanovic-98}
\bibinfo{author}{\bibfnamefont{M.}~\bibnamefont{Franz}} \bibnamefont{and}
  \bibinfo{author}{\bibfnamefont{Z.~B.} \bibnamefont{Tesanovic}},
  \bibinfo{journal}{Phys. Rev. Lett.} \textbf{\bibinfo{volume}{80}},
  \bibinfo{pages}{4763} (\bibinfo{year}{1998}).

\bibitem[{\citenamefont{S{\'a}~de Melo}(1999{\natexlab{b}})}]{sdm-99}
\bibinfo{author}{\bibfnamefont{C.~A.~R.} \bibnamefont{S{\'a}~de Melo}},
  \bibinfo{journal}{Phys. Rev. B} \textbf{\bibinfo{volume}{60}},
  \bibinfo{pages}{10423} (\bibinfo{year}{1999}{\natexlab{b}}).

\bibitem[{bou()}]{boundaries}
\bibinfo{note}{The two fluids are not strictly independent, because vortex core
  states depend on properties (like the order parameter behavior) outside and
  at the boundaries of the core.}

\bibitem[{\citenamefont{Shivaram et~al.}(1986)\citenamefont{Shivaram,
  Rosenbaum, and Hinks}}]{shivaram-86}
\bibinfo{author}{\bibfnamefont{B.~S.} \bibnamefont{Shivaram}},
  \bibinfo{author}{\bibfnamefont{T.~F.} \bibnamefont{Rosenbaum}},
  \bibnamefont{and} \bibinfo{author}{\bibfnamefont{D.~G.} \bibnamefont{Hinks}},
  \bibinfo{journal}{Phys. Rev. Lett.} \textbf{\bibinfo{volume}{57}},
  \bibinfo{pages}{1259} (\bibinfo{year}{1986}).

\bibitem[{\citenamefont{Choi and Sauls}(1991)}]{sauls-91}
\bibinfo{author}{\bibfnamefont{C.~H.} \bibnamefont{Choi}} \bibnamefont{and}
  \bibinfo{author}{\bibfnamefont{J.~A.} \bibnamefont{Sauls}},
  \bibinfo{journal}{Phys. Rev. Lett.} \textbf{\bibinfo{volume}{66}},
  \bibinfo{pages}{484} (\bibinfo{year}{1991}).

\bibitem[{\citenamefont{Mortensen et~al.}(1982)\citenamefont{Mortensen,
  Tomkiewicz, and Bechgaard}}]{mortensen-82}
\bibinfo{author}{\bibfnamefont{K.}~\bibnamefont{Mortensen}},
  \bibinfo{author}{\bibfnamefont{Y.}~\bibnamefont{Tomkiewicz}},
  \bibnamefont{and}
  \bibinfo{author}{\bibfnamefont{K.}~\bibnamefont{Bechgaard}},
  \bibinfo{journal}{Phys. Rev. B} \textbf{\bibinfo{volume}{25}},
  \bibinfo{pages}{3319} (\bibinfo{year}{1982}).

\bibitem[{\citenamefont{Lee and Naughton}(1998)}]{lee-98}
\bibinfo{author}{\bibfnamefont{I.~J.} \bibnamefont{Lee}} \bibnamefont{and}
  \bibinfo{author}{\bibfnamefont{M.~J.} \bibnamefont{Naughton}},
  \emph{\bibinfo{title}{The Superconducting State in Magnetic Fields: Special
  Topics and New Trends}} (\bibinfo{publisher}{World Scientific, Singapore},
  \bibinfo{year}{1998}), chap.~\bibinfo{chapter}{14}, pp.
  \bibinfo{pages}{272--295}, \bibinfo{note}{(Edited by C. A. R. S{\'a} de
  Melo)}.

\end{thebibliography}

\end{document}